\documentstyle[12pt,epsf]{article}
\input epsf

\begin{document}
\begin{flushright}
{IOA.300/93}\\
\end{flushright}
\newcommand{\newc}{\newcommand}
\newc{\ra}{\rightarrow}
\newc{\lra}{\leftrightarrow}
\newc{\beq}{\begin{equation}}
\newc{\eeq}{\end{equation}}
\newc{\barr}{\begin{eqnarray}}
\newc{\earr}{\end{eqnarray}}
\def\ap{a^\prime}
\def\bp{b^\prime}
\def\al{\alpha}
\def\alp{\alpha^\prime}
\def\be{\beta}
\def\bep{\beta^\prime}
\def\gm{\gamma}
\def\dl{\delta}
\def\la{\lambda}
\def\ka{\kappa}
\def\gmp{\gamma^\prime}
\def\np{\nu^\prime}
\def\Gm{\Gamma}
\def\Sh{\hat S}
\def\mue{({\mu^-},{e^-})}
\def\nop{\nu_1^{\prime}}
\def\nwp{\nu_2^{\prime}}
\def\no{\nu_{1}}
\def\nw{\nu_2}

\def\fpt{F^{\prime2}}
\def\sF{sinF}
\def\sFt{{sin^2}F}
\def\cF{cosF}
\def\cf{cos{F\over2}}
\def\sf{sin{F\over2}}
\def\Wp{w^\prime}
\def\Wpt{w^{\prime2}}
\def\Ft{F_\pi^2}
\def\ft{{\Ft\over2}}
\def\gp{G^\prime}
\def\gpt{G^{\prime2}}
\def\go{\gamma_1}
\def\gw{\gamma_2}
\def\gt{\gamma_3}
\def\Wo{\Omega}
\def\wp{\omega^\prime}
\def\wpt{\omega^{\prime2}}
\def\bet{(\beta^\prime/2)}
\def\pa{\partial}
\def\ur{\underline r}
\def\ut{\underline \tau}
\def\bpp{\beta^{\prime\prime}}


\baselineskip 15pt

\vspace*{10mm}
\begin{center} { \bf LEPTON  FLAVOR  NON-CONSERVATION } \\
\vspace{15mm}
T. S. KOSMAS$^{1,2}$,  G. K. LEONTARIS$^{1}$  AND  J. D. VERGADOS$^{1}$\\
\vspace{8mm}
1  Division of Theoretical Physics, University of Ioannina, GR 451
10, Ioannina, Greece \\
\vspace{3mm}
2  Institute of Theoretical Physics, University of T\"ubingen,
D-72076 T\"ubingen, Germany\\
\end{center}

\vspace{10mm}
\hspace{27.4mm} {\bf ABSTRACT}\\
\vspace{10pt}

In the present work we review the most prominent lepton flavor
violating processes ($\mu \ra e\gamma$, $\mu \ra 3e$, $(\mu , e)$
conversion, $M-\bar M$ oscillations etc), in the context of
unified gauge theories. Many currently fashionable extensions of the
standard model are considered, such as:
{\it i)} extensions of the fermion sector (right-handed neutrino);
{\it ii)} minimal extensions involving additional Higgs scalars
(more than one isodoublets, singly and doubly charged isosinglets,
isotriplets with doubly charged members etc.);
{\it iii)} supersymmetric or superstring inspired unified models
emphasizing the implications of the renormalization group equations in
the leptonic sector.
Special attention is given to the experimentaly most interesting
$(\mu - e)$ conversion in the presence of nuclei. The relevant nuclear
aspects of the amplitudes are discussed in a number of fashionable
nuclear models. The main features of the relevant
experiments are also discussed,  and detailed predictions of the above models
are compared to the present experimental limits.

\newpage
\begin{center}
{ \bf TABLE  OF  CONTENTS }
\end{center}

\bigskip
\bigskip
\noindent
{\bf 1. INTRODUCTION }

\medskip
\noindent
{\bf 2. LEPTON FLAVOR VIOLATING PROCESSES }

\medskip
\noindent
\hspace{5.0mm}{ {\bf 2.1. The $\mu\ra e\gamma $ process}}\\
\smallskip
\hspace{5.0mm}{ {\bf 2.2. The $\mu\ra ee\bar e$ decay}}\\
\smallskip
\hspace{5.0mm}{ {\bf 2.3. The $(\mu, e)$ conversion in the presence of
nuclei}}\\
\smallskip
\hspace{5.0mm}{ {\bf 2.4. Muonium-antimuonium oscillations}} \\
\smallskip
\hspace{5.0mm}{ {\bf 2.5. Lepton flavor violating meson decays}}\\

\medskip
\noindent
{\bf 3. LEPTON FLAVOR VIOLATION  IN GAUGE THEORIES}

\medskip
\noindent
\hspace{5.0mm}{ {\bf 3.1. Minimal extensions of the standard model}}\\
\smallskip
\hspace{10.0mm}{ {\bf 3.1.2.  The extended  higgs sector.} }\\
\smallskip
\hspace{5.0mm}{ {\bf 3.2. The flavor violating decays}}\\
\smallskip
\hspace{10.0mm}{ {\bf 3.2.1. The $\mu\ra e\gamma $ and $\mu\ra 3e $ decay
rates}}\\
\smallskip
\hspace{10.0mm}{ {\bf 3.2.2. $(\mu^- - e^-)$ conversion decay rates.}}\\
\smallskip
\hspace{10.0mm}{ {\bf 3.2.3. $\beta \beta_{0\nu}$ and $(\mu^- - e^+)$
processes}}\\
\smallskip
\hspace{10.0mm}{ {\bf 3.2.4. Other lepton flavor violation mechanisms}}\\
\smallskip
\hspace{5.0mm}{ {\bf 3.3.  Supersymmetric extensions of the standard model}}\\
\smallskip
\hspace{10.0mm}{ {\bf 3.3.1.  Minimal Supersymmetric Standard Model}}\\
\smallskip
\hspace{10.0mm}{ {\bf 3.3.2.  Flavor violation in Superstring Models}}\\
\smallskip
\hspace{10.0mm}{ {\bf 3.3.3.  Flavor violating processes}}\\

\medskip
\noindent
{\bf 4. EXPRESSIONS FOR THE BRANCHING
  RATIO OF $\mue$}

\medskip
\noindent
\hspace{5.0mm}{ {\bf 4.1. The effective Hamiltonian of $(\mu^{-}, e^{-})$
conversion}}\\
\smallskip
\hspace{5.0mm}{ {\bf  4.2. Expressions for the branching ratio of $(\mu^{-},
e^{-})$ conversion}}\\
\smallskip
\hspace{10.0mm}{ {\bf  4.2.1.  Coherent $(\mu^{-}, e^{-})$ conversion} }\\
\smallskip
\hspace{10.0mm}{ {\bf  4.2.2.  Incoherent $(\mu^{-}, e^{-})$ conversion}}\\

\medskip
\noindent
{\bf   5.  EVALUATION OF THE NUCLEAR MATRIX ELEMENTS}

\medskip
\noindent
\hspace{5.0mm}{ {\bf  5.1.  The  coherent  $\mu-e$  conversion
matrix elements}}\\
\smallskip
\hspace{5.0mm}{ {\bf 5.2. Total $\mue$ Conversion branching ratios}}\\
\smallskip
\hspace{10.0mm}{ {\bf 5.2.1. RPA calculations envolving the final states
explicitly}}\\
\smallskip
\hspace{10.0mm}{ {\bf 5.2.2.  Sum-Rules in the context of QRPA} }\\
\smallskip
\hspace{10.0mm}{ {\bf 5.2.3. Sum-rules in the context of shell model}}\\
\smallskip
\hspace{10.0mm}{ {\bf 5.2.4. The $\mue$ conversion in the Primakoff's
method} }\\
\smallskip
\hspace{5.0mm}{ {\bf 5.3.  Discussion of nuclear matrix elements} }\\

\medskip
\noindent
{\bf 6. CONCLUDING REMARKS }

\newpage
\hspace{27.4mm}{ \bf 1. INTRODUCTION }

\bigskip
\noindent
All currently known experimental data are consistent with the standard
model of weak and electromagnetic interactions (SM).  Within the
framework of the SM, baryon and lepton quantum numbers are seperately
conserved.  In fact one can associate an additive lepton flavor quantum
number with each lepton generation which appears to be conserved. There
are thus three such conserved quantum numbers $L_e$, $L_{\mu}$ and $L_{\tau}$
 each  one associated with the lepton
generations $(e^-,\nu_e)$, $(\mu^-,\nu_{\mu})$,
$(\tau^-,\nu_\mu)$, with
 their antiparticles having opposite lepton flavor. It is
in fact these quantum numbers which distinguish between
the three neutrino species if they are massless.

\bigskip
\noindent
Most theorists, however, view the SM not as the ultimate theory of
nature but as a successful low energy approximation.  In possible
extensions of the SM it is legitimate to ask whether lepton flavor
conservation still holds.  In fact in such gauge models (Grand
Unified Theories, Supersymmetric extensions of the SM,
Superstring inspired models) such quantum numbers  are
associated with global (non local) symmetries and their
conservation must be broken at some level.

\bigskip
\noindent
Motivated in part by this belief the search for lepton flavor
violation, which began almost  half a century ago (Hincks and
Pontecorvo, 1948 \cite{HING}, Lagarigue and Peyrou, 1952 \cite{LAGAR},
 Lokanathan and Steinberger, 1955 \cite{LOKA},
see also Frankel, 1975 \cite{FRAN}) has
been revived in recent years and is expected to continue in the near
future.  In the meantime the number of possible reactions for testing
lepton flavor has been increased.  The most prominent such reactions are
\beq
\mu \rightarrow e\gamma\label{eq:mueg1}
\eeq

\beq
\tau \rightarrow e\gamma \qquad and \qquad \tau \rightarrow \mu\gamma
\label{eq:taueg1}
\eeq

\beq
\mu \rightarrow e e^+ e^- , \label{eq:mu3e}
\eeq

\beq
\tau \rightarrow e e^+ e^- , \qquad    \tau \rightarrow \mu e^+ e^-
\eeq

\beq
\tau \rightarrow e \mu^+ \mu^-, \qquad \tau \rightarrow \mu \mu^+\mu^-
\label{eq:tau4}
\eeq

\beq
K_L \ra \mu^{\pm} e^{\mp}, \qquad  K^+ \rightarrow \pi^+ \mu e \label{eq:kmu}
\eeq

\beq
(\mu^+ e^-) \leftrightarrow (\mu^- e^+) \quad
 muonium-antimuonium \quad oscillations
\label{eq:mu6}
\eeq

\beq
 \mu^- (A,Z) \rightarrow e^- (A,Z) \quad (muon-electron \quad conversion)
\label{eq:Az7}
\eeq

\noindent
Finally one could have both lepton and lepton flavor violating
processes like

\beq
(A,Z) \rightarrow (A,Z{\pm}2) + e^{\mp} e^{\mp} \qquad
 ( \beta \beta_{o\nu}
-decay) \label{eq:Az8}
\eeq

\beq
 \mu^- (A,Z) \rightarrow  e^+ (A,Z-2) \qquad
(muon - positron\,\, conversion)
\label{eq:Az9}
 \eeq

\noindent
 From an experimental point of view the most interesting reactions
are (\ref{eq:mueg1}), (\ref{eq:mu3e}), (\ref{eq:Az7}), (\ref{eq:Az8})
and (\ref{eq:Az9}). In this report we will only briefly be concerned
about the last two reactions.

\bigskip
\noindent
The problem of lepton flavor non-conservation is connected with the family
mixing in the lepton sector.  Almost in all models the above process
can proceed at the one loop level via the neutrino mixing.  However,
due to the GIM mechanism in the leptonic sector, the amplitude
vanishes in the limit in which the neutrinos are massless. In some
special cases the GIM mechanism may not be completely operative even
if one considers the part of the amplitude which is independent of
the neutrino mass (Langacker and London, 1988 \cite{GUTS}, Valle, 1991
 \cite{WAL},
Conzalez-Garcia and Valle, 1992 \cite{GWAL}). Even then, however, the process
is suppressed if the neutrinos are degenerate. It should be  mentioned that
processes (\ref{eq:mueg1})-(\ref{eq:Az7}) cannot distinguish between Dirac
and Majorana neutrinos.  Processes (\ref{eq:Az8}) and (\ref{eq:Az9}) can
proceed only if the neutrinos are Majorana particles.

\bigskip
\noindent
In more elaborate models one may encounter additional mechanisms for
lepton flavor violation. In Grand Unified Theories (GUT's) one may
have additional Higgs scalars which can serve as intermediate
particles at the one or two  loop level leading to processes
(\ref{eq:mueg1})-(\ref{eq:Az7}).  In supesymmetric extentions
of the standard model one may encounter as intermediate particles the
superpartners of the above. Lepton flavor violation can also occur in
composite models, e.g. technicolor \cite{DIM}.  In fact,
 such models have already been ruled out by the present experimental
bounds (see next section).

\bigskip
\noindent
The observation of any of the above processes (eqs.
(\ref{eq:mueg1})-(\ref{eq:Az9})) will definitely signal new physics
beyond the standard model.  It will severely restrict most models.  It
may take, however, even then much more experimental effort to unravel
specific mechanisms responsible for lepton flavor violation or fix the
parameters of the models.  The question of lepton flavor
non-conservation has been the subject of several review papers (Scheck,
1978 \cite{SCH}, Costa and Zwirner, 1986 \cite{COS}, Engfer and Walter,
1986 \cite{ENGF}, Vergados, 1986 \cite{VER},
Melese, 1989 \cite{MELE}, Heusch, 1990 \cite{HEN}, Herczeg, 1992 \cite{HER},
Schaaf, 1993 \cite{SCHA}). In the present
review we will focus our attention on recent theoretical developments of
the subject.  We will only cover the essential points of the experimental
situation since we do not intend to dublicate the recent experimental
review which appeared in this journal (Shaaf, 1993 \cite{SCHA}).
Furthermore, the reader
can find an interesting account of the early experiments by Di Lella
\cite{LELL,LELL2}.

\bigskip
\bigskip
{ \bf 2. LEPTON FLAVOR VIOLATING PROCESSES }

\bigskip
\noindent
We have seen in the previous section that lepton flavor  violation, if
it occurs, can be demonstrated  by many  reactions (see eqs.
(\ref{eq:mueg1})-(\ref{eq:Az7})). In this  section we are going to
examine the most basic features of the experimentally most important
processes.

\bigskip
\bigskip
\noindent
{\underline {\bf 2.1.  The $\mu\ra e\gamma $ process}}

\bigskip
\noindent
As we have already mentioned this is the oldest and perhaps the best
studied process.  It was expected to proceed quite fast since the muon
and electron, with the exception of their mass, are identical and
possess identical electromagnetic and weak interactions (for a
historical review see Di Lella, 1993 \cite{LELL}, Vergados,
1986 \cite{VER}). In such early estimates the branching ratio was
(Feinberg, 1958 \cite{FEI})

 \beq
R = \frac{\Gamma(\mu^+ \rightarrow e^+ \gamma)
}{\Gamma(\mu^+ \rightarrow e^+ \nu_e \bar{\nu}_\mu) }
\simeq
\frac{\alpha}{24\pi} \simeq 10^{-4}
\label{eq:gammaeg2.1}
 \eeq

\noindent
which was puzzling since it was already an order of
magnitude larger than the experimental limit at that time.  We will see,
however, in sect. 3.1.1 that we have additional suppression due to the
leptonic analogue of the GIM mechanism.  In purely left-handed theories
the branching ratio takes the form (Vergados, 1986 \cite{VER})

\beq
R = \frac{3\alpha}{32\pi} \,\, [ {\eta}^{(L)}_\nu +
{\eta}^{(L)}_N]^2\label{eq:Reg2.1.1}
 \eeq

\noindent
with

\beq
{\eta}^{(L)}_\nu = \sum_j U^{(11)}_{ej}  U^{*(11)}_{\mu j}
\frac{m^2_j}{m^2_W} , \quad  m_j \ll m_W  \label{eq:Reg2.1.2}
 \eeq

\beq
{\eta}^{(L)}_N = \sum_j U^{(12)}_{ej}  U^{*(12)}_{\mu j}
\left( \frac {m^2_W}{M^2_j} \right) \left [a ln \left( \frac
{M^2_j}{m^2_W} \right) + b \right], \quad M_j \gg  m_W
\label{eq:Reg2.1.3}
 \eeq

\noindent
where $ U^{(11)}_{ej} $  $ \left ( U^{(11)}_{\mu j} \right)$
is the amplitude for producing the light eigenstate with
mass $ m_j \ll m_W$ in the weak interaction of $e^- \, (\mu^-)$.
$ U^{(12)} $ provide the corresponding amplitudes for the heavy neutrino
components. For neutrinos less than $1 MeV$ the term
${\eta}^{(L)}_\nu$ is negligible. The term  ${\eta}^{(L)}_N$
also becomes negligible for very heavy neutrinos.

\bigskip
\noindent
It is clear, therefore, that lepton flavor violating processes are
suppressed. Precisely how suppressed one does not know. Thus,
experimentalists have not been detered from pursuing such hard
experiments. For purely experimental reasons only positive muons have
been considered since, among  other things, they do not undergo capture
by a nucleus as negative muons do (see sect. 2.2).  The experiment
consists in the simultaneous detection of a photon and a positron moving
essentially back to back with momentum $p \sim  m_{\mu} c/2$.

\bigskip
\noindent
In addition to accidental $ e \gamma $ events, which can be minimized by
shielding, the main source of background is radiative muon decay i.e.

\beq
\mu^+\ra e^+ \nu_e {\bar \nu}_\mu \gamma
 \label{eq:Reg2.2}
 \eeq

\noindent
in the kinematic regime in which the neutrinos carry away very little energy.
In spite of the heroic experimental efforts, this process has not been
observed.  The best experimental limit

\beq
R = \frac{\Gamma(\mu^+ \rightarrow e^+ \gamma)}
{\Gamma(\mu^+ \rightarrow e^+ \nu_e \bar{\nu}_\mu)} \,
< \,4.9\times 10^{-11} \qquad (90\% \,\,\, CL)
\label{eq:Reg2.3a}
 \eeq

\noindent
has been set by LAMPF (Bolton {\it et al.}, 1988 \cite{LAMPF}) using the
crystal box  detector. This is almost an order of magnitude improvement over
the previous record, $R \leq 1.7 \times 10^{-10}$  (Kinnison
{\it et al.}, 1982 \cite{KIN}). The proposed limit by the new LAMPF
detector MEGA (magnetic spectrometer with large solid angle) is

\beq
 R \, \sim \, 10^{-12} \label{eq:Reg2.4}
 \eeq

\noindent
the final sensitivity depending on the availability of LAMPF beam time
(MEGA collaboration \cite{AMMA}, Cooper {\it et al.}, 1985 \cite{COOP}).

\bigskip
\noindent
In the same experiment a limit on the branching ratio for two photon
emission was indirectly set

\beq
 R \, \leq \, 7.2 \times 10^{-11} \qquad (90{\%} \,\,\, CL)  \label{eq:Reg2.5}
 \eeq

\noindent
For comparison we mention the corresponding limits for $\tau$
decay obtained by the CLEO collaboration (Bean {\it et al.}, 1993
\cite{BEA})

\beq
R\, = \, \frac{(\tau \rightarrow \mu \gamma) }
{(\tau \rightarrow all) } \, < \, 4.2 \times 10^{-6}
\qquad (90\% \, \,\, CL)
\label{eq:Reg2.6}
 \eeq

\noindent
In the intermediate neutrino mechanism the
branching ratio is still given by eq. (\ref{eq:Reg2.1.1}) with the
modification $U_{ej} \rightarrow U_{\tau j}$ provided of course that we
ignore all chanells other than $ \tau^{\pm}\ra \mu^{\pm}\nu_{\mu}\nu_{\tau}
 $ in  the denominator (otherwise the
$\tau$-mass dependence is complicated).  Even in this case, however, the
branching ratio is expected to be larger crudely speaking by a factor
$(m_{\nu_{\tau}} /m_{\nu_{\mu}})^2$
due to the large coupling of the $\tau$ to
the heaviest neutrino mass eigenstate.  The branching ratio for $\tau
\ra e \gamma$  is $R \leq 1.2 \times 10^{-4}$ obtained by Argus
(Albercht {\it et al.}, 1992 \cite{ARGUS}). For a complete list see
Depommier and Leroy, 1993 \cite{DELE}.

\bigskip
\bigskip
\noindent
{\underline {\bf 2.2. The $\mu\ra ee\bar e$ decay}}

\bigskip
\noindent
In principle, every model, which allows $\mu \ra e\gamma$ to proceed, will
also allow  $\mu \ra e e^+ e^-$, the only difference being that now the
photon can be virtual decaying to an $e^+ e^-$ pair.  In such models one
expects $\mu \ra 3e$ to be suppressed by a power of $\alpha$ i.e.

 \beq
\frac{R(3e)}{ R(e\gamma)} \, \approx \, \frac{\alpha}{24\pi} \,
\approx 10^{-4}  \label{eq:Reg2.7}
 \eeq

\noindent
One can construct models, of course, in which the opposite is true
(see Bilenky and Petcov, 1987 \cite{BILE}). Furthermore, since the photon
is now virtual, one may have a contribution from the $E0$ and
$M0$  form factors \cite{VER}, which vanish at $q^2 = 0$, i.e. for real
photons. We should emphasize, however, that there exist models which allow
$\mu \ra 3e$ but forbit $\mu \ra e \gamma$.
Such are e.g. models in which lepton flavor can be mediated by Higgs
scalars which are doubly charged \cite{VER}. In fact, in such cases
$\mu \ra 3e$ can proceed even at the tree level. In short,
if lepton flavor violating mechanisms exist,
$\mu \ra 3e$ has a better
chance of  being allowed.

\bigskip
\noindent
In addition to the above theoretical considerations, $\mu \ra 3e$
offers a number of experimental advantages as well. One can take
advantage of the three charged particles appearing in the  final
state to reduce the background by a variety of timing and kinemetic
constraints. The chief required qualities for the detector are:  good
energy resolution, time resolution and precise vertex construction.
A detector which meets well these specifications is the SINDRUM at
PSI \cite{BELL}. The best upper limit obtained by such a detector is

 \beq
R = \frac{\Gamma(\mu^+ \rightarrow e^+e^-e^+ )
}{\Gamma(\mu^+ \rightarrow all)} \, < \, 1.0 \times10^{-12}
\qquad (90\% \,\,\, CL)  \label{eq:Reg2.8}
 \eeq

\noindent
It is worth noting that, this limit is a two order of magnitude
improvement over the limit obtained by the same group (Bertl {\it et al.},
1985 \cite{BERT}), which in turn was an order of magnitude improvement  over
that of the Dubna group  (Korechenko {\it et al.}, 1976 \cite{KORE}), which had
stood for about 10 years. It is obvious that, such experiments should be
encouraged to continue.  At least they will provide supplemental
information to  $\mu \ra e \gamma$.

\bigskip
\bigskip
\noindent
{\underline {\bf 2.3. The $(\mu, e)$ conversion in the presence of nuclei}}

\bigskip
\noindent
When a negative muon stops in matter, it finally forms a muonic atom
captured by a nucleus with a radius of about 200 times smaller than
that of the usual atom and a binding energy in the $KeV$ regime
for light and medium nuclei (for heavy nuclei the binding energy
is of the order of a few $MeV$ see below sect. 4.2.1).
After it cascades down to the  $1s$  orbit by emitting x-rays, it
eventually disappears by decay in flight

\beq
\mu^- \rightarrow e^- {\bar \nu}_e \nu_\mu
 \label{eq:Reg2.9}
 \eeq

\noindent
or by capture by the nucleus

\beq
\mu^- (A,Z) \rightarrow  \nu_\mu (A,Z-1)
 \label{eq:Reg2.10}
 \eeq

\noindent
The former dominates for light nuclei but it is negligible for heavy
nuclei $(Z > 10)$.  If lepton flavor is not conserved, one can
encounter the process

\beq
 \mu^- (A,Z) \rightarrow e^-(A,Z)\, \quad (muon-electron\quad conversion)
\label{eq:Reg2.11}
 \eeq

\noindent
The $(\mu^- ,e^-)$ conversion in the presence of nuclei is the most interesting
lepton flavor violating process from an experimental point of view.  The
reason is that, the detection of one particle is sufficient.  No
coincidence is needed.  The energy region of the produced electron is
almost background free. There are, however, some sources of backgound which
have to be taken into account. The first  is the muon disintegration
in orbit mentioned above eq. (\ref{eq:Reg2.9}).  The electron energy in free
muon decay is below $53 MeV$.  The bound muon decay, unfortunately,
yields electron in the high energy tail.  These electrons can be
mistaken with the interesting $( \mu ,e)$ conversion produced electrons due
to the finite energy resolution.  Good energy resolution is thus a
critical and crucial feature of a detector.

\bigskip
\noindent
The second source of backgound is radiative muon capture

\beq
 \mu^- (A,Z) \rightarrow \nu_\mu\gamma (A,Z-1)  \label{eq:Reg2.12}
\eeq

\noindent
followed by pair production. This source of background can be
eliminated if one considers only $(\mu , e)$ conversion leading to the ground
state of the final nucleus. Then, by a judicious choice of the target
nucleus the maximum electron energy following  pair production in
(\ref{eq:Reg2.12}) can be much smaller than that of the monochromatic
electrons of $(\mu ,e)$
conversion  \cite{KoVe}.  The backgound free region can
be as big as $2.5 MeV$.

\bigskip
\noindent
The third source of background, which can be minimized by shielding,
comes from electrons produced by cosmic rays. Fortunately, this kind
of backgound can be studied experimentally, since it is easy to
accumulate good high statistics data on cosmic rays during the beam
off periods.

\bigskip
\noindent
Another troublesome background comes from radiative pion capture which
is followed by pair production. This can lead to copious electrons in
the interesting energy region. Fortunately, the life times for muonic
atoms are quite long (70ns or longer).  So, conversion events can be
seperated from the prompt background.  It is, however, important to
eliminate most pion contamination in the beam with electromagnetic
seperators.

\bigskip
\noindent
It is clear from the above  discussion that, the detectors must have
good energy resolution and large solid angle.  Such detectors with
good detection efficiency must meet conflicting requirements.  One
usually tries to strike a compromise between the expected sensitivity,
which rises with Z, and the signal to background ratio which drops with Z.
It is also important to ensure that the coherent production exhusts a
large fraction of all the conversion electrons.  Fortunately this
happens to be the case (see sect. 5.2.2).

\bigskip
\noindent
The first most efficient such detector for $(\mu , e)$ conversion is
the Time Projection Counter (TPC) developed at TRIUMF (Bryman
{\it et al.}, 1993 \cite{BRY1}). Using a target of $Ti$, dictated by the
above requirements as well as additional experimental constraints, the
TRIUMF collaboration has obtained

\beq
R \, = \, \frac{\Gamma(\mu^-Ti \rightarrow e^-Ti)}{ \Gamma(\mu
\rightarrow all)} \, < \, 4.6 \times10^{-12} \qquad (90\% \,\,\, CL)
\label{eq:Reg2.13}
 \eeq

\noindent
(Ahmad {\it et al.,} 1988 \cite{AHM1}). Using a $Pb$ target the same group
has obtained

\beq
R \, = \, < 4.9 \times10^{-10} \qquad (90\% \,\,\, CL)   \label{eq:Reg2.14}
 \eeq

\noindent
At the same time the experiment found no positron candidate for process
(\ref{eq:Az9}) with excitation energy in the final nucleus  below a few
$MeV$. This has led to the  following limits

\beq
R = \Gamma(\mu^-Ti\rightarrow e^+Ca(gs))
\, < \, 0.9 \times10^{-11} \quad (90\% \,\, CL) \label{eq:Reg2.15}
 \eeq

\noindent
The  total branching ratio depends, of course, on the final nucleus
excitation energy.  For a giant  resonance distribution with a mean
value of $20 MeV$ the authors deduced the limit

\beq
R \, = \, \Gamma(\mu^-Ti\rightarrow e^+Ca (E < 20MeV) )
\, < \, 1.7 \times10^{-10}\quad (90\% \,\, CL)  \label{eq:Reg2.16}
 \eeq

\bigskip
\noindent
Another detector with high sensitivity is the SINDRUM II Spectrometer
(Badertscher {\it et al.}, 1991 \cite{BAD1}).  With this detector, during the
test run, a marginal improvement was achieved for the $(\mu, e)$
conversion branching ratio

\beq
R \, = \, < 4.4 \times10^{-12} \quad (90\% \,\, CL)  \label{eq:Reg2172}
\eeq

\noindent
while the $(\mu^-,e^+)$ has improved by a factor of 2, i.e.

\beq
R (\mu^-Ti\rightarrow e^+Ca(gs)) \, < \, 5.5 \times10^{-12} \quad
(90\% \,\, CL)  \label{eq:Reg217.2}
\eeq

\noindent
The next goal set by SINDRUM \cite{BAD1} is

\beq
R \, \approx \, 3\times 10^{-14}  \label{eq:Reg2.18}
\eeq

\noindent
and by MELC (Djilkibaev and Lobashev, 1992 \cite{DJIL}) is

\beq
R \, \approx \, 10^{-16}
 \label{eq:Reg2.19}
 \eeq

\noindent
It is clear from the above discussion that, $(\mu, e)$ conversion has
definite experimental advantages. The proposed limits may meet the
predictions of realistic models. Furthermore, $(\mu, e)$ conversion,
like  $\mu \ra 3e$ discussed in the previous section, may occur in
a number of models  which do not lead  to  $(\mu \ra e \gamma)$ such
as e.g. those involving the box  diagrams (see sect. 3).  Due to the fact
that, the various mechanisms (discussed in sect. 3) lead to different A
and Z dependence, $(\mu, e)$ conversion, if it is ever observed, may be
able to shed light even on  the detailed mechanisms for lepton flavor
violation.  For this reason it will be discussed in  detail in
sects. 4 and 5.

\bigskip
\bigskip
\noindent
{\underline {\bf 2.4 Muonium-antimuonium oscillations}}

\bigskip
\noindent
The muonium atom $M = (\mu^+ e^-)$ has interesting electromagnetic
properties compared to positronium.  In the presence of lepton flavor
changing interactions transition between muonium and
antimuonium, $\bar{M} = (\mu^- e^+)$, can occur which are the analogue of
the well known oscillations in the $K^0 - \bar{K}^0$  system.  The only
difference is that the  expected oscillation period is much longer compared
to the life-time of $\mu^+$.  Additional complications occur due to
the fact that the  degeneracy of $M$ and $\bar{M}$ may be destroyed in
an external magnetic field  ($\ge 0.05 G$) or due to interactions
with matter.

\bigskip
\noindent
The amplitude for $M-\bar{M}$ oscillations takes the value of a
typical  four-fermion weak interaction multiplied by a scale factor
$n_x$ (Vergados, 1986 \cite{VER}), which depends on the gauge model, and
represents the lepton flavor violating parameter. It is this
parameter which we expect to extact from the oscillation
measurements. Even though the oscillation time is inversely
proportional to $n_x$,

\beq
\tau \simeq 1.3\times10^{-2}n^{-1}_x
 \label{eq:Reg2.20}
 \eeq

\noindent
in the actual experiment one measures the probability for $\bar{M}$
decay  which takes the form

\beq
P_{\bar M} \, \sim \, 2.6\times10^{-5} n^{-2}_x
 \label{eq:Reg2.21}
 \eeq

\noindent
Further reduction can occur in the presence of a magnetic field, as we
have already mentioned above.

\bigskip
\noindent
An important step towards experimental detection of $M- \bar{M}$
oscillation was achieved after the development of techniques to
produce thermal muonium in vacuum (Marshall {\it et al.}, 1982,
1988 and Huber et al., 1990 \cite{MAR}).
Another important step was the ability to measure a low energy atomic
positron in coincidence with the electron of $\mu^-$decay (Mudinger
{\it et al.}, 1988 \cite{MUD}) and the capacity of SINDRUM I (Jungmann
{\it et al.,} 1989 \cite{JUNG}) spectrometer to  increase the electron
solid angle by roughly a factor of 300.

\bigskip
\noindent
In spite of the above important steps, the sensitivity of $M-
\bar{M}$ oscillation experiments is limited not only by the
smallness of $n_x$, but by the additional factor of $2.6 \times
10^{-5}$ of eq. (\ref{eq:Reg2.21}) (Schaaf, 1991 \cite{SCHA}).  It may,
however, become of interest in some special models, especially those for
which the process can occur at tree level (Herczeg and Mohapatra,
1992 \cite{HER-MO}, Vergados, 1986 \cite{VER}).

\bigskip
\noindent
The highest sensitivity has thus far been achieved by LAMPF (Matthias
{\it et al.,} 1991 \cite{MATH}). No candidate events were found in the
experiment which led to the upper limit for $P_{\bar M}$

\beq
P_{\bar M} \, < \, 6.5 \times10^{-7} \quad (90\% \,\,\,CL) \label{eq:Reg2.22a}
\eeq

\noindent
This leads to

\beq
n_{x} \, < \, 0.16  \label{eq:Reg2.22b}
\eeq

\noindent
The SINDRUM collaboration plans to reach a sensitivity of $10^{-10}$
by 1993 with the ultimate of $10^{-11}$ one year later.

\bigskip
\bigskip
\noindent
{\underline {\bf 2.5.  Lepton flavor violating meson decays}}

\bigskip
\noindent
At first sight, the best such process seems to be $\pi ^0 \ra \mu^{\pm}
e^{\mp}$. It turns out, however, that the branching ratio for this
reaction is small, since it has to compete not against a weak competitor
but against the electromagnetic decay $\pi ^0 \ra 2 \gamma$. Thus, the
most prominent such decays are

\beq
K_L \ra \mu^{\pm} e^{\mp}
 \label{eq:Reg2.23}
 \eeq

\beq
K^+ \ra \pi^+\mu e
 \label{eq:Reg2.24}
 \eeq

\noindent
in spite of the fact that the experiments have  to be done in flight
with not so intense beams.  Such experimental efforts are expected to
intensify, if any of the planned meson factories are ever constructed
(TRIUMF, European, Moscow). Historically, such processes have been
favored, because of some kind of prejudice in favor of generation number
conservation.  In other words, a change of generation in the leptonic
sector may be compensated by a corresponding change of generation in
the quark sector.

\bigskip
\noindent
For reasons analogous to those mentioned in the discussion of $\mu \ra
3e$, the reaction (\ref{eq:Reg2.24}) appears to have some advantages.  In
other words, it has the advantage of abundant particle identification,
which may be used to discriminate against background i.e. $\pi^+ \pi^-$
pairs being mistaken as $\mu^+ \mu^-$ pairs.  The best limit ever set
for the branching ratio of process (\ref{eq:Reg2.24}) is

\beq
R = (K^+ \rightarrow \pi^+ \mu e)/(K^+ \rightarrow all)  < 2.1 \times
10^{-10} \quad  (90\% \,\,\, CL)
 \label{eq:Reg2.25}
 \eeq

\noindent
by Lee {\it et al.,} 1990 \cite{LEEL}.
 This is almost an order of magnitude improvement
over the previous limit $R < 1.1 \times 10^{-9}$ (BNL E777 experiment,
Campagnari {\it et al.,} 1988 \cite{CAP}). The branch $\mu^+ e^-$ rather
than the branch $\mu^- e^+$ was selected partly due to the generation
argument mentioned above $(\bar s\leftrightarrow \mu^+,d \leftrightarrow
e^-)$ but mainly due to the fact that the positron background is more
formidable than the electron background.

\bigskip
\noindent
At present, the most sensitive limit comes for the reaction
(\ref{eq:Reg2.23}), i.e.

 \beq
R \,= \, \frac{(K_L \rightarrow \mu^+ e^-)}{(K_L \rightarrow all)}
\, < \, 3.3 \times 10^{-11} \quad (90\% \,\, CL)
 \label{eq:Reg2.26}
 \eeq

\noindent
which has been set by Arisaka {\it et al.,} 1993 \cite{ARI}. Once again,
this is an order of magnitude improvement over the previous experimental
limit,

 \beq
 R \, < \, 3 \times 10^{-10}
 \eeq

\noindent
which has been set by the BNL E791 experiment (Cousins {\it et al.,} 1988
\cite{Cons}).

\bigskip
\bigskip
{\bf 3. LEPTON  FLAVOR  VIOLATION IN  GAUGE  THEORIES}

\bigskip
\noindent
We have pointed out in the introduction that, in the standard model of
electroweak and strong interactions neutrinos remain strictly massless
and lepton flavor is automatically conserved as a global symmetry of
the Lagrangian. In most of the extensions of the standard model, however,
there are various sources of lepton flavor mixing and processes like
$\mu\ra e\gamma$, $\mu\ra ee\bar e$,
$\mu\ra e$ conversion etc., which occur at the one-loop level.
In particular, lepton flavor non-conservation appears in the cases where
these extensions predict non-zero neutrino masses or an extended higgs
sector. Natural candidate theories are Grand Unified Models \cite{PLA} and
Supersymmetric Theories \cite{SUSY}.
In this section, we are going to summarize briefly the most important
extensions of the standard model. We start with a general discussion
of the neutrino mass mechanism when a right handed neutrino is included.
We further discuss models with additional scalar particles
(singly charged isosinglets, doubly charged isosinglets and
isotriplets), and finally we present a brief overview of flavor
violating effects in Supersymmetric and String motivated Grand Unified
models.

\bigskip
\bigskip
\noindent
{\underline {\bf 3.1.  Minimal extensions of the standard model}}

\bigskip
\noindent
{\underline {\bf 3.1.1.  The right handed neutrino.}}
The most obvious way to extend the standard model is to include the right
handed neutrinos. When the right
handed neutrino is present, a non-zero Dirac mass term is possible in
the theory and a  ``Kobayashi-Maskawa'' leptonic mixing matrix appears,
which gives rise to flavor violations.

\bigskip
\noindent
Moreover, since
neutrinos are electrically neutral, they can in principle have Majorana
masses, violating the {\it total} lepton number by two units.
In this latter case, new processes may also occur, namely the
neutrinoless double beta decay ($\beta\beta_{o\nu}$-decay) and
muon to positron conversion in the presence of nuclei
$\mu^{-}+(A,Z)\ra e^++(A,Z-2)$.

\bigskip
\noindent
Thus, in the general case one may have the following new mass terms
in the Yukawa sector of the theory

\beq
{\cal L}_{mass}=\bar \nu^0_Lm\nu^{0c}_R+\bar N^{0c}_Lm_NN^{0}_R
+{\bar \nu}^{0}_Lm_DN^{0}_R+\bar N^{0c}_Lm_D^T\nu^{0c}_R\label{eq:YL}
\eeq

\medskip
\noindent
where, $\nu^{0}_L=(\nu^{0}_e,\nu^{0}_{\mu},\nu^{0}_{\tau})_L$
and  $N^{0}_R=(N^{0}_e,N^{0}_{\mu},N^{0}_{\tau})_R$,
are the left and right handed neutrino weak eigenstates. With

\barr
\nu^{0c}_R=C(\bar \nu^0_L)^T&,&N^{0c}_L=C(\bar N^{0}_R)^T
\earr

\noindent
we denote the conjugate fields, while $m_{\nu},m_{D}, m^T_D, m_N$ are
$3 \otimes 3$ matrices. Thus, the most general neutrino mass matrix is the
$6 \otimes 6$ matrix

\beq
\pmatrix{m_{\nu}&m_D \cr
m_{D}^T & m_N}\label{eq:Reg2.3}
\eeq

\bigskip
\noindent
The matrix  (\ref{eq:Reg2.3}) can be diagonalized by separate left and right
unitary transformations.  Assuming that the masses of the left handed
neutrinos are much lighter than those of the right handed ones,
and labeling their eigenstates $\nu_{jL},$ and
$N_{jR}$, respectively, the transformation for the left
- handed fields is

\beq
\left(\begin{array}{c}\nu^0_L\\
 N^{0c}_L\end{array}\right)
=
\left(\begin{array}{cc}S^{11}_L e^{-i\Lambda_{\nu} }
 & S_L^{12} e^{-i\Lambda_N} \\
S_L^{21} e^{-i\Lambda_{\nu}} & S_L^{22} e^{-i\Lambda_N}\end{array}\right)
 \left(\begin{array}{c}\nu\\ N\end{array}\right)_L
\label{eq:Reg 2.4a}
\eeq

\noindent
For the right-handed fields the transformation is

\beq
\pmatrix{\nu^{0c}_R\cr N^0_R} = \pmatrix{S_L^{11*} e^{-ia}&
S_L^{12*} e^{-i\varphi} \cr
S_L^{21*} e^{-ia} & S_L^{22*} e^{-i \varphi}}
 \pmatrix{\nu \cr N}_R \label{eq:Reg 2.4b}
\eeq

\noindent
In eqs. (\ref{eq:Reg 2.4a}) - (\ref{eq:Reg 2.4b}),
 $\Lambda_{\nu}  ={\Lambda_j(\nu)}$ and $\Lambda_N  ={\Lambda_j(N)}$
are diagonal matrices of arbitrary phases,
while $a = {a_j}$ and $\varphi = {\varphi_j}$ are
phases related to the ${\cal CP}$- eigenvalues of the neutrino
eigenstates as

 \beq
({\cal CP}) \nu_j ({\cal CP})^{-1} = e^{-ia_j} \nu_j\,\, ,\,\,
 ({\cal CP}) N_j ({\cal CP})^{-1} = e^{-i \varphi} N_j
\label{eq:Reg2.55}
 \eeq

\noindent
Due to the presence of neutrino mass, the charged leptonic currents remain
no longer diagonal. Thus the left handed current becomes \cite{VER},

\beq
 j^L_\mu = - 2 \bar{e}^0_L \gamma_{\mu} \nu^0_L + h.c. = -
 2 (\bar{e}_L \gamma_{\mu} U^{11}\nu_L + \bar{e}_L \gamma_{\mu}
U^{12}N_L) + h.c. \label{eq:Reg2.68}
 \eeq

\noindent
with

\beq
 U^{11} = e^{-i \Lambda_e} (S^e_L)^{\dagger}
 S^{11}_L e^{-i \Lambda_{\nu}}
 \eeq

\beq
U^{12} = e^{-i \Lambda_e} (S^e_L)^{\dagger}
 S^{12}_L e^{-i \Lambda_N}
 \label{eq:Reg2.77}
\eeq

\noindent
where $S^e_L$ is the charged lepton mixing matrix.  The charged  current
involving the conjugate fields becomes

 \beq
 (j^L_\mu)^c = - 2 [\bar{\nu}_R \gamma_{\mu} e^{ia} ( U^{11})^T e^c_R +
\bar{N}_R \gamma_{\mu} e^{i \varphi} ( U^{12})^T e^c_R]
\label{eq:Reg2.88}
\eeq

\noindent
Thus, the phases $\alpha_j$, $\varphi_j$ are in principle measurable.  They
appear in processes where the conjugate fields are present $(\beta
\beta_{0\nu}$ - decay and $\mu^- \ra e^+$ conversion).
The right-handed current is modified  analogously

\beq
j^R_{\mu} = 2(\bar e_R \gamma_{\mu}  U^{21} \nu_R +
\bar{e}_R \gamma_{\mu} U^{22} N_R + h.c.) \label{eq:Reg2.90}
\eeq

\noindent
where

\beq
 U^{21} = e^{i(\lambda_e + \Lambda_e)} (S^e_L)^T S^{21*}_L e^{-i
\alpha}
 \eeq

\beq
U^{22} = e^{i(\lambda_e + \Lambda_e)} (S^e_L)^{\dagger} S^{22*}_L e^{-i
\varphi}
 \label{eq:Reg2.100}
\eeq

\noindent
Notice that, in the presence of right-handed currents the $S^{21}$ and
$S^{22}$ matrices are involved.
If the right-handed neutrinos are absent, in the above formulae
$U^{12},U^{21}$ and $U^{22}$ are zero. $U^{11}$ is the leptonic
analogue of the Kobayashi-Maskawa mixing matrix.

\bigskip
\noindent
The above considerations apply in most Grand Unified models
(GUT's) \cite{PLA} where non-zero neutrino masses arise  naturally.
In Supersymmetric GUT's in particular,
motivated by the observed merging of the
Standard Model gauge coupling constants,  there has been a
revived interest in determining the low energy parameters of
the theory, including the unknown neutrino masses and mixing angles
\cite{NEUT}, in terms of a few inputs at the GUT scale.
The general strategy in these approaches is to use the minimal
number of parameters at the GUT scale, so as to have the
maximum number of predictions at $m_W$. Ultimately, one hopes
that this minimal set of parameters at the GUT scale may be
justified in terms of a more fundamental theory, such as the String
Theory. The advandage of such a procedure is that
there are many direct or indirect constraints on the neutrino masses
 from the rest of fermions.
For example, the Dirac neutrino mass $m_{\nu _D}$ is usually
related to the up-quark masses in most of the GUT models.
 It thus appears challenging to
utilize all possible such constraints in the neutrino mass matrix,
in order to make definite predictions for the as yet elusive
neutrinos, which will then be checked by experiment, this way
supporting or excluding such GUT scenarios.
In the next sections we are going to use particular predictive models for
fermion masses in our estimations of the flavor violating branching
ratios.

\bigskip
\bigskip
\noindent
{{\underline {\bf  3.1.2.  The extended  higgs sector.} }
 In this subsection, we are going to review the basic features of minimal
extensions of the standard model based on the introduction of new scalar
particles \cite{ZEE,PET82,LTV,slh}
which are consistent with the gauge symmetry. Neutrino masses are
generated in this case without introducing any right handed neutrino.
 Possible scalars,
which couple in a renormalizable way to leptons, are additional higgs
doublet fields $H^{(k)}=(2,\frac{1}{2})$, $k=1,2,...$, a simply charged
scalar singlet field $S^-=(1,-1)$ \cite{ZEE,PET82,LTV},
a doubly charged state $\Delta^{++} =(1,-2) $ \cite{LTV}
as well as a triplet $T=(3,-1)$. All the above states can couple to
leptons and create diagrams leading to lepton flavor non-conservation.
Their Yukawa  couplings are \cite{LTV}

\beq
 \delta {\cal L} = \lambda_{ij} \bar {\ell}_{iL} \ell^c_{jR} S + d_{ij}
\bar{\ell}_{iL} \frac{{\bf{\tau }}\cdot {\bf T}}{\sqrt{2}} \ell^c_{jR}
+ f_{ij} \bar{e}_{jL}^ce_{iR} \Delta^* +h.c\label{eq:Reg 3.1}
 \eeq

\noindent
It is possible to conserve lepton number
in the above Yukawa Lagrangian, if in the above fields we assign
the following  lepton numbers:

\beq
  L(S) = 2, \,\, L(T) = 2, \,\, L(\Delta) = 2\label{eq:Reg 3.2}
 \eeq

\noindent
But lepton number can be violated explicitly
by cubic as well as quartic terms of the above fields \cite{LTV}

\beq
 \mu (H^{0*},-H^{-*}) \pmatrix{H^{\prime +}\cr H^{\prime
0}} S^- +
  \mu^{\prime} (H^{0*},-H^{-*})
\frac{{\bf{\tau }}\cdot {\bf T}}{\sqrt{2}}
\pmatrix{H^{\prime +}\cr H^{\prime 0}} +...  \label{eq:Reg 3.3}
\eeq

\noindent
where $H'$ is a second doublet and $\mu, \mu',...$ are the qubic
coupling mass parameters. (Notice that $S^-$ couples antisymmetrically to
isospin $1/2$ particles.)

\bigskip
\noindent
A generalization of the above model involves, one
additional higgs doublet $H^{\prime}$, the additional isosinglet fields
$S^{ab}$ and the singlet scalars $\Phi ^{ab}$,  where $a,b$ stand for
$e,\mu ,\tau $. Here, rather than breaking
the lepton flavor explicitly \cite{ZEE,PET82,LTV},
we prefer to follow the approach of references \cite{BH,LVV}
and consider models that above some scale
$V$,  have a global
abelian  lepton flavor symmetry $G=U(1)_e\times U(1)_{\mu } \times
U(1)_{\tau }$. At the scale $V$, the group $G$ is broken spontaneoulsy by
vacuum expectation values
(vev's) of the singlet scalars $\Phi ^{ab}$ giving rise to  Goldstone bosons:
$\Phi ^{e\mu }=e^{{iF_{e\mu } } /{V_{e\mu }}} , \Phi ^{e\tau }=e^{iF_{e\tau
}/V_{e\tau }}$ and  $\Phi ^{\mu\tau }=e^{iF_{\mu\tau }/
V_{\mu\tau }}$.
Thus, the new interactions in the Lagrangian are \cite{BH,LVV}

\beq
 g^{\prime }_al_a e^{ca} H^{\prime }+\lambda
_{ab}l_al_b S^{ab}+ \tilde g_{ab}HH^{\prime } S^{ab}\Phi
^{*}_{ab}.\label{eq:(2)}
\eeq

\bigskip
\noindent
Neutrino masses are generated by the one--loop diagram
shown in fig. 1. The loop calculation for this graph gives:
\beq
m_{ab}={1\over
{16\pi^2}}\lambda_{ab} \, \tilde g_{ba}<\Phi^{ba}>
{\upsilon (g^{\prime }_a m_a +g^{\prime }_b
m_b)\over {M^2_S-m_W^2}}\,ln{M^2_S\over m_W^2}\label{eq:mnu)}
\eeq
\begin{figure}
\begin{center}
\unitlength 1.0cm
\leavevmode
\epsfxsize=13.0 cm
\vspace{5mm}
\caption{ One loop contribution to the neutrino masses in the model with
an extended Higgs sector.}
\end{center}
\end{figure}

\noindent
In the above, $m_a$, $m_b$ are the masses of the charged leptons,
 $M_S$ is the mass of the
heavy charged singlet $S^{ab}$ which appears in the loop, while
$\upsilon$ is the vev of
the standard higgs doublet. We have also assumed that,
the second doublet circulating in the loop, which does not develop
a {\it vev}, has a mass of order $m_W$.

\bigskip
\noindent
In the most general case, where all the singlet fields
$\Phi ^{ab}$ acquire vev's, the  neutrino mass matrix can be
parametrized as follows \cite{LVV}:

\beq
m_{\nu }=m_0\pmatrix{0&tan\theta &cos\phi\cr
tan\theta &0&sin\phi\cr
cos\phi&sin\phi&0}\label{eqno:mnup}
\eeq

\noindent
In the above matrix $m_0$ sets the mass scale and is
given in terms of the various parameters entering eq. (\ref{eq:mnu)})
from the formula

\beq
m_0={1\over{16\pi^2}}{\upsilon (V_{e \tau}^2 +V_{\mu \tau}^2)^{1\over
2}\over {M^2-m_W^2}}m_{\tau} ln{M^2\over m_W^2}\label{eqno:m0}
\eeq

\noindent
where we have made the approximations $m_{\tau } +m_{\mu} \approx
m_{\tau }$  and $m_{\mu} +m_e \approx  m_{\mu}$, while

\beq
V_{ab}=\lambda_{ab} \, \tilde g_{ba} \, g^{\prime}_a<\Phi ^{ba}>
\label{eqno:vab}
\eeq

\noindent
Furthermore, since we know nothing about the couplings $g^{\prime}_a$,
in the last equation we have used the approximation $g^{\prime }_e \approx
g^{\prime }_{\mu} \approx g^{\prime }_{\tau} \approx g^{\prime }$.
Finally, we have defined

\beq
tan\phi={V_{\mu \tau}\over
V_{e \tau}}\label{eqno:tanph}
\eeq

\noindent
and

 \beq
tan\theta = {m_{\mu }V_{e \mu} \over { m_{\tau }
V_{\mu \tau}}}sin\phi\label{eqno:tanth}
\eeq

\noindent
In order to diagonalize the above matrix, guided by phenomenological
reasons, we make the natural assumption  that $sin2\phi <<1$.
In this case, the eigenmasses are found to be

\barr
m_{\nu_1}\approx -\frac{1}{2}m_{0}sin2\theta sin2\phi \\
m_{\nu_2}\approx -{m_0\over cos\theta}+\frac{1}{4}m_{0}sin2\theta
sin2\phi \\
 m_{\nu_3}\approx {m_0\over cos\theta}+\frac{1}{4}m_{0}sin2\theta
sin2\phi \label{eqno:meig}
\earr

\noindent
Furthermore, we get two different  sets of eigenstates, depending upon
whether $sin\phi <<1$ or $cos\phi <<1$.  Thus, in the case where
$cos\phi <<1$, the eigenstates are related to the weak
states as follows \cite{LVV}:

\barr
\nu_e &\approx &
cos{\theta}(sin{\phi}-cos{^2\theta}cos{\phi}sin{2\phi})\nu{_1}\nonumber
\\
&+&
\frac{1}{\sqrt{2}}(sin{\phi}sin{\theta}-cos{\phi})\nu_{2}+
\frac{1}{\sqrt{2}}(sin{\phi}sin{\theta}
+cos{\phi})\nu_{3}\nonumber
\\
\nu_{\mu} &\approx &
cos{\theta}(cos{\phi}-cos{^2\theta}
sin{\phi}sin{2\phi})\nu{_1}\nonumber
\\
&+&
\frac{1}{\sqrt{2}}(cos{\phi}sin{\theta}-
sin{\phi})\nu_{2}+\frac{1}{\sqrt{2}}(cos{\phi}sin{\theta}
+sin{\phi})\nu_{3}\nonumber
\\
\nu_{\tau} &\approx &
-sin{\theta}\nu{_1}+
\frac{1}{\sqrt{2}}cos{\theta}(1-sin{2\phi}sin{\theta})\nu_{2}\nonumber
\\
&+&
\frac{1}{\sqrt{2}}cos{\theta}(1+sin{\theta}sin{2\phi})\nu_{3}
\label{eq:nueig}
\earr

\noindent
A simpified version of the above model arises when one considers
the particular value $cos\phi =0$.  This case
corresponds to the particular choice $<\Phi _{e\tau }>=0$,
which exibits an exact $L_{e-\mu +\tau }$ symmetry. In this latter
case, one finds a zero mass for the first neutrino and two completely
degenerate states for the other two, i.e. \cite{BH,LVV}

$$m_{\nu_1}=0, \quad  m_{\nu_2}=-{m_0\over cos\theta},
\quad m_{\nu_3}={m_0\over cos\theta}\label{eq:311.15}$$
The diagonalizing matrix is given by

\barr
U=\pmatrix{cos\theta&sin\theta\over \sqrt2&sin\theta\over \sqrt2\cr
0&$-$1\over \sqrt2&1\over \sqrt2\cr
$-$sin\theta&cos\theta\over \sqrt2&cos\theta\over \sqrt2}
\label{eq:um}
\earr

\bigskip
\noindent
In the last two equations there are only two parameters, namely, the
mass scale $m_0$ and the angle $\theta $.  If we adopt the MSW-solution
for the solar neutrino problem, and a natural hierarchy between the
$m_{\nu_1}$ and $m_{\nu_2}$ neutrino mass eigenstates, both parameters
can be fixed uniquely. ( We mention here that
an alternative solution to the solar neutrino ``puzzle'' is based on
the assumption of a large neutrino magnetic moment;
for a recent review see Pulido \cite{Pul}.})

\bigskip
\noindent
The model under consideration gives the following formulae for the
various oscillation probabilities

$$P(\nu _e\leftrightarrow \nu _\tau ) \approx {1\over 2 }sin^22\theta
(1-{1\over 2} sin^2{\pi {L\over l_{23}}}) $$

 $$ P(\nu _{\tau}\rightarrow \nu _{\mu }) \approx
cos^2\theta sin^2{\pi {L\over l_{23}}}$$

$$ P(\nu _e\rightarrow \nu _{\mu }) \approx
sin^2\theta sin^2{\pi {L\over l_{23}}}$$

\noindent
Here, the short oscillations have been averaged out. We notice that, the
above oscillation probabilities are expressed solely in terms of two
parameters, $\theta $ and $l_{23}$.
If the MSW-effect is interpreted through $\nu_e\lra \nu_{\mu}$
oscillations, then, from the experimental data \cite{SOLAR} one finds
that

\beq
m_{\nu_2}\approx (1.79-3.46)\times 10^{-3}eV,
\quad sin\theta\approx
(0.71-1.10)\times 10^{-1}\label{eq:sol}
\eeq

\noindent
(see also \cite{SPet}).
We notice, however, that the above model can also accommodate a
relatively ``large'' neutrino mass without creating any particular
problem in low energy phenomenology. As an example, we mention
the solution given by the proposed model \cite{BH,LVV}
to the $17 KeV$ neutrino ``puzzle''.
%

\bigskip
\noindent
Other neutrino mass mechanisms have also been proposed in the context of
Grand Unified Theories. Of particular interest is the Witten mechanism
\cite{Witten}, which is possible \cite{LV91} in all GUT-models
which include the right hand neutrino in a larger representation.
The advandage of this mechanism lies in the fact that the Higgses
need not belong to large representations of the corresponding symmetry.
Other mechanisms giving mass at the two loop level have also been
discussed in models predicting large magnetic moments \cite{2loop}.

\bigskip
\bigskip
\noindent
{\underline {\bf 3.2. The flavor violating decays}}

\bigskip
\noindent
We are now in a position to discuss the various violating processes and
obtain numerical expressions for the corresponding branching ratios. The
processes under consideration are the flavor violating processes $\mu
\ra3\gamma$, $\mu \ra 3e$, ($\mu^- , e^-$) conversion
as well as the lepton number
violating processes ($\mu^- , e^+$)  and  $\beta \beta_{o\nu}$-decay.

\bigskip
\bigskip
\noindent
{\underline {\bf 3.2.1. The $\mu\ra e\gamma $ and $\mu\ra 3e $ decay rates.}}
We start with the familiar $\mu\ra e\gamma$ decay. From the
theoretical point of view, this decay, due to its importance, has been
discussed extensively  in the literature the last two decades \cite{MEG}.
The essential features and experimental limits of this process
have been discussed in sect. 2.1. This decay violates lepton
flavor and can proceed through diagrams which involve massive neutrinos or
new scalars predicted in various extensions of the standard theory.

\bigskip
\noindent
The most general form for the on-shell
$(q^2=0)$  amplitude for  $\mu\ra e\gamma$ is given by

\beq
 {\cal M} (\mu \ra e\gamma)\, = \, {\bf{\bar e}} \,(f_{E1} + f_{M1} \gamma_5)
\imath em_{\mu} \sigma_{\rho\nu} q^{\rho}\epsilon^{\nu}
 \bf{\mu }\label{eq:Reg5.1}
 \eeq

\noindent
where $\epsilon^{\nu}$  is the photon polarization vector and
 $\sigma_{\rho\nu} = \frac {\imath}{2} [\gamma_{\rho}, \gamma_{\nu}]$.
Once the form
factors $f_{E1}$, $f_{M1}$  are given, in a certain gauge model one can
easily obtain the branching ratio with respect to $\mu \ra e + \bar{\nu}_e
+\nu_\mu$  decay i.e.

\beq
 R(\mu\ra e\gamma) = 24\pi^2 (2\pi \alpha)\,
\frac {|f_{E1}|^2 + |f_{M1}|^2}{G^2_Fm_{\mu}^4}\label{eq:Reg5.2}
 \eeq

\noindent
In the mass mechanism there are basically four diagrams contributing to the
amplitude \cite{VER}. We distinguish the following cases:

\bigskip
\noindent
{\it i) Left-handed currents only:}

\noindent
The form factors in this case are

\beq
 |f_{E1}| = |f_{M1}| = \frac {G_F}{\sqrt{2}} \, \frac {m_{\mu}^2}{32\pi^2}
 \left ({\eta^{LL}_\nu + \eta^{LL}_N} \right) \label{eq:Reg5.3}
 \eeq

\noindent
where $\eta^{LL}_\nu $, $\eta^{LL}_N$ are lepton violating parameters defined
for light and heavy neutrinos, respectively, as

\begin{eqnarray}
\eta^{LL}_\nu &=& \sum U^{(11)}_{ei} U^{(11)*}_{\mu i}
 \frac{m^2_i}{m^2_W}\\
 \eta^{LL}_N
 &=& \sum U^{(12)}_{ei} U^{(12)*}_{\mu i} \frac{m^2_W}{M^2_i}
\left( a ln \frac{ M^2_i}{m^2_W} +b \right) \label{eq:Reg5.4}
\end{eqnarray}

\noindent
with $a=2$ and $b=-3$.  The above parameters in eq. (\ref{eq:Reg5.4})
include the mixing which is model dependent.  In the case of a particular
mass matrix  ansatz for the $SO(10)$ model, given for example
in ref. \cite{LV88}, we have

\beq
 \eta^{LL}_\nu \approx 5\times 10^{-20} ,\quad  \eta^{LL}_N \approx 0
\label{eq:Re5.4a}
 \eeq

\noindent
which lead to a suppressed branching ratio

 \beq
 R \sim  10^{-46}   \label{eq:Re5.66}
 \eeq

\noindent
Thus in left-handed theories $\mu \ra e\gamma$ is unobservable.

\bigskip
\noindent
{\it ii) Right-handed currents (R-R couplings):}

\noindent
In this case, we have further suppression due to the presence of the
intermediate boson $W_R$, which is assumed to be much havier than its
left-handed partner

 \beq
 \left ( \frac{m_{W_L}}{m_{W_R}} \right)^2 \, = \, k \leq \, \frac{1}{10}
   \label{eq:Re5.6}
 \eeq

\noindent
Assuming that the mixing $\zeta$ of the two bosons $W_L$, $W_R$ is small enough
$(\zeta \sim 0.1k)$, one obtains the following lepton violating parameters

\begin{eqnarray}
\eta^{RR}_\nu &=&  k^2 \,\sum \, U^{(21)}_{ei} U^{*(21)}_{\mu i}
\, \frac{m^2_i}{m^2_W}\label{eq:Reg5.7a}
\\
\eta^{RR}_N
& =& k^2 \, \sum \, U^{(22)}_{ei} U^{(22)*}_{\mu i} \, \frac{m^2_W}{M^2_i}
\left(a ln \frac{ M^2_i}{m^2_W} +b \right) \label{eq:Reg5.7}
\end{eqnarray}

\noindent
which again lead to an unobservable branching ratio
 $R \le 10^{-36}$ depending on the precise value of $k$.

\bigskip
\noindent
{\it iii) Left - Right mixing:}

\noindent
This is the most flavorable case in the mass
mechanism, since the obtained amplitude has different structure  from the
previous ones.  We get

\beq
 f_{E1,M1}^{LR(RL)} = 6 \,\frac{G_F}{\sqrt{2}} \, \frac{m_{\mu}^2}{32\pi^2}
\left\{ \eta^{LR(RL)}_\nu + \eta^{LR(RL)}_N \right\}
   \label{eq:Re5.69}
 \eeq

\noindent
where $ (\eta^{LR(RL)})_{\nu,N} $ refer to the L(R)-handed coupling for the
$\mu\nu W(e\nu W)$ vertex and to the R(L)-handed coupling for
the $ e\nu W(\mu\nu W) $ vertex. Their expressions are given by

\barr
\eta^{LR}_\nu &=& \zeta \sum U^{(21)}_{ei} U^{(11)*}_{\mu i}
\frac{m_i}{m_\mu} \label{eq:Reg5.9a}
\\
\eta^{RL}_\nu &=& \zeta \sum U^{(11)}_{ei} U^{(21)*}_{\mu i}
\frac{m_i}{m_\mu}\label{eq:Reg5.9b}
\\
 \eta^{LR}_N
 &=& \zeta \sum U^{(12)}_{ei} U^{(22)*}_{\mu i} \frac{m^2_W}{m_\mu M_i}
\left(a ln \frac{ M^2_i}{m^2_W} + b\right)\label{eq:Reg5.9c}
\\
 \eta^{RL}_N
 &=& \zeta \sum U^{(22)}_{ei} U^{(12)*}_{\mu i} \frac{m^2_W}{m_\mu M_i}
\left(a ln \frac{ M^2_i}{m^2_W} + b\right)\label{eq:Reg5.9}
\earr

\noindent
However, even in this most favourable case, in the mass mechanism the
branching ratio turns out to be very small:

 \beq
 R^{L,R} \sim R^{L,R}_N \simeq 10^{-32}   \label{eq:Re5.10}
 \eeq

\noindent
{}From the above discussion it is clear that, the neutrino mass mechanism
is in all cases negligibly small due to the tiny mixing and since
 $m_j\ll m_W$ or $M_j\gg m_W$.

\bigskip
\noindent
{\it iv) extended higgs sector:}

\noindent
It is possible to avoid the suppression mass mechanism by using an
expanded Higgs sector.  In sect. 3.2, we discussed such a possibility by
introducing the isosinglet $S^-$. With the diagrams given in fig. 2 one
can compute the branching ratio for the $\mu\ra e\gamma$ decay \cite{LTV},
which is given in terms of the mass $M_S$ of the isosinglet.
One finds \cite{MEX}

 \beq
 R_S=\frac{\alpha}{48\pi}\frac{1}{G_F^2}
\left(\frac{\lambda_{\mu \tau} \lambda_{\tau e}}{M_S^2}\right)^2
\leq \, 4.9 \times 10^{-11}
\label{eq:Re5.11}
 \eeq

\noindent
which results to a bound for the singlet mass $M_S$

\beq
 M_S\, \ge \,94 GeV \times [ 10^{2}(\lambda_{\mu \tau}
 \lambda_{\tau e})^{1/2} ]
 \label{eq:Re5.12}
 \eeq


\begin{figure}
\begin{center}
\unitlength 1.0cm
\leavevmode
\epsfxsize=16.0 cm
\vspace{5mm}
\caption{
$\mu \ra e \gamma$ via the singlet field $S$. }
\end{center}
\end{figure}

\noindent
$\bullet$
Another interesting process, which can occur with more mechanisms than
$\mu \ra e \gamma$, is the process $\mu \ra {\bar e} e e$.  Firstly, it
can occur in all mechanisms which allow $\mu \ra e \gamma$ decay,
provided that the emitted photon is virtual.
Secondly, this process can be mediated by the neutral boson $Z$, which
finally decays to an $e^+ e^-$ pair. Finally, we can include now box
diagrams \cite{VER}. Similar diagrams are also generated in the case
of the Higgs singlet S.

\bigskip
\noindent
In the mass mechanism we can have contributions from all classes of
diagrams.  We discuss here the most important ones.

\bigskip
\noindent
{1).}  For purely left handed
theories by comparing $\mu \ra {\bar e} e e$ branching ratio
 with the corresponding one for
$\mu \ra e \gamma$ decay, one obtains \cite{VER}

\beq
 R \left( \frac {\mu \ra 3e} {\mu \ra e \gamma}\right)\approx 1.4 \times
10^{-2}\label{eq:Re5.12a}
 \eeq

\noindent
The $Z$ and $W$ diagrams give a larger contribution than that of the
photonic ones in the case of light neutrinos
$R_Z(\mu \ra 3e) \sim 9 \times 10^2 \,R \gamma (\mu \ra 3e)$,
$R_W(\mu \ra 3e) \sim 1.8 \times 10^3\, R \gamma (\mu \ra 3e)$,
but still far from the experimental limit.

\bigskip
\noindent
{2).} With the presence of the right handed currents, the only
important contribution comes from the left - right mixing in the
photonic diagrams.  We get

\beq
 R \, \sim \, 2.7 \times 10^{-5} \, |\eta^{RL}_{\gamma}|^2
\label{eq:Re5.12b}
 \eeq

\noindent
with

\beq
 |\eta^{RL}_{\gamma}| \, = \, |\eta^{LR} + \eta^{RL}|^2 + |\eta^{LR} -
\eta^{RL}|^2 \label{eq:Re5.12c}
 \eeq

\noindent
where

\beq
 \eta^{LR(RL)} \, = \, \eta^{LR(RL)}_{\nu} + \eta^{LR(RL)}_N
\label{eq:Re5.12d}
 \eeq

\noindent
Using the values of the previous sections for the neutrino masses and
mixings one obtains

\beq
 R(\mu \ra 3e) \, \le \, 10^{-35} \label{eq:Re5.12e}
 \eeq

\bigskip
\noindent
{3).} We can also introduce here the singlet $S^-$ to avoid the mass
mechanism suppression.  Again three classes of diagrams are
generated \cite{VER}.  The most important contribution comes from the photonic
ones.  The decay rate is given by

\beq
 \Gamma_S(\mu \ra 3e) = \frac {25}{2}\, m^3_{\mu}\, \frac
{\alpha^3}{\pi} \, |f_{M1}|^2
 \label{eq:Re5.12f}
 \eeq

\noindent
and the branching ratio by

\beq
 R_S(\mu \ra 3e)\, \sim \,0.73
\times 10^{-2} R_S (\mu \ra e \gamma) \label{eq:Re5.12g}
 \eeq

\bigskip
\bigskip
\noindent
{\underline {\bf 3.2.2.  $(\mu^- - e^-)$ conversion decay rates.}}
The next flavor changing process we are going to consider is the
$(\mu^- , e^-)$ conversion in the presence of nuclei.
The diagrams contributing to
this process are similar to those we have considered for $\mu \ra {\bar e}
e e$ decay  (see ref. \cite{VER}).

\bigskip
\noindent
Starting again with the photon diagrams, we write down the decay rate which is

\beq
 \Gamma = 8m_{\mu}^3 \, \frac {Z^4_{eff}}{Z} \alpha^5 \,\frac{E_e p_e}
{m^2_{\mu}} \, \frac{1}{8 \pi} \,\xi^2_0 \,|ME|^2 \label{eq:Re5.12ga}
 \eeq

\noindent
where $ \xi^2_0 = |f_{E0} + f_{M1}|^2 + |f_{M0} + f_{E1}|^2 $ and
$|ME|$ is the nuclear matrix element.
(A detailed discussion will be presented in sects. 4 and 5.)
For a wide range of nuclei and assuming only ground state
transitions the matrix element
 $|ME|\ra M_{gs \ra gs}$ lies in the range

\beq
 |ME| \ra M_{gs \ra gs} \, \simeq \, (0.2 - 0.9)Z^2 \label{eq:Re5.12gb}
 \eeq\

\noindent
In the rest of this section we assume an average value $|ME| \approx 0.5 Z^2$.
The branching ratio is obtained with respect to ordinary muon capture $\mu^-
+ (A,Z) \ra \nu_{\mu} + (A,Z-1)$ decay rate which is \cite{Prima,G-P}

\beq
 \Gamma(\mu \ra \nu_{\mu})= m^5_{\mu} \,
\frac{\alpha^3}{2\pi} \, G^2_F Z^4_{eff}{Z}
 \,[F^2_V + 3F^2_A + F^2_P - 2F_A F_P]\, f(A,Z) \label{eq:Re5.12gc}
 \eeq

\noindent
where $F_V, F_A, F_P$ are form factors and the Primakoff's function
$f(A,Z)$ for nuclei with
$A\approx 2Z$ has the value $f(A\approx 2Z,Z) \approx 0.16$.  Thus

\beq
 R_{\gamma} \left(\frac {\mu^- \ra e^-}{\mu \ra \nu_{\mu}} \right)
\sim 10^{-3}\alpha^2 Z
\frac{E_e p_e} {m^2_{\mu}} \, |n_{\gamma}|^2
\label{eq:Re5.12ge}
 \eeq

\noindent
where $n_{\gamma}$ is related to the lepton violating parameters already
defined in previous processes.
It is obvious that, even in the L-R mixing where
$|n_{\gamma}|^2$  gets its maximum value $\sim 9 \times 10^{-26}$, the
contribution remains small.  Thus, for light neutrinos

\beq
 R_{\gamma} (\mu^- \ra e^-) \sim 10^{-33} Z \,
 \frac{E_e p_e}{m^2_{\mu}} \label{eq:Re5.12gd}
 \eeq

\noindent
For the $Z$-diagrams the basic contribution arises in
the L-L case.  We thus get the branching ratio

\beq
 R_Z (\mu^- \ra e^-) \approx \frac {g^2}{32 \pi^2}
 \frac{E_e p_e} {m^2_{\mu}} \frac {A^2}{8Z} \frac {|F_{ch}(q^2)|^2} {f(A,Z)}
|n^L_Z|^2 \\
\approx 0.34 \times 10^{-2} |n^L_Z|^2 Z^2 \frac{E_e p_e} {m^2_{\mu}}
 \label{eq:Re5.12gf}
 \eeq

\noindent
where

\beq
\eta ^L_Z \approx \left ( \frac {3}{2} + ln \frac {<m^2_q>}{m^2_W}
\right ) \eta^{LL}_{\nu} + \eta^{LL}_N \approx 4 \times 10^{-23}
 \label{eq:Re5.12gg}
 \eeq

\noindent
The quantity $<m^2_q>$ represents the effective quark mass.
In the case of box diagrams, the L-L currents are also the most important
 and the branching ratio is found to be

\beq
R_W (\mu^- \ra e^-) \approx 0.6 \alpha^2 Z\frac{E_e p_e} {m^2_{\mu}}
 (\eta^{\kappa}_W - 4 \eta^a_W)^2
 \label{eq:Re5.12gh}
 \eeq

\noindent
where

\beq
\eta^{\kappa (a)}_W =  \left (1+ ln \frac {<m^2_q>}{m^2_W} \right)
  \eta^{LL}_{\nu} + \eta^{LL}_N \label{eq:Re5.12gi}
\eeq

\noindent
Here $\kappa$  $(a)$ stand for the down (up) quarks.  If we ignore the mixing
in the quark sector and take  $m_{u} \sim m_d \sim 4 \times 10^{-2} GeV$,
we get $R_W \sim 10^{-24} Z E_e p_e/m^2_{\mu}$ for light neutrinos, which is
better that the previous cases.  The presence of the right-handed currents
in the box diagrams again appear to give negligible contribution.

\bigskip
\noindent
Although $\mu - e$ conversion turns out to be negligible in the
above context, as will be discussed in the following sections, one
finds that this process is enhanced in supersymmetric GUTs, when
renormalization group corrections are taken into account.
Let us, however, mention here
that, in theories with extended fermion sector and  a  new neutral gauge
boson $Z'$, one may also have a significant impact in the $\mu -e$
conversion. It is argued \cite{Bern} that in such theories, the present
experimental limits on the above process give a nuclear model independent
bound on the $Z-e-\mu$ vertex, which is twice as strong as that obtained
from $\mu \ra eee$ decay.
In particular, in the case of the $E_6$ models \cite{LepQ},
these limits provide stringent constraints \cite{Bern}
on the mass ($M_Z\ge 5 TeV$) and the mixing angle of  $Z$ and $Z'$.

\bigskip
\bigskip
\noindent
{\underline {\bf 3.2.3.  $\beta \beta_{0\nu}$ and $(\mu^- - e^+)$ processes.}}
We discuss now briefly the lepton number violating processes
eqs. (\ref{eq:Az8}), (\ref{eq:Az9}) (see also ref. \cite{VER,LV83}).

\bigskip
\noindent
$\bullet$ The oldest lepton violating process, intimately related to the
nature of the neutrino, is the neutrinoless ${\beta} {\beta}_{0\nu}$ decay

\beq
(A,Z) \ra (A,Z \pm 2) + e^{\mp} + e^{\mp},\,\, e^-_b +(A,Z) \ra (A,Z - 2)
+e^+ \label{eq:Re5.12gj}
\eeq

\noindent
which together with the allowed $ \beta \beta_{2\nu}$ decay

\beq
(A,Z) \ra (A,Z \pm 2) + e^{\mp} + \left\{ \begin {array}{ll}
2 {\bar \nu}_e\\
2 \nu_e
\end{array} \right\},\,\,
 e^-_b +(A,Z) \ra
(A,Z - 2) + e^+ + \nu_e + {\bar \nu}_e \label{eq:Re5.12g1}
\eeq

\noindent
are the  only modes of some otherwise absolutely stable nuclei.
One finds that \cite{VER}, the life time of these processes is
given by

\beq
T_{1/2} (o\nu) = K_{0\nu}/|n|^2 |ME|^2_{o\nu}
\eeq

\beq
T_{1/2} (2\nu) = K_{2\nu}/ |ME|^2_{2\nu} \label{eq:Re5.12gk}
\eeq

\noindent
where $\eta$ is the relevant lepton violating parameter and $|ME|$ the
corresponding nuclear matrix elements.  The quantities $K_{0\nu}, K_{2\nu}$
are functions of $(A,Z)$ which can take the following values \cite{VER}

\beq
1.5 \times 10^{13} y < K_{0\nu} < 2.5 \times 10^{17} y,
\eeq

\beq
2.5 \times 10^{19} y < K_{2\nu} < 3.3 \times 10^{26} y \label{eq:Re5.12gl}
\eeq

\noindent
Thus,  $\eta$, which contains all information about the gauge models, should
not be much smaller that $10^{-6}$  (the present experimental limit) in
order $ \beta \beta_{o\nu}$ decay to be within the capabilities of present
experiments.  In the $SO(10)$ model \cite{LV88} discussed above,
the best value for $\eta$  is given in
the case of light neutrinos $\eta^{LL}_{\nu} \sim 4 \times 10^{-9}$.
It is not difficult, however, to invent scenarios where
$\beta \beta_{o\nu}$  decay is not far from the experimental capability
of the near future experiments.

\bigskip
\noindent
In particular, there have been recently discussed cases \cite{PeSm}
with three light neutrinos predicting  $\beta \beta_{o\nu}$ decay half-life
in the range of the sensitivity of future experiments, induced by the
exchange of
majorana neutrinos with an effective neutrino mass $\sim (0.1-1.0) eV$.
The interesting feature of these scenarios is that they also provide a
solution to the solar neutrino problem and in
some particular cases, they  can also accommodate a solution to the
atmospheric neutrino problem.

\bigskip
\noindent
$\bullet$ The other interesting lepton number violating process is the
$(\mu^-,e^+)$ conversion

\beq
\mu^-_b +(A,Z) \ra e^+ + (A,Z-2)^*
\eeq

\noindent
The half-life time is given by

\beq
T_{1/2} (\mu^-,e^+) = K'_{0\nu}/ |n'|^2|ME'|^2_{0\nu}
\label{eq:Re5.12gm}
\eeq

\noindent
where, $ K'_{0\nu}(\mu^-,e^+) = 2.2 \times 10^{10} y A^{2/3} /
(Z^4_{eff}/Z) $ or $ 5 \times 10^7 y < K'_{0\nu} < 6 \times 10^8
y$,  for nuclei  ranging from $^{58}Ni$ to $^{12}C$ \cite{LV83}.
  Even though this process is
$10^{10}$ faster than its sister $(e^-,e^+)$, unfortunatelly it must
compete against ordinary muon capture $\mu^- \ra \nu_{\mu}$.  Thus, one
gets the branching ratio

\beq
R = \frac {\Gamma (\mu^-,e^+)}{\Gamma (\mu^-,\nu_{\mu})} \approx
1.5 \times 10^{-21} \frac {|\eta'| |ME'|^2}{Z(1.62 Z/A -0.62)}
\label{eq:Re5.12gn}
\eeq

\noindent
where $\eta'$ is the corresponding lepton violating parameter. Thus, for
$|\eta'|\approx |\eta| \leq 10^{-6}$, this process is unobservable.  Even
if transitions to all final nuclear states are considered \cite {LV83},
$|ME'|^2
\approx 0.1 Z^2 = 100$.  The present experimental limit is
$R<9\times10^{-12}$ (see sect. 2.3).

\bigskip
\bigskip
\noindent
{\underline {\bf 3.2.4. Other lepton flavor violation mechanisms.}}
We mention here some other mechanisms of lepton flavor violation
which are also possible.
Of particular interest is the Bjorken-Weinberg \cite{BW} mechanism,
based on a two loop contribution when additional Higgs bosons are present.
The result is sensitive to the various  unknown parameters
(Higgs mass etc. \cite{Chang}), but it can in
principle fall close to the experimental limit.

\bigskip
\noindent
Another possibility of lepton flavor violation arises, if we extend the scalar
part of the theory with the introduction of a doubly charged singlet
$\Delta^{--}$ already mentioned in the sect. $3.1.2$. The most important
effect of this scalar is in the $\mu \ra 3e$ decay, which is mediated at the
tree level \cite{LTV}. One finds that the branching ratio is

\beq
R_{\Delta}(\mu \ra 3e)=
\frac{1}{2}\left(\frac{c_{e\mu}c_{ee}}{g^2}\right)^2
\left(\frac{m_W}{M_{\Delta}}\right)^4 \eeq

\noindent
which, combined with the experimental limit, gives the following bound

\beq
\frac{M_{\Delta}}{\sqrt{c_{e\mu}c_{ee}}} \, \ge \, 3.2\times 10^4GeV
\eeq

\bigskip
\noindent
Before closing this section, we should recall that,
lepton flavor violating processes play a crucial role in  theories
proposed to solve the hierarchy problem. One such example is given by
the theories
of dymamical symmetry breaking (Technicolor (TC) \cite{WS74} and extended
Technicolor (ETC) \cite{DS79}.)  Flavor changing reactions, both in leptonic
as well as in the quark sector, are found to be incompatible with the
experimental limits in these theories.  It has been shown
recently \cite{AKW86} that,
fixed point or walking technicolor theories can solve the problem of large
flavor violation and bring it down to the experimentally allowed region.

\bigskip
\noindent
The most interesting theory, which can solve the hierarchy problem
of cource, is the theory of Sypersymmetry. Lepton flavor violating
reactions are always present in this case, receiving new contributions
from the supersymmetric partners and renormalization effects. These,
will be discussed in the next section.

\bigskip
\bigskip
\noindent
{\underline {\bf 3.3.  Supersymmetric extensions of the standard model}}

\bigskip
\noindent
One of the most important extensions of the standard theory is the minimal
supersymmetric standard model (MSSM).  Although there is no experimental
evidence of supersymmetry as yet, it is a common belief that supersymmetry
and supergravity play an important  role in the theory of elementary
particles. The main motivation for the incorporation of supersymmery in
the fundamental theory of interactions is the natural solution of the
hierarchy problem \cite{SUSY}.
Moreover, supersymmetry seems to play important role in other issues of the
unification program.  SUSY models predict a longer life-time for
proton, they  play crucial role in models with inflation, while they
predict exact unification \cite{UNI} of the three gauge couplings consistent
with the LEP-data.  Furthermore, superstring theories \cite{GSW},
 which appear today
as the only candidates for a consistent theory unifying  all
fundamental interactions, result to an effective superymmetric theory
in low energy.

\bigskip
\noindent
Flavor changing neutral currents
\cite{fvs,LTV86,camp,KR85,RGFCNC}  present one of the
most important tests of all these low energy supersymmetric theories.
In the following subsections we are going to present a brief review of
the effects of the new sources of flavor violation in the context of the
above theories.

\bigskip
\bigskip
\noindent
{\underline {\bf 3.3.1.  Minimal Supersymmetric Standard Model.}}
In the minimal supersymmetric extension of the standard model one can write
down the following Yukawa couplings,

\beq
 {\cal W} = \lambda_u Q \bar{H} U^C + \lambda_d Q H D^C
+ \lambda_e L H E^C \label{eq:Reg3.1}
 \eeq

\noindent
where $Q$, $D^C$, $E^C$, are the usual superfields which accommodate quarks and
leptons. The potential is given by

\beq
{\cal V}=\sum_i{\mid \frac{\partial {\cal W}}{\partial \varphi_i}\mid}^2
+m_{3/2}^2\sum_i{\mid \varphi_i\mid}^2+A({\cal W}+{\cal W}^*)
+B(\mid \frac{\partial {\cal W}}{\partial \varphi_i}\mid {\varphi_i}+c.c.)
\eeq

\noindent
where $m_{3/2}$  the gravitino mass and $A,B$ are the scalar mass
parameters depending on the details of the superymmetry breaking.
Then, the s-lepton
$6\otimes6$ matrix in the basis $(\tilde e,\tilde e^*)$ takes the form

\barr
\pmatrix{m^2_{3/2}I+m^{\dagger}_e m_e  &\bar A^* m^{\dagger}_e\cr
\bar A m_e & m^2_{3/2}I+m^{\dagger}_e m_e}
\label{eq:3.3}
\earr

\noindent
where $m_e$ is the usual lepton mass matrix and $\bar A=A+2B$.
 In this approximation, the lepton and
s-lepton mixing mass matrices are similar which, in turn, implies that
$S_e^{\dagger} S_{\tilde{e}} = 1$.  This means that, there  is no lepton
flavor violation induced.

\bigskip
\noindent
One can also show that, in this model there is no contribution from neutral
intermediate particles, since at this level either they remain massless
(neutrinos) or degenerate (s-neutrinos).

\bigskip
\noindent
The inclusion of the isosinglet right-handed neutrino,
$N^C =(N^C_L;\tilde{N}_C)$,
can lead to additional terms in the superpotential of the form

\beq
 W_1 = \lambda_N L {\bar H} N^C + \frac{1}{2} M_N N^C N^C \label{eq:Reg3.4}
 \eeq

\noindent
Now, the $6 \otimes 6$  neutrino mass matrix can have both Dirac and
isosinglet majorana mass terms. The corresponding $12\otimes 12$
neutral s-lepton mass matrix becomes analogous to that of the neutrinos.
Even though $S_e^{\dagger} S_{\nu}$ is non-diagonal, lepton flavor violating
 processes are suppressed due to the fact that s-neutrinos are degenerate
\cite{LTV86}.

\bigskip
\noindent
The above results are modified if one goes beyond the tree level and
includes radiative corrections arising from the first term of the
superpotential (eq. (\ref{eq:Reg3.4})) taking into account renormalization
effects \cite{KR85,RGFCNC,KLV89} and considering radiative contributions
to the scalar masses (Squarks, Sleptons etc,) at the one loop level.
Indeed, let us assume that $L_i$ are the quark and lepton superfields and
that the MSSM is extended with the inclusion
of $X$, $Y$, $Z$ additional superfields \cite{KR85}.
Then, one gets the additional Yukawa couplings

\beq
 W' = \lambda_{ij} L_i L_j X + \lambda'_i L_i Y Z
\label{eq:Reg3.4a}
 \eeq

\noindent
The additional terms create the diagrams of fig. 3,  which lead to
radiative contributions of the type

\beq
 \Delta M^2_{L_i} \propto \lambda_{ij}\lambda_{ij} +
 \lambda^{'}_i  \lambda^{'}_i
 \label{eq:Reg3.4b}
 \eeq

\noindent
where $ \lambda_{ij}$ is a $3 \otimes3$ matrix in generation space,
while $\lambda'_i$ is a column vector.

\noindent
Now, if the fields $X$, $Y$, $Z$ are light compared to the Plank scale,
then the above corrections are proportional to a large logarithmic
factor $ln (M_{Pl}/M_{X,Y,Z})$
and the contributions to the scalar masses are
significant.  The origin of flavor violation here lies in the fact that,
due to these contributions, it is no longer possible to diagonalize fermion
and s-fermion mass matrices simultaneously.


\begin{figure}
\begin{center}
\unitlength 1.0cm
\leavevmode
\epsfxsize=10.0 cm
\vspace{5mm}
\caption{ Radiative contributions to the $L_i$ masses.}
\end{center}
\end{figure}

\bigskip
\noindent
Let us consider the above effects in the case of s-leptons. In
the presence of the right handed neutrino, the mass matrices discussed in
the previous section, receive contributions which modify the tree level
results as follows

 \beq
 m_{\tilde {e}}\, m^{\dagger}_{\tilde {e}}\, =\, m^2_{3/2} I +
 m_e m_e^{\dagger} + c\,m_{\nu_D} m_{\nu_D}^{\dagger}
\label{eq:Re3.5}
 \eeq

\noindent
where $m_{\nu_D}$ is the Dirac type neutrino mass matrix as before.  The
coefficient $c=c(t)$ is scale dependent ($t=ln {\mu}$) and
includes the running from the Plank scale down to the scale
where the right-handed neutrino acquires a mass $M_N \sim {\cal
O}(M_{GUT})$, thus $c$ is proportional to  $ln ({M_{Pl}}/{M_{GUT}})$.
Because the dependence of the corrections on the scale is logarithmic, and
the range $M_{Pl} - M_{GUT}$ is not very  large,
 the corrections inducing the flavor mixing in the charged
s-lepton mass matrix can be written as

\beq
 (\Delta m^2_{\tilde {e}})_{ij} \approx \left [ \frac{3m^2_{3/2} + A}{(2\pi
\upsilon sin \beta)^2} ln \frac{M_{Pl}}{M_{GUT}}\right ]\times
(V^{*}m_{\nu_D}^{\delta}V^T)_{ij}
\label{eq:Reg2.29b}
 \eeq

\noindent
where $\upsilon = 246 GeV$ and $\beta = tan^{-1}{(<\bar {H}>/<H>)}$.
 $m_{\nu _D}^{\delta}$
is the diagonalized Dirac neutrino mass matrix at the GUT scale.

\bigskip
\noindent
The most natural candidate models, that the above analysis can apply,
are the models derived from the superstring \cite{MB,ant,al,flny}.
Some particular cases will be considered in the next section.

\bigskip
\bigskip
\noindent
{\underline {\bf 3.3.2.  Flavor violation in Superstring Models.}}
One of the most promising avenues beyond local supersymmetry is
the theory of superstrings \cite {GSW}. Indeed, superstring theory is the best
candidate for a consistent unification of all fundamental forces including
gravity. On the other hand, Grand Unified Theories are incorporated
naturally in the superstring scenario.
Realistic string models suggest that, all gauge interactions should unify
within a simple gauge group,  with a common gauge coupling $g_{String}$
at the String unification scale $M_{String}$, which is found
relatively high and close to the Plank scale \cite{MSTR}

\beq
M_{String}\, \approx \, g_{String}\times M_{Pl} \, \sim \,5.\times 10^{17}GeV
\eeq

\noindent
On the other hand, renormalization group calculations have indicated
that minimal supersymmetric Grand Unified Theories (SUSY-GUTs)  are in
agreement with the precision LEP data when the SUSY-GUT scale $M_G$ is
taken close to $M_G\approx 10^{16}GeV$ \cite{UNI}. Thus, the minimal
supersymmetric standard model (MSSM) cannot probably be derived directly
at the string scale; the above descrepancy between the two scales,
would rather suggest that, either the MSSM should be obtained
through the spontaneous breaking of some intermediate GUT-like gauge group
which breaks at the scale $M_G$, or some extra matter fields are needed
to modify the gauge coupling running.

\bigskip
\noindent
Thus, string motivated non-minimal extensions of the MSSM, contain
additional Yukawa interactions, which may in principle lead to
interesting enhancement of flavor changing neutral processes.
In particular, the radiatively induced lepton flavor violations, discussed
in the previous section, occur naturally in this kind of models.

\bigskip
\noindent
In order to be specific and give some quantitative results, we will
concentrate on some realistic string constructions discussed
extensively in the literature.
These attempts have been made in the context of the free fermionic formulation
of the four dimensional superstrings, and led to the construction of three
types of models. Two of them are characterized
by an intermediate GUT scale
based on the symmetries $SU(5)\otimes U(1)$
\cite{ant} and $SU(4)\otimes O(4)$ \cite{al},
while there is the third type of models \cite{flny}, where the original
string symmetry breaks down to the standard model symmetry times some
additional $U(1)$ factors.

\bigskip
\noindent
In the case of flipped SU(5) model \cite{ant}, the superpotential of the
minimal model (assuming invariance under $H \ra - H$) reads

\barr
{\cal W }= \lambda_1^{ij}F_iF_jh + \lambda_2^{ij}F_i{\bar f}_j\bar h +
\lambda_3^{ij}\bar f_il^c_jh + \lambda_4 HHh \nonumber
   \\
+
 \lambda_5 \bar H \bar H \bar h
+ \lambda_6^{ia}F_i\bar
H \phi_a  +  {\lambda_7} h \bar h \phi_0  +  \mu_{ab}
\phi_a \phi_b
\label{eq:su5sup}
\earr

\noindent
The $F_i+ {\bar f}_i+{l^c_i}$ (i = 1,2,3) are the three generations
of ${\bf 10}$, ${\bf {\bar 5}}$ and singlet representations of
 SU(5) that accommodate the
light matter particles of the Standard Model, $H$ and ${\bar H}$
 are ${\bf 10}$ and ${\bf {\bar {10}}}$ Higgs representations,
 $h$ and ${\bar h}$ are ${\bf 5}$ and ${\bf {\bar 5}}$ Higgs
representations, and the $\phi_0,\phi_a$ ($a = 1,2,3$) are auxiliary
singlet fields.
The first 3 terms in the superpotential eq. (\ref{eq:su5sup}) give masses
to the charge 2/3 quarks $u_i$,
charge -1/3 quarks $d_i$ and charged leptons $l_i$,
respectively. The next two terms split the light Higgs doublets
from their heavy colour triplet partners in a natural way. The sixth
term provides a large element in the see-saw neutrino mass matrix, and
the term $\lambda_7h \bar h \phi_0$ gives the traditional Higgs mixing
parameter.
Under the same assumptions discussed in the previous sections, the induced
mixing in the s-lepton mass matrix, due to renormalization group running
in the range $M_{Pl}-M_{GUT}$, is \cite{feln,hkt}

\beq
 (\delta m^2_{\tilde {e}})_{ij} \approx \left [ \frac{3m^2_{3/2} + A}{(\pi
\upsilon sin \beta)^2} ln \frac{M_{SU}}{M_{GUT}}\right ]\times
(V^{*}m_{u}^{\delta}V^T)_{ij}
\label{eq:Reg2.2b}
\eeq

\noindent
where, due to the fact that up-quark and Dirac neutrino  masses arise from
the same superpotential term, we have substituted $m_{\nu_D}^{\delta}$ with
$m_{u}^{\delta}$, which is the diagonal up-quark mass matrix at the GUT scale.

\bigskip
\noindent
The diagonalizing unitary matrix $V\equiv S_e^{\dagger}S_{\nu}$
is unknown, due to the fact that $m_e$ and $m_d$ mass
matrices are unrelated in the flipped $SU(5)$.
Thus, it is possible that, the mixing may enhance the flavor changing
reactions in this model.

\bigskip
\noindent
In the case of the $SU(4)\otimes O(4) \equiv SU(4)\otimes SU(2)_L\otimes
SU(2)_R$ string
model,  the lepton and KM-mixing matrices are related due to the
fact that leptons and down quarks receive masses from the same
superpotential term.
The superpotential of the model under a $Z_2$ symmetry $H\lra -H$
can be written as follows

\barr
\cal W &=&\lambda_1F_L\bar F_Rh +\lambda_2\bar F_RH\phi_i+\lambda_3HHD+
\lambda_4\bar H\bar HD\nonumber \\
&{}&
+\lambda_5\bar F_R \bar F_R D +
\lambda_6F_L F_LD+\lambda_7\phi_0hh
+\lambda_8\phi^3\label{eq:supe}
\earr

\noindent
The phenomenological implications of the above superpotential
terms have been discussed extensively in previous works \cite{lt}.

\bigskip
\noindent
The minimal supersymmetric version of the model includes
three generations of quarks and leptons, which are accommodated in
$F_L+\bar F_R\equiv (4,2,1)+(\bar 4,1,2)$ representations of the
$SU(4)\otimes SU(2)_L\otimes SU(2)_R$ symmetry. One needs at least
one pair of $H+\bar H\equiv (4,1,2)+(\bar 4,1,2)$ Higgses to realize
the first symmetry breaking $SU(4)\times SU(2)_R\ra SU(3)\times
U(1)_{B-L}$, and one $h\equiv (1,2,2)$ higgs field to provide
the two Weinberg-Salam doublets after the first symmetry breaking.
In addition, a sextet field $D\equiv (6,1,1)$ is needed to produce
a pair of colored triplets $3+\bar 3$, which are going to combine
with the uneaten $d^c_H$ and $\bar d^c_{\bar H}$, to form two
superheavy massive states and avoid fast proton decay.

\bigskip
\noindent
Thus, in the case of the minimal version of this model, one
obtains the mass relations $m_u=m_{\nu_D}$ and
$m_e=m_d$ at the GUT scale, which in turn imply that
the lepton and KM mixing matrices are also equal at $M_{GUT}$.
This fact leads to a definite prediction for the above
processes. Due to the small mixing angles, however, there is
a significant suppresion of flavor violating processes.

\bigskip
\noindent
One novel feature of both models is the generalized see-saw
mechanism \cite{ant,al,GSS}. Indeed, three of the singlet fields
introduced, are used to realize the
see-saw mechanism together with the left and right handed
neutrinos. The see-saw matrix in the basis $(\nu_i, N_i^C, \phi_m)$
takes the form

\barr
m_{\nu }=
\left(\matrix{0&m_{\nu_D}&0\cr
m_{\nu_D}&M^{rad}&M_{\nu ^{c},\phi }\cr
0&M_{\nu ^{c},\phi }&\mu _{\phi }\cr}\right) \label{eq:(3)}
\earr

\noindent
where it is understood that, all entries in eq. (\ref{eq:(3)})
represent $3\times 3$ matrices.

\bigskip
\noindent
The $m_{\nu_D}$, $M_{\nu ^{c},\phi }$, $\mu _{\phi }$ submatrices arise
from the trilinear superpotential terms of the above models.
The contribution $M^{rad}$ is a  majorana type mass matrix for
the right handed neutrino and usually arises from some higher order
non-renormalizable terms. In principle, there are many arbitrary
parameters in  eq. (\ref{eq:(3)}), but within the context of
some recently proposed fermion mass matrix Ans\"{a}tze, as well as
 under some natural assumptions {\cite{ENO,LV93}},
it is possible to reduce the arbitrariness and obtain definite
predictions. In what follows, we are going to use the results of
a fermion mass matrix Ansatz proposed in the context of the above
string derived models to estimate the renormalization effects on the
various flavor violating processes.

\bigskip
\bigskip
\noindent
{\underline {\bf 3.3.3.  Flavor violating processes.}}
{\it i)} We start  again with the simplest decay $\mu \ra e\gamma$. The
diagrams are presented in fig. 4.  We can parametrize the
amplitude with the functions $f_{M1}, f_{E1}$ in the same way as
in sect. 3.1.  Here they are given by

 \beq
f_{M1} = -f_{E1} = - \frac{1}{2} \, \tilde{\eta} \alpha^2 \,
\frac{m^2_{\mu}}{\tilde m^2_{\alpha}} f(x) \label{eq:Reg13}
 \eeq

\noindent
where $\tilde{\eta}$ is the corresponding flavor violating
quantity defined as follows

\beq
\tilde{\eta} = - \frac{(\delta {\tilde m}^2_{{\tilde e} {\tilde
e}^*})_{12}}{\tilde{m}^2_a} \label{eq:Reg13a}
 \eeq

\noindent
and $(\delta {\tilde m}^2_{{\tilde e} {\tilde
e}^*})_{12}$ is determined by the analysis of the previous
section while $\tilde{m}_a$ is the mass of the haviest sparticle
circulating in the loop. If ${\tilde m}_a^2 = m^2_{\tilde e}$
for example, it is given by \cite{CK,GKL}

\beq
{\tilde m}^2_{\tilde e} = m^2_{3/2} + C_{\tilde e}(t)
m^2_{1/2} \label{eq:Reg13b}
\eeq

\noindent
where $C_{\tilde e}(t)$ is a function of the  scale $t = ln \mu$,
and for $\mu \sim m_W$, $C_{\tilde e} \approx 0.50$.
The function $f(x)$ of eq. (\ref{eq:Reg13}),
depends on the ratio $x = {m_{\tilde \gamma}}/{{\tilde m}_a}$,
$m_{\tilde \gamma}$ being the photino mass, and is given by

 \beq
f(x) = \frac {1}{12(1-x)^4} \{ 1+ 2x^3 + 3x^2 - 6x - 6x^2 lnx \}, x =
\frac{m^2_{\tilde{\gamma}}}{\tilde m^2_{\alpha}} \label{eq:Reg14}
 \eeq

\noindent
Now, the branching ratio $R_{e \gamma}$ takes the form

\beq
R_{e \gamma } = \frac {6 \pi} {\alpha}  \frac {|f_{E1}|^2 + |f_{M1}|^2}
{(G_F m^2_{\mu})^2} = |\tilde{\eta}|^2 R_0 \label {eq:Reg15a}
 \eeq

\noindent
where $R_0$ is defined through

\beq
R_0 = \frac {3 \pi \alpha^3 |f|^2}
{(G_F m^2_{\alpha})^2}  \label {eq:Reg15b}
 \eeq

\noindent
As an application, we use the results of ref. \cite{LV93} for the
leptonic mixing angles.
 Taking as an example the initial condition at the
GUT scale, $m_{3/2} = m_{1/2}$, we can obtain the bound
$m_{3/2} \geq 25 GeV$.

\begin{figure}
\begin{center}
\unitlength 1.0cm
\leavevmode
\epsfxsize=14.0 cm
\vspace{5mm}
\caption{
$\mu \ra e \gamma$ in supersymmetric theories.}
\end{center}
\end{figure}

\bigskip
\noindent
{\it ii)} The decay  $\mu \ra 3e$ is treated similarly. The corresponding
diagrams are shown in fig. 5.  We find

\beq
R_{3e} = \frac {|{\tilde \eta}|^2 \alpha^4}{(G_F m^2_{\alpha})^2}
 \, \frac{1}{2} \left \{ (16 ln \frac {m_{\mu}} {m_e} -
\frac{26}{3} ) f^2 - 12fg + 3g^2+ 2f_b^2 + 4gf_b - 8ff_b\right \}
\label{eq:Reg22}
 \eeq

\noindent
with

\beq
g(x) = \frac {1}{36(1-x)^4} \{ 2 - 11x^3 + 18x^2 - 9x + 6x^2 lnx \}
 \label{eq:Reg18c}
 \eeq

\beq
f_b(x) = \frac {1}{8(1-x)^4} \{ 1 - 5x^2 + 4x+2x(x+2) lnx \} ,
x = \frac{m^2_{\tilde{\gamma}}}{\tilde m^2_{\alpha}} \label{eq:Reg18d}
 \eeq

\noindent
Comparing $\mu \ra e \gamma$ with $\mu \ra 3e$ we get

\beq
\frac{R_{3e}}{R_{e \gamma}} \, \simeq \, \frac{a}{24\pi}
\{ 16 ln \frac{m_{\mu}} {m_e} - \frac{26} {3} - \frac{61} {6}
\} \, \sim \, 6.4
\times 10^{-3}  \label{eq:Reg18k}
 \eeq

\begin{figure}
\begin{center}
\unitlength 1.0cm
\leavevmode
\epsfxsize=14.0 cm
\vspace{5mm}
\caption{Diagrams for the $\mu \ra 3 e $ process in
supersymmetric theories.}
\end{center}
\end{figure}

\bigskip
\noindent
{\it iii)} Let us now discuss the $(\mu-e)$ conversion. The amplitude for
this process is given by

\beq
{\cal M} = \left \{ \frac {j^{\lambda}_{(1)} J^{(1)}_{\lambda}} {q^2} +
\frac {j^{\lambda}_{(2)} J^{(2)}_{\lambda}} {m^2_{\mu}}  \, \zeta
 \right\}  \label{eq:Reg16}
 \eeq

\noindent
where the first term corresponds to the photonic and the second
to the non-photonic contributions arising from the box diagrams
(see fig. 6) and $\zeta = {m^2_{3/2}/m^2_{\tilde u}}$.  We find that

\beq
j^{\lambda}_{(1)} = {\bar u}(p_1) (f_{M1} + \gamma_5 f_{E1}) i
\sigma^{\lambda \nu}
\frac{q_{\nu}}{m_{\mu}} + \frac {q^2}{m^2_{\mu}}
 \left (f_{E0} + \gamma_5 f_{M0}\right) \gamma^{\nu}  \left (
 g_{\lambda \nu} - \frac {q^{\lambda} q^{\nu}}{q^2}\right )
\label{eq:Reg17a}
 \eeq

\beq
J^{(1)}_{\lambda} = {\bar N} \gamma_{\lambda} \frac
{1 + \tau_3}{2} N, \,\,\, N = Nucleon
\label{eq:Reg17b}
 \eeq


\begin{figure}
\begin{center}
\unitlength 1.0cm
\leavevmode
\epsfxsize=14.0 cm
\vspace{5mm}
\caption{
Diagrams for the ($\mu^- \ra e^- $) conversion in supersymmetric theories.}
\end{center}
\end{figure}

\noindent
For the box diagrams we obtain
\beq
j^{\lambda}_{(2)} = {\bar u}(p_1)\gamma^{\lambda}
 \frac {1}{2} ({\tilde f}_V +
{\tilde f}_A \gamma_5) u(p_{\mu}) \label{eq:Reg17c}
 \eeq

\beq
J_{\lambda}^{(2)} = {\bar N}\gamma_{\lambda} \frac{1}{2}\left[
(3 + \beta f_V\tau_3) - (f_V +f_A \beta \tau_3)\gamma_5 \right ] N
\label{eq:Reg17d}
 \eeq

\noindent
where $\beta =\beta_0/\beta_1$, with

\begin{eqnarray}
\beta_0 &=&\frac{4}{9}+\frac{1}{9}\frac{m_{\tilde u}^2}{m_{\tilde d}^2}
\\
\beta_1 &=&\frac{4}{9}-\frac{1}{9}\frac{m_{\tilde u}^2}{m_{\tilde d}^2}
\label{eq:Reg18a}
\end{eqnarray}

\noindent
Furthermore,

\begin{eqnarray}
f_{E0} = -f_{M0}& = &- \frac{1}{2}
\tilde{\eta} \alpha^2 g(x) \frac{m^2_{\mu}}{m^2_{3/2}} \\
{\tilde f}_V = - {\tilde f}_A& =& - \frac{\beta_0}{2} \tilde{\eta} \alpha^2
f_b(x) \frac{m^2_{\mu}}{m^2_{3/2}}
\label{eq:Reg18b}
 \end{eqnarray}

\bigskip
\noindent
It is quite hard to write down general expressions containing both photonic
as well as non-photonic diagrams (see sect. 4.1). For the coherent process,
however, one can write down the  ($\mu -e$) conversion rate as follows

  \beq
R_{eN} = \frac {1} {(G_F m^2_{\mu})^2} \, \{ \,|\frac{ m^2_{\mu}}{q^2}f_{M1}
+ f_{E0} + \frac {1}{2} \kappa {\tilde f}_V|^2
+ |\frac{ m^2_{\mu}}{q^2}f_{E1} + f_{M0} +
\frac {1}{2} \kappa {\tilde f}_A|^2 \, \}\, \gamma_{ph} \label{eq:Reg19}
 \eeq

\noindent
where

\beq
\kappa =\left(1+\frac{N}{Z} \, \frac{3-\beta}{3+\beta} \,
\frac{F_N(q)^2}{F_Z(q)^2}\right) \zeta \label {eq:Reg19a}
\eeq

\noindent
and

\beq
\gamma_{ph} = \frac {Z |F_Z(q^2)|^2
}{6f(A,Z)}  \label{eq:Reg20}
 \eeq

\noindent
$F_{Z,N}(q^2)$ are the nuclear form factors to be discussed later and
$f(A,Z)$ is the Primakoff's function (see eq. (\ref{eq:Re5.12gc})).
Under some plausible approximations, we can write the ($\mu - e$)
branching ratio as

\beq
R_{eN} = \frac {1}{2} \frac {|{\tilde \eta}|^2 \alpha^4}{(G_F m^2_{3/2})^2}
\,  \{f - g + \frac {1}{2} f_b \kappa \}^2 \gamma_{ph} \label {eq:Reg220}
 \eeq

\noindent
Assuming that the photino is the lightest supersymmetric particle,
we can take $x\ll 1$, to obtain

\beq
f=\frac{1}{12}, \quad g=\frac{1}{18}, \quad f_b=\frac{1}{8}
\eeq

\noindent
Thus, the two terms from the photonic contribution tend to
cancel and the box diagram  dominates. Comparing
the $\mue$ conversion with the $\mu\ra e\gamma$
decay, we get

\beq
R_{eN}\approx \frac{\alpha}{6\pi}\left (\frac{1}{3}+
\frac{3\kappa}{4}\right)^2\gamma_{ph.}R_{e\gamma}
\label{eq:Reg155}
\eeq

\noindent
Analytic results of the abovebranching ratios are presented in sect. 5.3.

\bigskip
\bigskip
\noindent
\hspace{27.4mm}
{\bf   4.  EXPRESSIONS FOR THE BRANCHING RATIO OF $\mue$}

\bigskip
\noindent
In this section we will discuss the construction at nuclear level
of the effective operators which are responsible for the semileptonic process
$\mu^{-}(A,Z) \rightarrow e^{-} (A,Z)^*$ and derive the expressions
for the branching ratios of this process. If we assume that
the A nucleons of the nucleus (A,Z) interact individually with the muon field
(impulse approximation), then the needed nuclear matrix elements can be
obtained from an effective Hamiltonian $\Omega$ which arises from that of
the free particles, as it is described in the next section.

\bigskip
\bigskip
\noindent
{\underline {\bf 4.1.  The effective Hamiltonian of $(\mu^{-}, e^{-})$
conversion}}

\bigskip
\noindent
As we have seen in sect. 2, the effective amplitude for $(\mu^{-}, e^{-})$
conversion is given by eq. (\ref{eq:Reg16}).
It is not easy to separate the dependence on the nuclear physics from the
leptonic form factors. This can only be done for the coherent mode.
For the general case however, we will discuss separately the photonic and the
non-photonic contributions.
For compactness of our notation, we will write the hadronic currents
given in eqs. (\ref{eq:Reg17b}), (\ref{eq:Reg17d}) in the general form

\beq
J_{\lambda}^{(2)} = {\bar N}\gamma_{\lambda} \left[\tilde g_V
(3 + \beta f_V\tau_3) - \tilde g_A(f_V +f_A \beta \tau_3)\gamma_5 \right ] N
\label{eq:41.a}
 \eeq

\noindent
where

\beq
{\tilde g}_V=\frac{1}{6}, \quad {\tilde g}_A = 0, \quad f_V=1,
 \quad \beta =3 \quad (photonic \,\,\, case) \label{eq:41.b}
\eeq

\beq
{\tilde g}_V={\tilde g}_A= \frac{1}{2},\quad f_V=1,\quad
f_A=1.24   \quad (non-photonic \,\,\, case)  \label{eq:41.c}
\eeq

\noindent
The parameters $\beta$, $f_{E0}$, $f_{E1}$, $f_{M0}$, $f_{M1}$, ${\tilde f}_A$,
${\tilde f}_V$ and $\zeta$ depend, of cource, on the model. Thus, in the
case of the neutrino mediated processes in left handed theories one finds

\beq
\zeta = \frac{G_Fm_{\mu}^2}{\sqrt 2}, \quad
\beta_0 = \left\{
\begin{array}{l}30\\1\end{array}\right\}, \quad
 \beta_1 = \left\{
\begin{array}{ll} 25&(light \,\, neutrinos)\\
5/6&(heavy \,\, neutrinos) \\
\end{array} \right\}
\label{eq:41.1}
\eeq


\bigskip
\noindent
i.e. $\beta =5/6$. Furthermore

\begin{equation}
{\tilde f}_V=\beta_0 f_1, \quad {\tilde f}_A= \beta_0 f_2
\label{eq:41.2}
\end{equation}

\bigskip
\noindent
The quantities $f_{1}$, $f_{2}$, depend on the specific gauge model
and are given in the literature \cite{VER}. For the supersymmetric models,
the corresponding expressions have already been given in sect. 3.2.2.

\bigskip
\noindent
At the nuclear level the relativistic expression of the hadronic current
is not needed. Using the standard non-relativistic limits the relevant nuclear
matrix elements involve the operators

\begin{eqnarray}
\Wo_0 = \tilde g_V{\sum^{A}_{j=1}}\Big (
 {3+f_V\be\tau_{3j}}\Big)
e^{-i{\bf p}_e \cdot {\bf r}_j} ,\quad \quad
{\bf \Wo} =-\tilde g_Af_A{\sum^{A}_{j=1}}\Big (
{\xi +\be\tau_{3j}}\Big) {{{\bf \sigma}_j}\over
\sqrt 3}\,e^{-i{\bf p}_e \cdot {\bf r}_j} \label{eq:41.3}
\end{eqnarray}

\noindent
where the summation is over all nucleons and ${\bf p}_e$ is the momentum of
the outgoing lepton. The factor $ 1/\sqrt{3} $ is introduced to make the two
matrix elements in the total rate expression equal and compensated
by suitable factors elsewhere in the expression. In eq. (\ref{eq:41.3})

\begin{eqnarray}
\xi= f_V/f_A = 1/1.24 \label{eq:41.4}
\end{eqnarray}

\noindent
Assuming that the kinetic energy of the
final nucleus is negligible and taking $m_e \approx 0$, we can write the
magnitude of the momentum ${\bf p}_e$ of the outgoing electron approximately as

\beq
p_e \, = \,\mid {\bf p}_e \mid \,\,\approx\,\, m_\mu -\epsilon_b -(E_f-E_{gs})
\label{eq:41.5}
\eeq

\noindent
where $E_f$, $E_{gs}$ are the energies of the final and ground state of the
nucleus, respectively. $m_{\mu}$ is the muon mass and $\epsilon_b$ is the
binding energy. For coherent processes $E_f=E_i$ and since $\epsilon_b$
is relatively small (the biggest value occurs in lead region where
$\epsilon_b \approx 10 MeV$) \cite{Ford}, we have $p_e \approx m_\mu
\approx 0.53 fm^{-1}$. For incoherent processes $E_f\ne E_i$ and for a
sum-rule approach one can assume a mean energy \cite{Prima,G-P}
for the emitted electron corresponding to a mean excitation energy of the
nucleus ${\bar E}$. In this case $p_e \approx m_\mu - {\bar E}-\epsilon_b$.

\bigskip
\noindent
By expanding the exponential of the operators in eq.(\ref{eq:41.3}) in
terms of spherical Bessel functions $j_l(x)$, we obtain the multipole
expansion of the $\mue$ operator, i.e. the following two types of operators
${\hat T}^{(l,\sigma)J}$

 \beq
{\hat T}_M^{(l,0)J}= \tilde g_V \dl_{lJ} \, \sqrt{4\pi}\,
{\sum_{i=1}^A}
 (3+  \be \tau_{3i} )
 j_l (q r_i)Y_M^l ({\bf {\hat r}}_i)  \qquad
 \label{eq:41.6}
\eeq

\noindent
for the vector part, and

\beq
{\hat T}_M^{(l,1)J}=\tilde g_A \sqrt{\frac{4\pi}{3}}\,\,
{\sum_{i=1}^A} (\xi+ \be \tau_{3i}) j_l(q r_i)
\Big[ {Y^l({\bf {\hat r}}_i) {\bf \otimes} {\bf \sigma}_i} \Big] _M^J
\qquad \label{eq:41.7}
\eeq

\noindent
for the axial vector part. In eqs. (\ref{eq:41.6}), (\ref{eq:41.7}) $q$
represents the magnitude of the momentum transfered $\bf q$ to the nucleus
during the $\mue$ process. In a good appromimation $q \approx p_e$ and
thus,  $q$ is given by the energy conservation eq. (\ref{eq:41.5}).

\bigskip
\bigskip
\noindent
{\underline {\bf  4.2.  Expressions for the branching ratio of $(\mu^{-},
e^{-})$ conversion}}

\bigskip
\noindent
The probability density for converting the bound $\mu^-$ of a muonic atom to
an $e^-$ with momentum ${\bf p}_e$ is given by the Fermi's golden rule

\begin{eqnarray}
\Gamma_{i \ra f} & =& \frac{2\pi}{\hbar} \int d{\hat {\bf p}_e} \left(
\frac{p_{e}}{m_{\mu}} \right)^2 \mid <f \mid \Omega \mid i,\mu> \mid ^2
\Xi
\label{eq:42.1}
\end{eqnarray}

\noindent
where
$\mid i,\mu >$ is the initial state of the system: nucleus $(A,Z) +
\mu^-$ and $\mid f>$ the final state of the system: nucleus $(A,Z)^* + e^-$.
The factor $\left({p_e}/{m_{\mu}}\right)^2$
involves the density of the final states appropriate for normalization of the
wave packet representing the outgoing $e^-$ and ${\hat {\bf p}_e}$
is the unit vector in the direction of the electron momentum.
The quantity $\Xi $ depends on the gauge model (in the photonic
case, for example, it coinsides with $\xi_0$ of  eq. (\ref{eq:Re5.12ga})).
The effect of nuclear recoil has been neglected.
In the above expression

\beq
\mid <f \mid \Omega \mid i, \mu> \mid ^2 =
\mid <f \mid \Omega_0 \mid i,\mu> \mid ^2
+3 \mid <f \mid \vec \Omega \mid i,\mu> \mid ^2
\label{eq:42.2}
\eeq

\noindent
Of special interest, as we shall extensively discuss below, is the partial rate
for the $gs \ra gs$ transitions i.e. the coherent rate.
It has been previously assumed that,
 the $1s$ muon wave function varies very little inside the
light and medium nuclei \cite{Ford}, i.e.  the following approximation
has been used

\beq
\mid <f \mid \Omega \mid i,\mu> \mid ^2
=< \Phi_{1s} >^{2}\mid <f \mid \Omega \mid i> \mid ^2
\label{eq:42.3}
\eeq

\noindent
where

\begin{equation}
< \Phi_{1s}>^{2} \equiv \frac{ \int d^{3}x \vert \Phi_{\mu} ({\bf
x})
\vert^{2} \rho ({\bf x})}{ \int d^{3}x \rho ({\bf x})}
\label{eq:42.4}
\end{equation}

\noindent
In the latter definition,  $\Phi_{\mu} ({\bf x})$ is
the muon wave function and $\rho ({\bf x})$ is the nuclear charge density.
To a good approximation it has been found \cite{Ford}

\begin{equation}
< \Phi_{1s} >^{2} = \frac{\alpha^{3} m_{\mu}^{3}}{\pi} \;
\frac{Z_{eff}^{4}}{Z}
\label{eq:42.5}
\end{equation}

\noindent
($\alpha$ is the fine structure constant)
i.e. the deviation from the behaviour of the wave function at the origin has
been taken into account by $Z_{eff}$.
The above approximation was first used
for the $(\mu^-,\nu_{\mu})$ \cite{Prima,G-P} and
afterwards in analogy to muon capture in the
$\mue$ process \cite{WeiFei,Shank,KoVe}.
Recently \cite{Oset1}, it had been found that eq. (\ref{eq:42.3}) is
not very accurate in ordinary muon capture.
We will therefore present results, both, with and without this
approximation.

\bigskip
\noindent
The branching ratio of the total $\mue$ conversion rate divided
by the total muon capture rate, assuming the approximation of eq.
(\ref{eq:42.3}) for both $(\mu ,e)$ and $\mu ,\nu)$ processes
and only photonic or non-photonic mechanisms, takes the simple
form

\begin{equation}
R_{eN}\, = \,
\frac{\Gamma (\mu,e)}{\Gamma (\mu,\nu)} \, = \, \rho \, \gamma
\label{eq:42.6}
\end{equation}

\noindent
where the quantity $\rho$ contains all the nuclear dependence of $R_{eN}$
and $\gamma$ contains the elementary particle physics. We mention that
$R_{eN}$ is the quantity provided by experiments. Obviously, the effect
of the approximation eq. (\ref{eq:42.3}) on the branching ratio is
expected to be negligible.

\bigskip
\noindent
Another interesting quantity in the study of the $\mu - e$ process is the
ratio of the coherent $\mue$ rate, $\Gamma_{coh} = \Gamma_{gs \ra gs}$,
to the total $\mue$ rate, $\Gamma_{tot}= \sum_f \Gamma_{i
\ra f}$  for all final states $\mid f>$, i.e.

\begin{equation}
\eta \, = \,
\frac{\Gamma _{coh}(\mu^- \ra e^-)}{\Gamma _{total}(\mu^-
\ra e^-)}
\label{eq:42.7}
\end{equation}

\noindent
This will be discussed in detail below.

\bigskip
\bigskip
\noindent
{\underline {\bf  4.2.1.  Coherent $(\mu^{-}, e^{-})$ conversion} }.

\bigskip
\noindent
In the case of the coherent process, i.e. ground state to ground state
($0^+\ra 0^+$) transitions only the vector component of eq. (\ref{eq:42.2})
contributes and one obtains

\begin{equation}
<f \mid \Omega_0 \mid i,\mu>\, = \tilde g_V \,(3 + f_V \beta)
 \, \tilde F(q^2)
\label{eq:421.1}
\end{equation}

\noindent
where ${\tilde F}(q^2)$ is the  matrix element involving the ground state

\begin{equation}
\tilde F(q^2) = \int d^{3}x \{ \rho_{p} ({\bf x}) +
\frac{3-f_V\beta}{3+f_V\beta}
\rho_{n} ({\bf x}) \} e^{- i {\bf q} \cdot {\bf x}}
\Phi_{\mu}({\bf x})
\label{eq:421.2}
\end{equation}

\noindent
($\rho_{p}({\bf x})$, $\rho_{n}({\bf x})$ represent the proton,
neutron densities normalized to Z and N, respectively).
$\tilde F(q^2)$ can also be written as

\begin{equation}
\tilde F (q^2) = \tilde F_{p} (q^2) +
\frac{3-f_V\beta}{3+f_V\beta}
\tilde F_{n} (q^2)
\label{eq:421.3}
\end{equation}

\noindent
where

\begin{equation}
\tilde F_{p,n} (q^2) = \int d^{3}x \; \rho_{p,n} ({\bf x}) \; e^{-
i{\bf q} \cdot {\bf x}}\; \Phi_{\mu} ({\bf x})
\label{eq:421.4}
\end{equation}

\noindent
In eqs. (\ref{eq:421.3}), (\ref{eq:421.4}) we have not made use of the
eq. (\ref{eq:42.3}).
Thus, the coherent rate can be obtained from eq.
(\ref{eq:421.1}) by  first calculating $\tilde F(q^2)$ from eq.
(\ref{eq:421.2}) for a given muon wave function and a given nuclear
density distribution. In ref.
\cite{Chiang} the muon wave function was obtained by solving numerically  the
Schr\"odinger equation taking into consideration the effects of vacuum
polarization and finite nuclear size. Weinberg and Feinberg \cite{WeiFei} used
for the coherent rate the expression of eq. (\ref{eq:42.3}) and estimated the
quantity $Z_{eff}$ as in ref. \cite{Ford}. Shanker \cite{Shank}
using Fermi nuclear distribution included an additional correction
interference term. Using eqs. (\ref{eq:42.3}) and
(\ref{eq:42.5}), for the coherent rate one obtains

\begin{equation}
\vert \tilde F (q^2) \vert^{2} \approx
\frac{\alpha^{3} m_{\mu}^{3}}{\pi} \,
\frac{Z_{eff}^{4}}{Z} \, \vert  ZF_Z(q^2)+
\frac{3-f_V\beta}{3+f_V\beta} \, NF_N(q^2) \vert^{2}
\label{eq:421.5}
\end{equation}

\noindent
with ${F}_{Z}$  $({F}_{N}$) the proton (neutron)
nuclear form factors

\beq
{F}_{Z} (q^2) = \frac{1}{Z} \int d^{3}x \rho_{p} ({\bf x}) e^{- i
{\bf q} \cdot {\bf x}},
\qquad
{F}_{N} (q^2) = \frac{1}{N} \int d^{3}x \rho_{n} ({\bf x}) e^{- i {\bf q}
\cdot {\bf x}}
\label{eq:421.6}
\eeq

\noindent
These nuclear form factors can be calculated by using various models i.e.
shell model \cite{KV88}, quasi-particle RPA \cite{KVCF} etc., or can be
obtained from experimental data whenever possible \cite{Vries}.
The branching ratio in this approximation takes the form

\beq
R_{eN}=\frac{\Xi}{(G_F^2m_{\mu}^2)^2}\, \tilde g_V^2 \,
\left[ \, 1 + \frac{3-f_V\beta}{3+f_V\beta}\,
\frac{F_N(q^2)}{F_Z(q^2)}\right ]^2\gamma _{ph.}
\label{eq:421.7}
\eeq

\noindent
with $\gamma _{ph}$ defined in eq. (\ref{eq:Reg20}). The validity of
the approximation of eq. (\ref{eq:42.3}) will be discussed in sect. 5.3.
The parameters $F_N$, $F_Z$, $\tilde F_p$, $\tilde F_n$ and $Z_{eff}$
 for various nuclear systems appear in
$table \,\, 1$. The corresponding widths in arbitrary units are given in
$table \,\,  2$.

\vspace{0.6cm}
\noindent
{\it Table 1. Parameters needed for calculations of the
coherent $(\mu^-,e^-)$ conversion matrix elements
with: i) the exact muon wavefunction, $\tilde F_p$ and $\tilde F_n$
of eq. (\ref{eq:421.4}), and ii) two-parameter Fermi distribution,
proton and neutron form factors, $ F_Z$ and $ F_N$. The electron momentum
$p_e$ and the effective charge $Z_{eff}$ of a set of nuclei covering
the hole periodic table are also shown. }

\vspace{0.5cm}
\begin{center}
\begin{tabular}{rrllccrr}
\hline
  &    &                            &     & &  &  & \\
A &  Z & $\tilde F_{p} [fm^{-3/2}]$ & $\tilde F_{n} [fm^{-3/2}]$ &
$ F_Z$ & $ F_N$ & $p_e (MeV/c)$ & $Z_{eff}$\\
\hline
  &    &                            &     & &  &  & \\
12  & 6  & 0.86 $10^{-2}$ & 0.86 $10^{-2}$ & 0.77 & 0.77 & 105.067 &  5.74\\
24  & 12 & 0.37 $10^{-1}$ & 0.36 $10^{-1}$ & 0.65 & 0.64 & 105.017 & 10.81\\
27  & 13 & 0.45 $10^{-1}$ & 0.45 $10^{-1}$ & 0.66 & 0.62 & 104.976 & 11.62\\
32  & 16 & 0.66 $10^{-1}$ & 0.63 $10^{-1}$ & 0.62 & 0.59 & 104.781 & 13.81\\
40  & 20 & 0.99 $10^{-1}$ & 0.92 $10^{-1}$ & 0.58 & 0.54 & 104.449 & 16.47\\
44  & 20 & 0.97 $10^{-1}$ & 0.11 $10^0$    & 0.57 & 0.55 & 104.464 & 16.43\\
48  & 22 & 0.11 $10^0$    & 0.13 $10^0$    & 0.55 & 0.52 & 104.258 & 17.61\\
63  & 29 & 0.17 $10^0$    & 0.19 $10^0$    & 0.49 & 0.46 & 103.474 & 21.22\\
90  & 40 & 0.26 $10^0$    & 0.29 $10^0$    & 0.42 & 0.38 & 101.951 & 25.69\\
112 & 48 & 0.29 $10^0$    & 0.36 $10^0$    & 0.36 & 0.33 & 100.749 & 27.86\\
208 & 82 & 0.42 $10^0$    & 0.57 $10^0$    & 0.25 & 0.22 &  95.125 & 33.81\\
238 & 92 & 0.40 $10^0$    & 0.57 $10^0$    & 0.20 & 0.18 &  93.591 & 34.36\\
  &    &                            &     & &  &  & \\
\hline
\end{tabular}
\end{center}

\bigskip
\vspace{0.6cm}
\noindent
{\it Table 2. Coherent widths (in arbitrary units)
for the photonic and non photonic mechanisms: (i) with the exact
muon wave function, $\Phi_{\mu}({\bf x})$ (ii) with $<\Phi_{\mu}({\bf x})>$
in the approximation of $Z_{eff}$. The
ratios of the width in case (i) to the width of ordinary muon capture
with arbitrary normalization are also given.}
\vspace{0.7cm}

\begin{center}
\begin{tabular}{rrlccccc}
\hline
  &    &                      &     & &  &  & \\
 & & \multicolumn{3}{c}{non-photonic mechanism ($\beta=5/6$})&
\multicolumn{3}{c}{photonic mechanism ($\beta=3$)}\\
\hline
  &    &                            &     & &  &  & \\
 &    &  With                          &  With    &With &With   &  & \\
A & Z &  $\Phi_{\mu}({\bf x})$ &  $<\Phi_{\mu}({\bf x})>$ & Ratio
      &  $\Phi_{\mu}({\bf x})$ &  $<\Phi_{\mu}({\bf x})>$ &
Ratio\\
\hline
  &   &   &    &            &   &    &       \\
12 &6 & 0.52 $10^{-4}$& 0.51 $10^{-4}$ & 0.14& 0.21 $10^{-4}$
& 0.21 $10^{-4}$ & 0.058\\
24 &12& 0.91 $10^{-3}$& 0.90 $10^{-3}$ & 0.19& 0.38 $10^{-3}$
& 0.37 $10^{-3}$ & 0.078\\
27 &13& 0.14 $10^{-2}$& 0.14 $10^{-2}$ & 0.20& 0.56 $10^{-3}$
& 0.55 $10^{-3}$ & 0.083\\
32 &16& 0.29 $10^{-2}$& 0.28 $10^{-2}$ & 0.22& 0.12 $10^{-2}$
& 0.12 $10^{-2}$ & 0.093\\
40 &20& 0.64 $10^{-2}$& 0.62 $10^{-2}$ & 0.26& 0.27 $10^{-2}$
& 0.26 $10^{-2}$ & 0.110\\
44 &20& 0.72 $10^{-2}$& 0.69 $10^{-2}$ & 0.40& 0.26 $10^{-2}$
& 0.25 $10^{-2}$ & 0.150\\
48 &22& 0.94 $10^{-2}$& 0.91 $10^{-2}$ & 0.36& 0.35 $10^{-2}$
& 0.34 $10^{-2}$ & 0.140\\
63 &29& 0.21 $10^{-1}$& 0.19 $10^{-1}$ & 0.36& 0.78 $10^{-2}$
& 0.72 $10^{-2}$ & 0.140\\
90 &40& 0.47 $10^{-1}$& 0.42 $10^{-1}$ & 0.54& 0.17 $10^{-1}$
& 0.16 $10^{-1}$ & 0.200\\
112&48& 0.62 $10^{-1}$& 0.54 $10^{-1}$ & 0.62& 0.22 $10^{-1}$
& 0.19 $10^{-1}$ & 0.220\\
208&82& 0.13 $10^0$   & 0.89 $10^{-1}$ & 0.98& 0.41 $10^{-1}$
& 0.29 $10^{-1}$ & 0.310\\
238&92& 0.12 $10^0$   & 0.73 $10^{-1}$ & 0.92& 0.35 $10^{-1}$
& 0.22 $10^{-1}$ & 0.280\\
  &   &   &    &            &   &    &       \\
\hline
\end{tabular}
\end{center}

\bigskip
\noindent
{\underline {\bf  4.2.2.  Incoherent $(\mu^{-}, e^{-})$ conversion.}}
As we have mentioned in sect. 2.3, from the experimental
point of view, the coherent contribution to the
$(\mu^-, e^-)$ conversion is the most interesting.
It is, however, important to know what fraction of the total
rate goes into the coherent mode and how this varies with A and Z.
In this section we will
evaluate the incoherent contribution and will compare it to that of the
coherent channel.

\bigskip
\noindent
For the calculation of the contributions leading to the exited states
many methods exist which involve various approximations.
Those which are based on the aproximation inserted by eq. (\ref{eq:42.3})
will be discussed in detail in sect. 5. In this section we
will elaborate on a method recently developed by Chiang {\it et
al.,} \cite{Chiang} and which uses the exact muon wave function in
eq. (\ref{eq:42.1}). This method, known as "nuclear matter mapped into
nuclei by means of a local density approximation", is accurate and leads to a
good reproduction of the $\mu^-$ capture rates over the periodic table
once the proper renormalization of the weak currents is considered.

\bigskip
\noindent
In the ordinary muon capture only the incoherent channel is
open. In this process the total rate $\Gamma _{\mu c}$
is given by \cite{Oset1}

\begin{equation}
\Gamma_{\mu c} = \int d^{3}r \vert \Phi_{\mu} ({\bf x}) \vert^{2} \quad
\Gamma (\rho_{n} ({\bf x}), \rho_{p} ({\bf x}))
\label{eq:422.1}
\end{equation}

\noindent
where $\Gamma (\rho_{n}, \rho_{p})$ is the width in an infinite
slab of nuclear matter with neutron and proton densities $\rho_{n},
\rho_{p}$, respectively. This width can be obtained by means of
the Lindhard function as

\begin{equation}
\Gamma (\rho_{n}, \rho_{p}) = -2 \int \frac{d^{3} p_{\nu}}{(2 \pi)^{3}}
\Pi_{i} \frac{2 m_{i}}{2 E_{i}} \bar{\Sigma} \Sigma \vert T \vert^{2} \;
Im \bar{U}_{p,n} (p_{\mu} - p_{\nu})
\label{eq:422.2}
\end{equation}

\noindent
where T is the transition matrix, i.e. the amplitude associated
with the elementary process $\mu^-p\ra n\nu_{\mu}$ and $m_{i}, E_{i}$
are the masses, energies of the particles involved in this reaction.
The Lindhard function  $\bar{U}$ corresponds to ph excitations of
the p-n type and is given by (see e.g. ref. \cite{Oset2,Oset3}
for its usage)

\begin{equation}
\bar{U}_{1,2} (q) = 2 \int \frac{d^{3} p}{(2 \pi)^{3}}
\frac{n_{1} ({\bf p}) [1 - n_{2} ({\bf q} + {\bf p})]}{q^{0} +
E_{1} ({\bf p}) - E_{2} ({\bf q} + {\bf p}) + i \epsilon}
\label{eq:422.3}
\end{equation}

\noindent
with $n_{1} ({\bf q}')$ and $n_{2} ({\bf q})$ the integral (0 or 1)
occupation probabilities of the neutron and proton, respectively.

\bigskip
\noindent
In an analogous formalism, the incoherent $(\mu^{-}, e^{-})$ conversion rate,
$\Gamma_{inc}(\mu^- \ra e^-) = \sum_{f \ne gs} \Gamma_{i \ra f}$, can
be expressed as follows

\begin{eqnarray}
\Gamma_{inc} (\mu^{-}  A \rightarrow e^{-} X )
& = &
- 2 \int d^{3}x \vert
\Phi_{\mu} ({\bf x}) \vert^{2} \int \frac{d^{3} p_{e}}{(2 \pi)^{3}}
\Pi_{i} \frac{2 m_{i}}{2 E_{i}}\nonumber
\\
&\times &
 [ \bar{\Sigma} \Sigma \vert T \vert^{2} (\mu^{-} p
 \rightarrow
e^{-} p) Im \bar{U}_{p,p} (p_{\mu} - p_{e})\nonumber
\\
&+ &\bar{\Sigma} \Sigma
\vert T \vert^{2} (\mu^{-} n \rightarrow e^{-} n) Im \bar{U}_{n,n}
(p_{\mu} - p_{e}) ]
\label{eq:422.4}
\end{eqnarray}

\noindent
The two terms in the brackets of eq. (\ref{eq:422.4}) result from the
character of the $\mue$ operator to be of charge conserving type.

\bigskip
\noindent
One can separate the
nuclear dependence from the dependence on the elementary sector
in the two processes, $(\mu^-,\nu_{\mu})$ and  $\mue$, by factorizing
outside the integrals in eqs. (\ref{eq:422.1}), (\ref{eq:422.4})
an average value of the quantity $\bar{\Sigma} \Sigma \vert T \vert^{2}$.
For the ratio of the incoherent $\mue$ rate divided by the total muon
capture rate, since $\bar{\Sigma} \Sigma \vert T \vert^{2}$ for the
first process is proportional to $E_{\mu} E_e $ while for the ordinary
muon capture it is proportional to $E_{\mu} E_{\nu}$, we get

\begin{eqnarray}
R \, =\, \frac{\Gamma_{inc}}{\Gamma_{\mu c}}
&=& \frac{m_{p}
m_{e} [({E_{\mu} E_{e}})^{-1} \bar{\Sigma} \Sigma \vert T \vert^{2}
(\mu^{-} p \rightarrow e^{-} p)]_{av}} { m_{n} m_{\nu}
[ ({E_{\mu} E_{\nu}})^{-1} \bar{\Sigma} \Sigma \vert T \vert^{2} (\mu^{-}
p \rightarrow n \nu_{\mu})]_{av}}  G_{p} (Z, N)\nonumber
\\
 &+& \frac{m_{n} m_{e} [ ({E_{\mu} E_{e}})^{-1} \bar{\Sigma}
\Sigma \vert T \vert^{2} (\mu^{-} n \rightarrow e^{-} n)]_{av} }{m_{p}
 m_{\nu} [ ({E_{\mu} E_{\nu}})^{-1} \bar{\Sigma} \Sigma
\vert T \vert^{2} (\mu^{-} p \rightarrow n \nu_{\mu})]_{av}}
\frac{N}{Z} G_{n} (Z, N)
\label{eq:422.5}
\end{eqnarray}

\noindent
The quantities $G_{p}, G_{n}$ are smooth functions of the
momenta and contain the nuclear dependence of the ratio R. They
are defined as

\begin{equation}
G_{p} (Z, N) = \frac{\int d^{3}x \vert \Phi_{\mu} ({\bf x}) \vert^{2}
{(2 \pi)^{-3}}\int {d^{3} p_{e}} Im \bar{U}_{p,p} (p_{\mu} - p_{e})}
{\int
d^{3}x \vert \Phi_{\mu} ({\bf x}) \vert^{2} {(2\pi)^{-3}}\int {d^{3} p_{\nu}}
Im \bar{U}_{p,n} (p_{\mu} - p_{\nu})}
\label{eq:422.6}
\end{equation}

\begin{equation}
G_{n} (Z, N) =  \frac{Z}{N} \frac{\int d^{3}x \vert \Phi_{\mu} ({\bf x})
\vert^{2}
{(2 \pi)^{3}} \int {d^{3} p_{e}} Im \bar{U}_{n,n} (p_{\mu} - p_{e})}
{\int d^{3}x \vert \Phi_{\mu} ({\bf x}) \vert^{2}
{(2 \pi)^{3}} \int {d^{3} p_{\nu}} Im \bar{U}_{p,n} (p_{\mu} - p_{\nu})}
\label{eq:422.7}
\end{equation}

\bigskip
\noindent
Note that $R$  differs from the branching ratio $R_{eN}$, because
R  does not include the coherent part of the $\mue$ process. The calculation
of R in various nuclei is based upon $G_{p}$ and $G_{n}$, since the
$\vert T \vert^{2}$ at the elementary level is in most models
independent of the nuclear parameters (see ref. \cite{Oset1} for values
of $\vert T \vert^{2}$ in ordinary muon capture). As an example,
we give here the ratio $R$ in the non-photonic case \cite{VER} involving the
box diagrams of sect. 3.3.3 which is

\begin{equation}
R \, = \,
\frac{\Gamma_{inc} (\mu^-,e^-)}{\Gamma (\mu^-,\nu)} =
\frac{f^{2}_{1} + f^{2}_{2}}{2} \,
\left(\frac{ \beta_0(3+\beta)}{2} \right)^2 \,G (A,Z)
\label{eq:422.8}
\end{equation}

\noindent
All the nuclear information is contained in the function $G(A,Z)$
defined as

\begin{eqnarray}
G (A,Z) &\equiv &G_{p}
(Z,N)\left [f_V^2+3f_A^2\left(\frac{\xi+\beta}{3+\beta}\right)^2\right ]
\nonumber
\\
 &+& G_{n} (Z,N)\frac{N}{Z}\left[f_V^2 \left (
\frac{3 - \beta}{3 + \beta} \right )^{2}+3f_A^2
\left (\frac{\xi - \beta}{3 + \beta}\right )^2\right ]
\label{eq:422.9}
\end{eqnarray}

\bigskip
\noindent
We should mention that the quantities $G_p$, $G_n$ contain
Pauli blocking corrections and for the ordinary $\mu^{-}$
capture reaction the Q value, which is significant for light nuclei,
is also considered in the argument of the Lindhard function
in eq. (\ref{eq:422.3}).
In heavy nuclei the Q value is approximately given in terms of
the difference between the neutron and proton Fermi energies
and no other correction is needed.

\bigskip
\noindent
As a summary we would say that, the use of an exact
muon wave function for the calculation of the coherent rate,
gives more accurate results in the medium and heavy nuclei region.
It involves an approximation (local density approximation) for the sum of
the non-coherent contributions. It uses neither closure nor explicit
summation over all final states. It makes a summation over
the continuum of excited states in a local Fermi sea.
 Its accuracy is tied to the number of states one can excite; the
larger the better. The muon mass provides the energy for the excitation of
such states. Furthermore, the method is quite simple to apply in actual
calculations.

\bigskip
\bigskip

{\bf   5.  EVALUATION OF THE NUCLEAR  MATRIX ELEMENTS}

\bigskip
\noindent
In this section we will discuss the methods of calculating the nuclear
matrix elements needed for the partial and total $\mue$ conversion rates. It
is known that the main feature of the (photonic) $\mu$-e conversion rates is
the strong Z-dependence (see below eq. (\ref{eq:51.1}) )
accounted for by assuming that all protons can interact
independently with the muon  and that this interaction is proportional to the
muon density in the 1s atomic orbit at the position of the nucleus.

\bigskip
\noindent
In this case, as we have explained in sect. 4, we can assume that,
the probability density for converting a bound muon in the 1s orbit to an
electron of momentum $p_e$, is approximately analogous to the muon
average probability density
over the nucleus, $<\Phi_{1s}>^2$, and the nuclear matrix elements $M^2_{fi}$
between the initial and final nuclear states. Thus, in  order to find the
partial rate $\Gamma_{i \ra f}$ or the total rate $\Gamma = \sum_f
\Gamma_{i \ra f}$,
one needs to evaluate the contribution of the nuclear matrix elements
$|<f| \Omega |i>|^2$ for the states $ |f> $.

\bigskip
\bigskip
\noindent
{\underline{\bf  5.1.  The  coherent  $\mu-e$  conversion
matrix elements}}

\bigskip
\noindent
In the approximation in wich
one factorizes outside the integral of eq. (\ref{eq:421.2}) the average
value of the muon wave function $\Phi_{\mu}(\vec x) $ (see eq.
(\ref{eq:42.3}) ), the nuclear dependence of the rate in the
coherent process, analogous to the matrix element $M^2_{gs \rightarrow gs}$,
is written in terms of the proton and neutron elastic nuclear form factors,
$F_Z$ and $F_N$ (see eq. (\ref{eq:421.5}), respectively, as

\beq
M^2_{gs \ra gs} =  \tilde g_V^2 \, (3+f_V\beta )^2 \, Z^2F_Z^2(q^2)
 \left [1+\frac{3-f_V\beta}{3+f_V\beta}\,
\frac{N}{Z} \, \frac{F_N(q^2)}{F_Z(q^2)}\right ]^2
\label{eq:51.1}
\eeq

\noindent
We note that in the photonic case ($f_V=1, \beta = 3$) the neutron
contribution vanishes and the nuclear matrix element becomes
$Z^2F_Z(q^2)^2$. We also mention that the nuclear form factors for the
coherent  $\mu -e$ process are calculated at $q \approx m_{\mu} = .534
fm^{-1}$.

\bigskip
\noindent
The nuclear form factors,  $F_Z(q^2)$ and $F_N(q^2)$, have been calculated
by using various models i.e. Fermi distribution \cite{Shank,Chiang}
shell model \cite{KV88}, quasi-particle RPA \cite{KVCF}.
In the framework of shell model for spherical nuclei it was found that
$F_Z(q^2)$ and $F_N(q^2)$ can be cast in tractable analytical forms
containing fractional occupation probabilities \cite{KV92},
which take into account a
significant part of the nucleon-nucleon correlations.
Recently \cite{Oset3},
the neutron form factors $F_N(q^2)$ have been also extracted
from the analysis of pionic atoms by means of a
two-parameter Fermi distribution for heavy and very heavy nuclei and a
harmonic oscillator density for light nuclei. The form factor $F_Z$ can
be obtained from the available electron scattering data \cite{Vries}.

\bigskip
\noindent
In the context of quasi-particle RPA one can calculate the
coherent $\mu -e$ matrix elements of eq. (\ref{eq:51.1}),
by using as ground state an uncorrelated or correlated vacuum. In the
first case the nuclear form factors, $F_Z$ and $F_N$ take the form

\beq
 F_Z (q^2)\,=
\,{1 \over Z} \,{\sum_{j}}\,\, (2 j+1)\, <j \mid \mid j_0(q r)
 \mid \mid j > \,\Big( V_j^{Z}\Big)^2  \qquad
\label{eq:51.2}
\eeq
\beq
 F_N (q^2)\,= \, {1 \over N} \,{\sum_{j}}\,\,  (2 j+1) \,
<j \mid \mid j_0(q r) \mid \mid j > \,\Big(
V_j^{N}\Big)^2  \qquad \label{eq:51.3}
\eeq
The quantities $V_j^{Z}$, $V_j^{N}$ are the amplitudes for the proton,
neutron single particle states to be occupied.
Their values are determined by solving the known BCS
equations iteratively \cite{Ring}, lie between one and zero and
differ from those involved in the independent particle shell model (0 or 1).
This is due to the consideration of pairing correlations in the RPA ground
state, which deforms the Fermi surface of the
nucleus, a picture described by the fractional occupation probabilities.
The corresponding shell model form factors $F_Z$ with fractional
occupation probabilities, have been determined  \cite{KV92}
by fitting to the electron scattering data.

\bigskip
\noindent
The accurate determination of the RPA ground state is of great
importance for the exact calculation of the coherent and
the total $\mue$ rate. The $g.s.$ wave function provides the
$gs \rightarrow gs$ transitions, which are the dominant channel
of the $\mue$ process and the total rate in the sum rule approach. A usual
correction inserted in the ground state is the $g.s.$ correlations
\cite{Row1,Row2,Sande,PElli,McNe}, which can be included in the
ground state by defining the correlated QRPA vacuum $\mid {\tilde 0} >$
in terms of the uncorrelated vacuum $\mid 0 >$. By using the Thouless
theorem the correlated vacuum $\mid {\tilde 0} >$ can be written as

\beq
 {\mid {\tilde 0 } >}\,=\,\, N_0\,e^{{\hat S}^+}{\mid 0>} \label{eq:51.4}
\eeq

\noindent
where  ${\hat S}^+$ the operator

\beq
 {\hat S}^+ \, =\,{1 \over 2}
{\sum_{ab,\tau,JM}}\,{1 \over{2 J+1}}
\,\,C_{ab}^{(J,\tau)}\, A^+(a,JM)\,A^+(b,{\overline {JM}})
\label{eq:51.5}
\eeq

\noindent
The operators $A^+(a,JM)$ denote the two quasi-particle (or pair)
creation operators, in the angular momentum coupled representation.
The indices $a$ and $b$, denote proton
($\tau=1$) or neutron ($\tau=-1$) configurations coupled to J, i.e.
$ a \equiv (j_2\ge j_1) $ (and similarly for $b$), with $j_i$ running over
the single particle states of the chosen model space:
$j_i\,\equiv \,(n_i,l_i,j_i) $.
The correlation matrix  $C$ (symmetric matrix)
is constructed for each multipole field $\lambda$
from the X and Y matrices i.e. from the RPA amplitudes for
forward and backward excitation. A usual approximation for C is the
following \cite{Sande}

 \beq
 C_{ab}^{(\lambda)} \,=\,
\Big(Y^{(\la)} \Big[X^{(\la)}\Big]^{-1}\Big)_{ab} \label{eq:51.6}
\eeq

\noindent
In eq. (\ref{eq:51.4}), $N_0$ is the normalization constant, which
by keeping terms of first order in the correlation matrix $C$ is given by

\beq
 N^2_0\,=\,\Big[ \, 1+  \,{1 \over 2}
{\sum_{ab,\la,\tau}}\,\,{\tilde C}_{ab}^{(\lambda,\tau)}
\,C_{ab}^{(\lambda,\tau)}\, \Big]^{-1}\label{eq:51.7}
\eeq

\noindent
By using as ground state the correlated RPA vacuum of eq. (\ref{eq:51.4}),
the coherent rate matrix elements take the form

\beq
< {\tilde 0} \mid {\hat T}\mid {\tilde 0} > \,
= \,{ N^2_0}\, < 0 \mid {\hat T} \mid 0 >  \label{eq:51.8}
\eeq
which means that the correlated matrix elements are a rescaling of the
uncorrelated ones (see a similar expression in sect. 5.2.2 for the total
rate matrix elements).

\bigskip
\bigskip
\noindent
{\underline {\bf 5.2. Total $\mue$ Conversion branching ratios}}

\bigskip
\noindent
To find the total $\mue$ conversion rate, one need evaluate
the matrix elements for both the vector and axial vector operators
(see eq. (\ref{eq:42.2}) ) and for all the final nuclear states $\mid f>$
i.e. the quantities

\beq
S_{\la} = {\sum_f}
\Big({ {q_f}\over {m_{\mu}}}\Big)^2\int { { d {\hat {\bf q}}_f}
\over {4 \pi}} {\mid <f \mid {\Omega}_{\la} \mid i > \mid}^2,\qquad \la=V,A.
\label{eq:52.1}
\eeq

\noindent
(${\bf {\hat q}}_f$ is the unit vector in the direction of the momentum
transfer ${\bf q}_f$). Then the total $\mue$ rate matrix elements are given by

\beq
 M^2_{tot} = S_V + 3 S_A
\label{eq:52.2}
\eeq

\noindent
For the calculation of $S_V$ and $S_{A}$, one can use the following general
methods:

\bigskip
\noindent
{\underline 1) $Summing$ $over$ $partial$ $rates$:}

\bigskip
\noindent
With this method we construct explicitly
the final nuclear states $\mid f>$ in the context of a nuclear model
e.g. shell model, random phase approximation, etc. The total $\mue$ rate can be
obtained by summing the  partial rates for all possible excited states in a
chosen model space. By using the multipole expansion of the
$\mue$ operators eqs. (\ref{eq:41.6}) and (\ref{eq:41.7}),
the total rate matrix elements $M^2_{tot}$ can be
written as

\begin{eqnarray}
 M^2_{tot} = {\sum_s}(2s+1)
f_s^2 \,\, \Big[ {\sum_{f_{exc}}} &\Big(
{{q_{exc}} \over {{m_{\mu}}}} \Big)^2& {\sum_{l,J}}\, {\mid <f_{exc} \mid
\mid {\hat T}^{(l,s)J}\mid \mid gs >\mid}^2+\nonumber
\\
&\Big({{ q_{gs}} \over {m_{\mu}}}\Big)^2& {\sum_{l,J}}\,
{\mid <gs
\mid \mid {\hat T}^{(l,s)J}\mid \mid gs > \mid}^2 \Big]
\label{eq:52.3}
\end{eqnarray}
(s=0 for the vector operator and s=1 for the axial vector one). The first
term in the brackets of eq. (\ref{eq:52.3})
contains the contribution coming from all the
excited states $\mid f_{exc}>$ of the final nucleus (incoherent rate) and the
second term contains the contributions
coming from the  ground state to ground state
channel (coherent rate).

\bigskip
\noindent
{\underline 2) $ Closure$ $ approximation $: }

\bigskip
\noindent
It is well known that for the description of the total strengths in many
processes the sum rule techniques are very useful
\cite{KoVe,Prima,G-P,Civi,KV89}.
In such an approach the
contribution of each final state $\mid f>$ to the total rate is approximately
taken into account without constructing this final state explicitly.
One assumes
a mean excitation energy of the nucleus ${\bar E } =<E_f>-E_{gs}$  and uses
closure for the final states $\mid f>$ i.e.
 $\sum _f|f><f| =1 $. This approximation requires the evaluation of
  one and two-body matrix elements involving only the initial
(ground) state. This way the explicit calculation of the final states
 $\mid f>$ is avoided.

\bigskip
\noindent
The mean excitation energy $\bar E$ of the nucleus, involved
in the needed matrix elements is defined as \cite{G-P}

\beq
 {\bar E} =\,\, { { {\sum_{f}} (E_f - E_{gs})
\, \Big({{\mid {\bf q}_f \mid} \over {m_{\mu}} }\Big)^2\, {\mid <f
\mid {\hat T}^J \mid gs > \mid}^2 } \over { {\sum_f}
\, \Big({{\mid {\bf q}_f \mid} \over {m_{\mu}} }\Big)^2\,
{\mid <f \mid {\hat T}^J \mid gs > \mid}^2 }} \label{eq:52.4}
\eeq

\noindent
The numerator of this definition represents the energy
weighted sum rule and the denomenator the non-energy weighted sum rule.
Though ${\bar E}$ is defined in  analogy with the ordinary muon capture
reaction, the value of the ``mean excitation energy'' in $\mue$  is different
from that in $(\mu^-,\nu_{\mu})$ \cite{KVCF},
because in the last process the coherent
channel doesn't exist. For the $\mue$ process the coherent channel ($gs
\rightarrow gs$) appears only in the denominator of eq. (\ref{eq:52.4}) and,
since this dominates the total $\mu-e$ conversion rate, the resulting mean
excitation energy $\bar E$ in this process is  much smaller
than that for the $(\mu^-,\nu_{\mu})$ reaction. Obviously, the mean
excitation energy $\bar E$ can be evaluated by constructing explicitly  all
the possible excited states $\mid f>$ in the context of a nuclear model.
In ref. \cite{KVCF}, for example, the QRPA method has
been used for the determination of the mean excitation energy of the $^{48}Ti$
nucleus in the process $\mu^- + ^{48}Ti \ra e^- + ^{48}Ti^*$.

\bigskip
\noindent
The method of closure approximation proceeds by defining the
operator taken as a tensor product from the single-particle operators ${\hat
T}$ of eq. (\ref{eq:41.6}) or  (\ref{eq:41.7}). For a $0^+$ initial
(ground) state the relevant tensor product is

\beq
{ \hat O }\,=\,\sum_{k,k^{\prime}} \Big[\,{\hat T}^k \times
{\hat T}^{k^{\prime}} \,\Big]^0_0 \label{eq:52.5}
\eeq

\noindent
The corresponding total matrix elements in a sum-rule approach are written as

\beq
 M_{tot}^2=  \,
\Big( {{\mid {\bf k} \mid}\over {m_{\mu}}}\Big)^2 \, \Big[
f^2_V<i \mid {\hat O}_V \mid i >  + \, 3 f^2_A<i \mid {\hat O}_A \mid i >
 \Big] \label{eq:52.6}
\eeq
where $\mid {\bf k} \mid$ is the average momentum and
the operators ${\hat O}_V$ (vector) and ${\hat O}_A$
(axial vector), which contain both one-body and two-body pieces, result
from the corresponding $\hat T $ operators of eqs.
(\ref{eq:41.6}), (\ref{eq:41.7}). Consequently, one has to evaluate the
matrix elements of the operators
 ${\hat O}_V$, ${\hat O}_A$
in a given model.

\bigskip
\noindent
{\underline 3) {\it Nuclear matter mapped into nuclei with local density
approximation:}}

\bigskip
\noindent
The method has been already discussed in sects. 4.2.1-4.2.2.
The obtained results will be presented in sect. 5.3.

\bigskip
\bigskip
\noindent
{\underline {\bf 5.2.1. RPA calculations envolving the final states
explicitly.}}
In actual shell model calculations it is quite hard to construct the final
states explicitly in medium and heavy nuclei. In such cases we can employ
the RPA approximation. In the context of the quasi-particle RPA the final
states entering the partial rate matrix elements are obtained by acting on the
vacuum $\mid 0 >$ with the phonon creation operator
\cite{Ring,Row2,Civi}

\beq
 Q^+(fJM) =
{\sum_{a,\tau}} \Big [ X_{a}^{(f,J,\tau)} A^+(a,JM)  -
  Y_{a}^{(f,J,\tau)} A(a,{\overline {JM} }) \Big ] \label{eq:521.1}
\eeq

\noindent
(angular momentum coupled representation) i.e.  $\mid f > = Q^+ \mid 0>$.
The quantities X and Y in eq. (\ref{eq:521.1}) are the
forward and backward scattering amplitudes. The index $a$, runs over proton
($\tau=1$) or neutron ($\tau=-1$) two particle configurations coupled to J.

\bigskip
\noindent
The nuclear matrix  element involved in the partial rate $\Gamma_{i \ra f}$
from an initial state $0^+$ to an excited state $\mid f>$  takes the form

\beq
 <f \mid \mid {\hat T}^{(l,S)J}\mid \mid 0^+ > =
{\sum_{a,\tau}} \,\, W_{a}^J\,
 \Big [\,\, X_{a}^{(f,J,\tau)} U^{(\tau)}_{j_2}V^{(\tau)}_{j_1}  +
    (-)^{\theta} Y_{a}^{(f,J,\tau)} V^{(\tau)}_{j_2}U^{(\tau)}_{j_1}
    \,\, \Big]
\label{eq:521.2}
\eeq

\noindent
The phase $\theta $=0, 1 for the vector, axial vector operator,
respectively \cite{BoMo}.
The probability amplitudes $V$ and $U$ for the single particle
states to be occupied and unoccupied, respectively, are determined from
the BCS equations and the X and Y matrices are provided by solving
the QRPA equations. The quantities $W_a^J \equiv W_{j_2j_1}^J$
contain the reduced matrix elements of the operator ${\hat T}$ between
the single particle proton or neutron states $j_1$ and $j_2$ as

 \beq
W_{j_2j_1}^J = \frac { {<j_2 \mid\mid {\hat T}^J\mid\mid
j_1>} }{ {2 J }+1}
\label{eq:521.3}
\eeq

\noindent
In  {\it  table} 3,  the results of $^{48}Ti$ are shown for various values
of the parameter $\beta$ in a model space consisting of all the
single particle levels included up to 3$\hbar \omega$ (same for protons
and neutrons). In the photonic mechanism  mechanism the axial
vector matrix elements are zero (see eq. (\ref{eq:41.3})).

\vspace{0.7cm}
\noindent
{\it Table 3. Incoherent $\mu -e$ conversion matrix elements:
vector $(S_V)$, for the photonic mechanism ($\beta = 3$) and vector
and axial vector $(S_A)$, for a non-photonic mechanism ($\beta =5/6$).
They are for all the excited states included in the up to
$3\hbar \omega$ model space for $^{48}Ti$.}
\vskip0.6cm
\begin{center}
\begin{tabular}{ccll}
\hline
       &              &                          &\\
 Mode  & \multicolumn{1}{c}{ photonic mechanism } &
       \multicolumn{2}{c}{ non-photonic mechanism }\\
\hline
          &           &              &          \\
$J^{\pi}$ & $S_V$     &   $S_V$      & $S_A$     \\
\hline
       &                    &             &         \\
$ 0^+ $ & 1.111             & 2.363       & 0.0     \\
$ 1^+ $ & 0.0               & 0.0         & 0.297   \\
$ 2^+ $ & 0.309             & 0.422       & 0.046   \\
$ 3^+ $ & 0.0               & 0.0         & 0.050   \\
$ 4^+ $ & 0.002             & 0.002       & 2. $10^{-4}$\\
$ 5^+ $ & 0.0               & 0.0         & 2. $10^{-6}$\\
$ 6^+ $ & 2. $10^{-6}$      & 2. $10^{-6}$& 2. $10^{-7}$\\
        &                   &             &             \\
$ 0^- $ & 0.0               & 0.0         & 0.818   \\
$ 1^- $ & 9.744             &17.853       & 0.795   \\
$ 2^- $ & 0.0               & 0.0         & 0.670   \\
$ 3^- $ & 0.052             & 0.068       & 0.011   \\
$ 4^- $ & 0.0               & 0.0         & 0.010   \\
$ 5^- $ & 8. $10^{-5}$      & 1. $10^{-4}$& 1. $10^{-5}$ \\
$ 6^- $ & 0.0               & 0.0         & 1. $10^{-5}$ \\

        &                   &             &         \\
Total   &11.217             & 20.708      & 2.697   \\
\hline

\end{tabular}
\end{center}

\bigskip
\bigskip
\noindent
{\underline {\bf 5.2.2.  Sum-Rules in the context of QRPA.} }
The  RPA sum-rules for the total $\mue$ rate, in the case when an
uncorrelated ground state vacuum $\mid i> = \mid 0>$ in eq.
(\ref {eq:52.6}) is used, can be easily obtained. Then
the matrix elements $<0 \mid {\hat O} \mid 0>$ are given by

\begin{eqnarray}
{<0 \mid \hat O  \mid 0> }&&\\ \nonumber
&= &
{\sum_J }  \, \Big\{\,\,  \Big[\,{\sum_{j,\tau}}\,{(2 j+1)} <j \mid \mid
{\hat T}^J \mid \mid j > \Big(\, V_j^{(\tau)}\Big)^2\,  \Big]^2
\\ \nonumber
&+&
\, {(2 J+1)}
\sum_{a,\tau }\,\,{\tilde p}(aJ,\tau) \, p(aJ,\tau)\,\,\Big\}
\label{eq:522.1}
\end{eqnarray}

\noindent
where the first term gives the one-body contribution
and the second the two-body contribution. The quantities $p$,
$\tilde p$ are given by

\beq
 p(\al J,\tau)= \, W_{j_2j_1}^J \, \Big[ \,
 U_{j_2}^{(\tau)} V_{j_1}^{(\tau)} \,+\, (-)^{\theta} U_{j_1}^{(\tau)}
 V_{j_2}^{(\tau)}
 \Big] \label{eq:522.2}
\eeq

\beq
{\tilde p}(\al J,\tau)=
\, W_{j_2j_1}^J \, \Big[
V_{j_2}^{(\tau)} U_{j_1}^{(\tau)}\,+\, (-)^{\theta} V_{j_1}^{(\tau)}
 U_{j_2}^{(\tau)}
\Big]
\label{eq:522.3}
\eeq

\noindent
$ a \equiv (j_2\ge j_1) $. The phase $\theta$ is the same as in eq.
(\ref{eq:521.2}).

\bigskip
\noindent
The correlated quasi-particle RPA sum-rule of the $\mue$ reaction, which
is the expectation value of the operator ${\hat O}$ of eq. (\ref{eq:52.5})
in the correlated ground state $\mid {\tilde 0}>$ of eq. (\ref{eq:51.4}),
is written as

\beq
 < {\tilde 0} \mid {\hat O}\mid {\tilde 0} >\,
= \,{ N^2_0}\,\Big( < 0 \mid {\hat O} \mid 0 > +
 <0\mid \{ [{\hat S},{\hat O}]\, +\,
[{\hat O},{\hat S}^+]\, \} \,\mid 0> \Big) \qquad \label{eq:522.4}
\eeq

\medskip
\noindent
where the second term in the brackets takes into account the 2p2h excitations.
Thus, in order to compute the total rate with the correlated matrix elements,
$ < {\tilde 0} \mid {\hat O}\mid {\tilde 0} >$, one should first  calculate the
uncorrelated matrix element $<0 \mid{\hat O}\mid 0 >$ and afterwards the
contribution coming from the ground state RPA correlations. The latter
contribution can be cast in the form

\begin{eqnarray}
 < 0 \mid  \{ [{\hat S},{\hat O} ]\, +\, [{\hat O},
{\hat S}^+] \} \,\mid  0 >  && \\
\nonumber
 &=&
{\sum_{ab,\la,\tau}}\,C_{ab}^{(\la,\tau)}\,
 { {\, W_{j_2j_1}^{\lambda} \, W_{{j'}_2{j'}_1}^{\lambda}}
\over {(1+\delta_{j_2j_1})(1+\delta_{{j'}_2{j'}_1}) } }\\
\nonumber
&\times & \Big[
 U_{j_2}^{(\tau)} V_{j_1}^{(\tau)} \, U_{{j'}_2}^{(\tau)}
V_{{j'}_1}^{(\tau)}  \\ \nonumber
+ V_{j_2}^{(\tau)} U_{j_1}^{(\tau)}\, V_{{j'}_2}^{(\tau)} U_{{j'}_1}^{(\tau)}
\Big]
\label{eq:522.5}
\end{eqnarray}
\bigskip
\noindent
The summation (\ref{eq:522.5}) gives the proton-proton (p-p) and
neutron-neutron (n-n) correlations. The results obtained for $^{48}Ti$ (see
{\it tables 4} and {\it 5}) show that the contribution from the two-body
correllations as expected  are small.

\vskip1.8cm
\noindent
{\it Table 4.
Total rate nuclear matrix elements and $gs \ra gs$ transition
for the photonic mechanism $\mue$ conversion rates in $^{48}Ti$
calculated: 1) with shell model sum-rule and 2) with various types of QRPA
sum-rules for different mean excitation energies.}
 \vskip1.0cm

\begin{center}
\begin{tabular}{lrrrc}
\hline
             &           &                 &             &  \\
 $Method$    & $M_{gs \rightarrow gs}^2 $ & ${\bar E}$  &
 $M_{tot}^2$ & $\eta$ ($ \% $) \\
\hline
             &           &                 &             &  \\
 $Shell\,Model (sum-rule)$  & 144.6 & 20.0  & 188.8 & 67.2  \\
 $QRPA \,(explicit)$        & 135.0 &  -    & 161.0 & 83.9  \\
 $QRPA  (sum-rule)$         & 135.0 &  1.7  & 138.3 & 97.6  \\
 $QRPA (sum-rule)$          & 135.0 &  5.0  & 140.6 & 96.0  \\
 $QRPA (sum-rule) $         & 135.0 & 20.0  & 141.7 & 95.3  \\
 $QRPA + Corr \,(sum-rule)$ & 87.8  &  1.7  & 90.4  & 97.1  \\
 $QRPA + Corr \,(sum-rule)$ & 87.8  &  5.0  & 91.8  & 95.6  \\
 $QRPA + Corr \,(sum-rule)$ & 87.8  & 20.0  & 92.6  & 94.8  \\
              &           &                 &             &  \\
\hline
\end{tabular}
\end{center}

\vskip2.5cm

\noindent
{\it Table 5. Total rate matrix elements and $gs \ra gs$ transitions
for $^{48}Ti$ given by various methods for the non-photonic mechanism
$\beta=5/6$
(see caption of table 4). }

\vskip0.4cm

\begin{center}
\begin{tabular}{lrrcc}
\hline
             &           &                 &             &  \\
 $Method$ & $M_{gs \rightarrow gs}^2 $ & ${\bar E}$ & $M_{tot}^2 $ &
 $\eta$ ($ \% $) \\
 \hline
              &           &                 &             &  \\
 $Shell\,Model (sum-rule)$ & 374.3 & 20.0  & 468.0 & 80.0  \\
 $QRPA \, (explicit)     $ & 363.0 & -     & 386.4 & 93.9  \\
 $QRPA (sum-rule)$         & 363.0 & 0.5   & 366.2 & 99.1  \\
 $QRPA (sum-rule)$         & 363.0 & 5.0   & 376.5 & 96.4  \\
 $QRPA (sum-rule)$         & 363.0 & 20.0  & 382.8 & 94.8  \\
 $QRPA+Corr \,(sum-rule)$  & 236.2 & 0.5   & 238.6 & 99.0  \\
 $QRPA+Corr \,(sum-rule)$  & 236.2 & 5.0   & 245.4 & 96.3  \\
 $QRPA+Corr \,(sum-rule)$  & 236.2 & 20.0  & 249.5 & 94.7  \\
              &           &                 &             &  \\
\hline
\end{tabular}
\end{center}


\bigskip
\bigskip
\noindent
{\underline {\bf 5.2.3 Sum-rules in the context of shell model.}}
The total $\mu$-e conversion rate matrix elements of eq. (\ref{eq:52.6})
can be conveniently evaluated by  using shell model closure approximation.
To obtain the needed matrix elements we assume that the initial nuclear wave
function $\mid i>$ is a Slater determinant with closed proton and neutron
(sub)shells constructed out of single particle harmonic oscillator
wave functions. The tensor product operator $\hat O$ of eq. (\ref{eq:52.5})
in this case (no second quantization) is written as
\cite{KV89}

\beq
{\hat O}_{\al} \, = \, \tilde g_{\alpha}^2 \, \sum_{ij} \left[
A_{\alpha} +  B_{\alpha} (\tau_{3i}+\tau_{3j}) +C_{\alpha}
\tau_{3i}\tau_{3j}  \right]\Theta^{\alpha}_{ij}
 ,\quad \al = V,A
\label{eq:523.1}
\eeq

\noindent
where

\begin{eqnarray}
 A_V=9,\,\,\, B_V=9,  \,\,\, C_V=9,        & \tilde g_V=\frac{1}{6}, \,\,
\tilde g_A=0  \label{eq:phot} \\
 A_V=9/f_V^2, \,\, B_V=(3/f_V)\beta ,\,\,  C_V=\beta^2,& \tilde
 g_V=\frac{1}{2}&{} \nonumber \\
 A_A=\xi^2/f_V^2,\,\, B_A=(\xi /f_V)\beta ,\,\, C_A= \beta^2,&\tilde
g_A=\frac{1}{2}
\label{eq:nonphot}
\end{eqnarray}

\noindent
for the photonic and non-photonic cases correspondigly. Furthermore,

\beq
\Theta^V_{ij}=j_0(q r_{ij}), \quad \Theta^A_{ij}=j_0(q r_{ij})
\frac{ {\vec \sigma}_i  \cdot   {\vec \sigma}_j }{3}
\label{eq:523.2}
\eeq

\noindent
 From eq. (\ref{eq:523.1}) we can
see that,  contrary to the $(\mu^-,\nu_{\mu})$ reaction, a distinct feature of
the operator $\hat O $, which is responsible for the process $\mue$, is the
presence of an isospin independent term in both the vector and axial vector
component. It has been shown \cite{KV89} that
this part of the $\mu -e$ operator, is the dominant one.
By using standard shell model techniques $\cite{KoVe}$ the matrix  elements
$M^2_{tot}$, which contain one body and two body terms, can be  cast in
compact analytical forms as

 \beq
 S_{\la} \,=\, g_{\la}(A,Z) \Big( {{\mid {\bf q} \mid}\over
{m_{\mu}}}\Big)^2 \, \Big(1-\sum^{N_{max}}_{\ka=1} \xi_\ka \,\,
\alpha^{2\ka}\Big ) e^{-\alpha^2/4},\quad \alpha =\sqrt2 \mid {\bf q} \mid b,
\qquad \la=V,A  \label{eq:523.3}
\eeq

\noindent
where b is the harmonic oscillator parameter, $N_{max}$ is the maximum
number of oscillator quanta occupied by the nucleons in the considered
nucleus and
$ \xi_\ka$ are appropriate coefficients which may depend on A and Z.
The functions $g_\la(A,Z)$ describe the total rate if its
dependence on the momentum transfer can be neglected. The definitions of
the coefficients $\xi_{\la}$  and the functions $g_{\la}(A,Z)$, are
given in ref. \cite{KoVe}.

\bigskip
\noindent
The total $\mue$ matrix elements using the shell model sum rules take the form

 \beq
 M_{\al}^2= \, {\tilde g}_{\al}^2 \,
\Big[ Z + \Big( { {3-f_V\be} \over {3+f_V\be}}  \Big)^2 N  \, + \,
 A_{\al} \,S_{\al}^{(0)}\, + \,
 B_{\al} \,S_{\al}^{(1)}\, + \,
 C_{\al} \,S_{\al}^{(2)} \Big], \qquad \al=V,A
\label{eq:523.4}
\eeq

\noindent
where $S^{\kappa}_{\al}, \,\, \kappa=0,1,2$ correspond to the three
operators entering  eqs. (\ref{eq:523.1})-(\ref{eq:523.2}). In $tables\, 3$
and $4$ the results obtained this way in $^{48}Ti$ for average exitation
energy $\bar E=20MeV$, are compared with those obtained by using
quasi-particle RPA.

\bigskip
\bigskip
\noindent
{\underline {\bf 5.2.4.  The $\mue$ conversion in the Primakoff's  method.} }
It is well known that success of the sum rule method hinges upon a good
choise of mean excitation energy $\bar E$. Then it requires only knowledge
of the structure of the ground state. Very early Primakoff \cite{Prima}
developed a phenomenological method for the ($\mu^-,\nu_{\mu}$) reaction,
which reproduces very well the experimental data for the total muon capture
rate and which does not contain the energy $\bar E$. It yields
the well-known Goulard-Primakoff \cite{G-P} function (see sect. 3.2.2).
which contains three parameters, the
strengths of the isoscalar, isovector and  isotensor parts of the muon-capture
operator which are determined by a fit  to the total muon capture
reaction data of the hole periodic table.

\bigskip
\noindent
By exploiting the common components of the ($\mu^-, \nu_{\mu}$) and $\mue$
conversion operators one can construct a phenomenological formula for the
$\mue$  process \cite{KoVe}. In this way the isospin dependent total
$\mu -e$ rate can be estimated by using the  values
for the isoscalar, isovector and isotensor parameters of the Goulard-Primakoff
function determined by the ordinary muon capture data. Even though the
results of this phenomenologilal method compare well with those of the
shell model closure approximation, the fraction of the total $(\mu\ra e)$
rate coming from the isospin dependent part of the $\mue$ operator is
tiny.  Furthermore, as we have stressed in sect. 4, the dominant channel
in the $\mue$ process is the coherent one which is not possible in the
$(\mu, \nu_{\mu})$ and therefore it cannot be obtained from the
Goulard-Primakoff method.

\bigskip
\bigskip
\noindent
{\underline {\bf 5.3.  Discussion of nuclear matrix elements} }

\bigskip
\noindent
As it can be seen from table 4, the ground state to ground state transition
(coherent contribution) exhausts a large portion of the sum rule given by the
parameter $\eta$. This is not entirely unexpected, since in this case the
contribution of all nucleons is coherent.  This is very encouraging  since,
as we have seen in sect. 2.3, the coherent process is of experimental
interest.  In order  to extract $\eta$ in addition to the coherent rate we
must have a reliable estimate of the total rate.  The value of $\eta$ depends,
of course, on the parameters ${\tilde g}_V,{\tilde g}_A$ and $\beta$  of the
elementary amplitude.

\bigskip
\noindent
The only nuclear information needed for the coherent mode are the
``muon-nuclear" form factors ${\tilde F}_p(q^2)$ and ${\tilde F}_n(q^2)$,
(see eq. (\ref{eq:421.4})), which can easily be calculated both in the shell
model and in RPA calculations.  Admittedly, however, these
calculations may not be very precise at the high momentum transfer
 $ q^2 = m^2_{\mu} c^2 \sim 0.25 fm^{-2}$.
If on the other
hand the muon wave function is assumed to be constant to be taken out
of the radial integrals, the nuclear matrix element essentially depends on the
proton  and neutron form factors $F_N(q^2)$ and $F_Z(p^2)$  which can be
taken from experiment (electron scattering data, pionic atoms etc.).

\bigskip
\noindent
Using the results of table 6 (Chiang {\it et al.}, 1993 \cite{Chiang}),
we can compute the parameter

\beq
\xi_{p,n}(A,Z) = \left[\frac{F^2_{Z,N}(p^2)}{{\tilde F}_{p,n}^2(p^2)}\right]
\, \frac{\alpha^3 m_{\mu}^3}{\pi} \, Z^4_{eff} Z
\label{eq:53.1}
\eeq

\noindent
This parameter gives us a measure of the deviation of the effective form
factor (involving and the muon) from the usual  nuclear form factor.  The
obtained results are presented in  {\it table  7}.

\vskip0.9cm
\noindent
{\it Table 7. Values of the parameter $\xi_{p,n}(A,Z)$ of eq. (\ref{eq:53.1})
in the text. The approximation of eq. (\ref{eq:42.3}) is not very
accurate for heavy nuclei.}

\bigskip
\bigskip
\begin{center}
\vglue 0.3cm
\begin{tabular}{ccccccc}
\hline
     &      &       &       &       &       &     \\
(A,Z)&(12,6)&(32,16)&(40,20)&(48,22)&(90,40)&(208,82)\\
\hline
     &      &       &       &       &       &     \\
$\xi_p$& 0.979&0.962&0.946&0.990&0.850&0.768\\
$\xi_n$& 0.979&0.956&0.951&0.639&0.560&0.285\\
     &      &       &       &       &       &     \\
\hline
\end{tabular}
\vglue 0.2cm
\end{center}
\vglue 0.4cm

\bigskip
\bigskip
\noindent
We thus see that, for heavy nuclei the muon wave function cannot be taken as a
constant. This is especially true for the neutron component.   This means
that in this region the experimental form factors must be used with caution,
even though the effect on the branching ratio may be less pronounced.

\bigskip
\noindent
All calculations of the coherent rate indicate that, in spite of
the earlier expectations (Weinberg and Feinberg, 1959 \cite{WeiFei},
the coherent branching ratio increases all the way up to the Pb region
where it starts decreasing.

\bigskip
\noindent
For the supersymmetric model discussed in sect. 3.3.3, we present our
results in $table\,\, 8$.

\noindent
{\it Table 8.  The nuclear form factors $F_Z$ and $F_N$ entering in the quasi
-elastic $(\mu,e)$ conversion (eq. (\ref{eq:Reg19})). The quantities
$\gamma_{ph}$ and $\kappa$ are defined in the text (see eqs.
(\ref{eq:Reg19a}), (\ref{eq:Reg20})), and are evaluated in the case
of supersymmetric models.}

\begin{center}
\vglue 0.2cm
\begin{tabular}{rrcccrr}
\hline
   &    &      &     &        &             &    \\
A  & Z  & $F_Z$&$F_N$&$\kappa$&$\gamma_{ph}$&$R_{eN}/R_{e\gamma}$\\
\hline
   &    &      &     &        &             &    \\
4  & 2  & 0.865&0.865&1.67&1.56&$1.51\,\,10^{-3}$\\
12 & 6  & 0.763&0.763&1.67&3.64&$3.53\,\,10^{-3}$\\
14 & 6  & 0.753&0.745&1.88&7.96&$9.36\,\,10^{-3}$\\
16 & 8  & 0.736&0.736&1.67&4.52&$4.39\,\,10^{-3}$\\
28 & 14 & 0.639&0.639&1.67&5.95&$5.78\,\,10^{-3}$\\
32 & 16 & 0.618&0.618&1.67&6.37&$6.19\,\,10^{-3}$\\
40 & 20 & 0.582&0.582&1.67&7.05&$6.85\,\,10^{-3}$\\
48 & 20 & 0.563&0.515&1.85&16.08&$1.84\,\,10^{-2}$\\
48 & 22 & 0.543&0.528&1.77&9.74 &$1.18\,\,10^{-2}$\\
60 & 28 & 0.489&0.478&1.74&9.24 &$9.58\,\,10^{-3}$\\
72 & 32 & 0.456&0.435&1.79&11.54&$1.25\,\,10^{-2}$\\
82 & 32 & 0.440&0.379&1.89&24.98&$3.00\,\,10^{-2}$\\
88 & 38 & 0.412&0.370&1.79&12.98&$1.40\,\,10^{-2}$\\
90 & 40 & 0.406&0.367&1.76&11.41&$1.20\,\,10^{-2}$\\
114& 50 & 0.335&0.306&1.77&10.35&$1.10\,\,10^{-2}$\\
132& 50 & 0.315&0.250&1.86&25.80&$3.00\,\,10^{-2}$\\
156& 64 & 0.263&0.207&1.76&11.96&$1.30\,\,10^{-3}$\\
162& 70 & 0.253&0.202&1.70&8.92&$8.94\,\,10^{-3}$\\
168& 68 & 0.249&0.191&1.76&12.47&$1.35\,\,10^{-2}$\\
176& 70 & 0.242&0.181&1.76&13.75&$1.49\,\,10^{-2}$\\
194& 82 & 0.198&0.168&1.77&7.20&$7.69\,\,10^{-3}$\\
208& 82 & 0.189&0.135&1.73&10.42&$1.07\,\,10^{-2}$\\
   &    &      &     &        &             &    \\
\hline
\end{tabular}
\end{center}
\vglue 0.2cm
\hspace{27.4mm}

The total rate can also be calculated in the context of the shell model.  In
this case it is not possible to construct all final states explicity in
realistic model spaces.  One is thus forced to invoke the closure
approximation (see sect. 5.2.3).  One, however, has no idea about the
proper average excitation energy ${\bar E}_{exc}$ to use.  In the earlier
calculations
(Kosmas and Vergados, 1990 \cite{KoVe})  the value used was the same
with that of the
common muon-capture.  Using this value of ${\bar E}_{exc}=20
MeV$, the value of $\eta$
ranged from  over 90{\%} in light nuclei to about 30{\%}
in heavy nuclei (ibid, Table
4).  We have seen, however, in sect. 5.2.2 that,
with the proper definition of eq.
(\ref{eq:52.4}), ${\bar E}_{exc}$ must be quite a bit lower.
 We should, therefore,
give more credibility to the RPA results discussed above
($tables$  $4$ and $5$).

\bigskip
\noindent
The calculation of the total rates is quite a bit harder.  It can, however,
easily be done in the context of QRPA.  In QRPA one can also apply the sum
rule techniques with an average energy defined by eq. (\ref{eq:52.4}),
which in this case can be calculated.  The agreement between  the two
methods is quite good (see
tables 4 and 5).  It is important to note that the ground state
correlations tend to
decrease all rates in our example \cite{KVCF} by 35{\%}.
 The obtained value of $\eta$  is quite high.

\bigskip
\noindent
 The total $(\mu,e)$ conversion rate has also
been computed by a recent new method
which utilizes nuclear matter mapped into  nuclei with the local density
approximation  (see sects. 4.2.1 and 4.2.2).  This method has the advantage
that both the incoherent rates of
$(\mu,e)$ conversion and ordinary muon capture can be computed in
the same way (see expressions for $G_p(N,Z)$ and $G_n(N,Z)$
of eqs. (\ref{eq:422.6} )
and (\ref{eq:422.7}) as well as eq. (\ref{eq:422.9}).  Using the results
of this calculation we see from {\it table} 6 that
the parameter $\eta$ is in all cases
greater than 80{\%} and keeps increasing from light to heavy nuclei.
 Furthermore,  the
nuclear matrix elements for the coherent mode
and the resulting branching ratios
increase all the way to  the  $Pb$ region.

\bigskip
\bigskip
\noindent
\hspace{27.4mm} {\bf 6. CONCLUDING REMARKS }

\bigskip
\noindent
As we have seen in sect. 3, the most popular scenario for lepton-flavor
violation
involves  intermediate neutrinos at the one -loop levels.
 We have  seen, however,
that due to the GIM mechanism we encounter an unfavorable
explicit dependence on
the neutrino  mass.  We have seen in sec. 3.2.1 that, for
leptonic currents of the
same helicity the amplitude for light neutrinos depends on the square of
the neutrino mass while for heavy neutrinos, on the inverse
neutrino mass squared  (see eqs. \ref{eq:Reg5.7a}, \ref{eq:Reg5.7}
\ref{eq:Reg5.9a}, \ref{eq:Reg5.9b}). Thus in this mechanism lepton  flavor
is unobservable for neutrinos
much lighter or much  heavier than the W-boson mass.
 This  unfortunately happens
to be the case with most  currently fashionable models.
   The explicit dependence
on the neutrino mass is somewhat  improved (linear for
 light neutrinos, inversely
proportional for heavy neutrinos) in the case of
L-R interference in the leptonic
sector (see eqs. \ref{eq:Reg5.9a}, \ref{eq:Reg5.9b}
 and \ref{eq:Reg5.9c}, \ref{eq:Reg5.9}).  However, the situation
essentially does not improve due to the presence of the mixings
$U^{(12)}$ or $U^{(21)}$ which in most fashionable models is negligible.

\bigskip
\noindent
{}From the phenomenological point of view,
therefore, we can add very little to the
discussion of a previous review \cite{VER} and we will not elaborate
further on this mechnism.  We will instead summarize the results obtained
using other intermediate particles.

\bigskip
\noindent
We will begin with an extended higgs sector
involving non-exotic particles, i.e.  two
doublets in the Bjorken - Weinberg mechanism.
The calculation of the branching
ratio is rather  complicated since the two loop
 contribution becomes dominant.  It
has recently been done by  Chang,
 Hou and Keung \cite{Chang}, and Barr and Zee \cite{BaZe}
only for the $\mu \ra e \gamma$ process.
The predictions of this model depend
essentially on four parameters:  $\Delta_{tt}$ (the ttH Yukawa coupling),
the combination $\Delta_{e\mu} cos \phi_{\alpha}$,
 ($\Delta_{e\mu}$  is the $e \mu H$
Yukawa coupling and $ cos \phi_{\alpha}$ (the $W^+W^-H$
 cubic coupling relative to that
of the stantard model), the top quark mass $m_t$ and
the Higgs scalar mass $m_H$.
The authors take the somewhat optimistic choice of
$\Delta_{tt} \simeq 1$  and
 $\Delta_{e\mu} cos \phi_{\alpha} = 1$ and plot
their obtained branching ratio as a
function of $m_t$ and $m_H$.  For $m_H$ very large
($m_H \gg 1 TeV$) or small ($m_H \ll
200 GeV$)  their results are essentially independent
 of $m_t$ with branching ratios
close to the present experimental limit.
 These authors, however, present
their results as though there is one Higgs Boson by assuming that the
contributions from the at least two needed Higgs Scalars do not cancel
each other.

\bigskip
\noindent
Unfortunately, the above elaborate calculation
does not  provide the form  factors
$f_{E0}$ and $f_{M0}$ which do not contribute to
$\mu \ra e \gamma$ but enter into
all the other lepton flavor violating processes
which involve virtual photons.

\bigskip
\noindent
The next extension of the standard model involves
singly and/or doubly charged
Higgs scalars.  Lepton flavor violation can be caused
 by introducing only the
singly charged isosinglet both for
$\mu \ra e \gamma$ and $\mu \ra 3e$.  The
branching ratio unfortunately depends on a big
power of the not accurately
constrained mass of the isosinglet (the  4th for
 $\mu \ra e \gamma$, the 8th
for $\mu \ra 3e$).  It also depends on the Yukawa
 couplings $\lambda_{\mu\tau}$ and
$\lambda_{e\tau}$ of the isosinglet.
 Thus, one cannot make  accurate predictions.  One
instead can use the present experimental limit for
 $\mu \ra e \gamma$ to  derive
the constraint \cite{LTV}

\beq
 M_S \ge 94 GeV (10^2(\lambda_{\mu\tau} \lambda_{e\tau})^{1/2})
 \label {eq:Reg1N}
\eeq

\noindent
The introduction of the doubly charged Higgs scalars
(isotriplet or isosinglet, see
sect. 3) can give rise to lepton flavor violation at
the tree level for
$\mu \ra 3e$ and muonium-antimuonium oscillations which
thus become favorable. In the case of the doubly charged
 isosinglet
$\chi^{++}$ the amplitude for the $\mu \ra 3e$ process
can be written \cite{VER} as

\beq
{\cal M}\, = \,{\tilde n}_x \frac{G_F}{\sqrt2}\, \frac{1-p_{12}}{\sqrt2} \,
{\bar u}(p_e) (1+\gamma_5) u(p_{\mu}){\bar u}(p_2) (1-\gamma_5) u(p_1)
\label{eq:4N} \eeq

\noindent
which leads to the branching ratio

\beq
 R = \frac{1}{8} |{\tilde n}_x|^2
\eeq

\noindent
with

\beq
 {\tilde n}_x = \frac{g_{e\mu} g_{ee}}{g^2} \frac{m_W^2}{m_x^2}
\label{eq:Reg5N}
\eeq

\noindent
Once again the branching ratio depends on the inverse 4th power of the
unknown mass $m_x$ of $\chi^{++}$.  Assuming that  $ g_{e\mu} \simeq g_{ee}
\simeq 0.1$ and  $m_x = 10^5 GeV$ we obtain

\beq
 {\tilde n}_x = 1.6 \times 10^{-10},\qquad  R \simeq 3.2 \times 10^{-17} \label
{eq:Reg6N}
\eeq

\noindent
which is many orders of magnitude away from the planned experiments.  The
situation with the doubly charged isosinglet is analogous except that
$\gamma_5 \ra -\gamma_5$ and ${\tilde n}_x \ra {\tilde n}_{\Delta}$.  Since,
neither the mass not the couplings of the  isosiglet are determined from
other experimental data, we can use the present experimental limit to
constrain ${\tilde n}_{\Delta}$. We obtain

\beq
 {\tilde n}_{\Delta} \leq 3 \times 10^{-6}
\label {eq:Reg7N}
\eeq

\noindent
which leads to the constrain

\beq
 \frac{c_{e\mu} c_{ee}}{M^2_{{\Delta}^{++}}} \leq 2\times 10^{-10} GeV^{-2}
\label {eq:Reg8N}
\eeq

\noindent
The amplitude for muonium-antimuonium oscillations is the same with that of
eq. (\ref{eq:4N})  except that, the antisymmetrization term
$({1-p_{12}})/{\sqrt2}$ is absent. Thus, for the isotriplet we see that the
calculated value of ${\tilde n}_x$ is much
smaller than the experimental limit $n_x
< 0.16$ obtained in sect. 2.4.  Similarly, the limit
${\tilde n}_{\Delta} \leq 3 \times 10^{-6}$ extracted
from the experimental limit of
$\mu \ra 3e$ is much smaller than 0.16.  This, of course,
 indicates that if the
doubly charged Higgs scalars exist it is much more likely
to observe lepton flavor
violation in  $\mu \ra 3e$ rather in $M - {\bar M}$ oscillations.

\bigskip
\noindent
We will finally discuss lepton flavor violation in supersymmetric
theories (see sects. 3.3.1 and 3.3.3).
For $\mu \ra e\gamma$ the relevant equations are
 (\ref{eq:Reg15a}) and (\ref{eq:Reg15b}).
Assuming that $m_{3/2} = m_{1/2} \simeq 150 GeV$, we obtain

\beq
  R_{e\gamma} \simeq 8 \times 10^{-15}
\eeq

\noindent
Similarly, for the $\mu \ra 3e$  the branching ratio  we obtain

\beq
  R_{3e} \simeq 5 \times 10^{-18}
\eeq

\noindent
i.e. $\mu \ra e\gamma$ is favorable in this model.

\bigskip
\noindent
Let us now discuss $(\mu,e)$  conversion in supersymmetric theories.
In addition to the parameter ${\tilde \eta}$ of (\ref{eq:Reg15a}),
which contains both the effect of renormalization and the mixing angles,
we encounter all three functions $f(x)$, $g(x)$ and $f_{b}(x)$ we met in
$\mu \ra 3e$. For the experimentally interesting coherent
contribution the branching ratio is given by eq. (\ref{eq:Reg19}). By noting
that $q^2 = -m_{\mu}^2$ and $m_{\tilde u} \sim m_{\tilde e} \sim m_{3/2}$
and assuming further that the photino is the lightest particle, one
is led to eq. (\ref{eq:Reg155}).  In summary then one can write

\beq
  \frac{R_{eN}}{R_{e\gamma}} =
\frac{\alpha}{6 \pi} ( \frac{1}{3} + \frac{3}{4}
\kappa)^2 \gamma_{ph}
 \eeq

\noindent
All the relevant nuclear physics is contained in the parameters $\kappa$ and
$\gamma_{ph}$.  These parameters are given in $table\,\, 7$.  For
 comparison purposes we mention that for the nucleus $^{48}Ti$, which is of
experimental interest, the form factors become $F_Z = 0.538,
F_N = 0.506$ (QRPA) and
$F_Z = 0.550$ and $F_N = 0.520$ (Fermi distribution).
 For the reader's convenience we
also present in $table\,\, 7$ the ratio $R_{eN}/R_{e\gamma}$.
 From  this table we see that
the coherent effect of all nucleons,  tends to enhance
the ratio $R_{eN}/R_{e\gamma}$
compared to the naive estimates mentioned in the introduction.
  We see that, $R_{eN}$
still remains about two orders of magnitude less that $R_{e\gamma}$.
This, however,
can be compensated by the desirable experimental signature
of $(\mu, e)$ conversion
mentioned in sec. 2.3.

\bigskip
\noindent
It is apparent that lepton flavor, unlike strangeness, may be absolutely
conserved if the Gods of physics bestowed upon the standard model of
electroweak interactions absolute authority.
Very few people, however, subscribe to this dogma. Lepton flavor
violation follows naturally in most extensions of the standard model. Its
observation, though, is not going to come easy. The experimental efforts
have reached limits which make further improvements extremely difficult.
Worst yet the predictions of currently fashionable models are not
encouraging them. So, barring unforseen developments, such efforts may be
classified in the pursuit of nothingness. This, under the difficult
conditions of present economies may be catastrophic. But the key lies in
``unforseen circumstances''. Historically, this has been not only the rule
but the main buty of science. In any case the theoretical predictions reflect
 present
biases and should not deter the continuation of experimental efforts$^*$.


\beq
^*{\nu \upsilon \nu }\,\,{\nu \upsilon \nu }\,\,{\tau o}\,\,
{\mu \eta \delta\epsilon \nu},\,\,
\kappa\alpha\iota \,\,
\alpha \iota \epsilon \nu \,\, o\,\, \kappa o \sigma \mu o \varsigma
\,\, o\,\, \mu \iota \kappa \rho o \varsigma \,\, o\,\,
\mu \epsilon\gamma \alpha \varsigma .
\nonumber
\eeq

\bigskip



\bigskip
\noindent
\begin{thebibliography}{99}
\bibitem{HING}E. P. Hincks and B. Pontecorvo, Phys. Rev. {\bf 73} (1948) 257
\bibitem{LAGAR}A. Lagarigue and C. R. Peyrou, Acad. Sc. (Paris)
{\bf 234} (1952) 1873
\bibitem{LOKA}S. Lokanathan and J. Steinberger,
 Phys. Rev. {\bf 98} (1955) 240
\bibitem{FRAN}S. Frankel, Rare and ultrarare Muon decay, in  Muon Physics
II, Weak interactions, ed. by V. W. Hughes and C. S. Wu,
Academic Press, New York 1975, p. 83.
\bibitem{GUTS}P. Langacker and D. London, Phys. Rev. {\bf D 38} (1988) 907
\bibitem{WAL}J. W. F. Valle, Prog. Part. Nucl. Phys. {\bf 26} (1991) 91
\bibitem{GWAL}M. C. Gonzalez-Garcia and J. W. F. Valle, Mod.
Phys. Lett. {\bf A 7} (1992) 477.
\bibitem{DIM}S. Dimopoulos, S. Raby and G. L. Kane, Nucl. Phys.
{\bf B 182} (1981) 77;
S. Dimopoulos and J. Ellis, Nucl. Phys. {\bf B 182} (1981) 505
\bibitem{SCH}F. Scheck, Phys. Rep. {\bf 44} (1978) 187
\bibitem{COS}G. Costa and F. Zwirner, Riv. Nuovo Cim. {\bf 9, 3} (1986) 1
\bibitem{ENGF}R. Engfer and H. K. Walter, Ann. Rev. Nucl.
 Part. Sci. {\bf 36} (1986) 327.
\bibitem{VER}J. D. Vergados, Phys. Rep. {\bf 133} (1986) 1.
\bibitem{MELE}P.L. Melese, Comments on Nucl. Part. Phys. {\bf 19} (1989) 117.
\bibitem{HEN}C.A. Heusch,  Nucl. Phys. (Proc. Suppl.) {\bf B 13} (1990) 612.
\bibitem{HER}P. Herczeg, Rare decays, Proc. 3rd Int. Symp.
on weak and EM int. in Nuclei (Wein 1992), Dubna, Russia, 1992,
ed. Ts. D. Vylov, World Scientific, Singapure 1993, p. 262
\bibitem{SCHA} A. van der Schaaf, Prog. Part. Nucl. Phys. {\bf 31} (1993) 1.
\bibitem{LELL}L. Di Lella, in Thierty-three Years of Physics of CERN
Synchro-Cyclotron, Proc. of the SC 33 Symposium of CERN, Geneve,
 Switzerland, 22 April 1991, ed. by Fidecaro.
\bibitem{LELL2}L. Di Lella, Phys. Rep. {\bf 225} (1993) 45
\bibitem{FEI}G. Feinberg, Phys. Rev. {\bf 110} (1958) 1482
\bibitem{LAMPF}R. D. Bolton {\it et al.}, Phys. Rev. {\bf D 38}
(1988) 2077
\bibitem{KIN}W. W. Kinnison {\it et al.}, Phys. Rev. {\bf D 25} (1982) 2846
\bibitem{AMMA}J. F. Amann {\it et al.}, MEGA Collaboration, A proposal to
measure the Michel parameter $\rho$ with the MEGA positron Spectrometer,
LAMPF proposal No 1240
\bibitem{COOP}M. D. Cooper, The MEGA experiment, A search for the $\mu \ra e
\gamma$, Proc. Workshop on future Dir. in Part. and Nucl. Phys. at Multi GeV
Hadron beam facilities, BNL, Upton, N.Y., March 4-6,1993
\bibitem{BEA}A. Bean {\it et al.}, Phys. Rev. {\bf Lett. 70} (1993) 138
\bibitem{ARGUS} H. Albercht {\it et al.}, Determination of the Michel
Parameter in tau decay, DESY-90-059;
\bibitem{DELE} P. Depommier and C. Leroy, searches for lepton flavor
violation, Rep. Prog. Phys., to appear
\bibitem{BILE}S. M. Bilenky and S. T. Petcov, Rev. Mod. Phys.
{\bf 59} (1987) 671
\bibitem{BELL}U. Bellgardt {\it et al.}, (SINDRUM Collaboration),
Nucl. Phys. {\bf B 299} (1988) 1
\bibitem{BERT}W. Bertl {\it et al.}, (SINDRUM Collaboration),
Nucl. Phys. {\bf B 260} (1985) 1
\bibitem{KORE}S. M. Korechenko {\it et al.}, [Zh. Eksp. Teor. Fiz.
70 (1976) 3], Sov. Phys. {\bf JETP 43} (1976) 1
\bibitem{KoVe}T.S. Kosmas and J. D. Vergados, Nucl. Phys. {\bf A 510}
 (1990) 641.
\bibitem{BRY1}D. A. Bryman, The Time Projection Chamber AIP Conference Proc. No
108, American Institute of Physics, 1993.
\bibitem{AHM1}S. Ahmad {\it et al.}, Phys. Rev. {\bf D 38} (1988) 2102
\bibitem{BAD1}A. Badertscher {\it et al.}, (SINDRUM Collaboration),
J. of Phys. {\bf G 17} (1991) S47
\bibitem{DJIL}K. M. Djilkibaev and V. M. Lobashev, MELC Experiment to search
for the $\mu^-A \ra e^-A$ process, Moscow Meson Factory Proposal
\bibitem{MAR} G. M. Marshall {\it et al.},  Phys. Rev. {\bf D 25} (1982) 1174;
G. Marshall {\it et al.},
   3d Conf. on Part. Phys. Rockport, May 14-19 (1988) TRI-PP-88-53;
T. M. Huber {\it et al.}, Phys. Rev. {\bf D 41} (1990) 2709.
\bibitem{MUD}H. J. Mudinger  {\it et al.}, Proc. Rare Decay Symbosium
(D. A. Bryman {\it et al.}, ed.) p. 434, World Scientific, Singapore.
\bibitem{JUNG}K. Jungmann {\it et al.},
(SINDRUM Collaboration), Search for Spontaneous Conversion of muonium to
antimuonium, PSI Proposal , R-89-06.1
\bibitem{HER-MO}P. Herczeg and R. N.
Mohapatra, Phys. Rev. {\bf Lett. 69} (1992) 2475
\bibitem{MATH}B. E. Matthias  {\it et al.}, Phys. Rev. {\bf Lett. 66}
(1991) 2716
\bibitem{LEEL}A. M. Lee {\it et al.}, Phys. Rev. {\bf Lett. 64} (1990) 165.
\bibitem{CAP}C. Campagnari {\it et al.}, Phys. Rev. {\bf Lett. 61} (1988) 2062.
\bibitem{ARI}K. Arisaka {\it et al.}, Rhys. Rev. {\bf Lett. 70} (1993) 1049.
\bibitem{Cons}R. D. Cousins {\it et al.}, Phys. Rev. {\bf D 38}
(1988) 2914.
\bibitem{PLA}P. Langacker, Phys. Rep. {\bf 72} (1981) 185
\bibitem{SUSY}P. Fayet and S. Ferrara, Phys. Rep. {\bf 32} (1977) 249;
H. P. Nilles, Phys. Rep. {\bf 110} (1984) 1;
H. E. Haber and G. L. Kane, Phys. Rep. {\bf 117} (1985) 75;
A. B. Lahanas and D. V. Nanopoulos,  Phys. Rep. {\bf 145} (1987) 1.
\bibitem{NEUT}J. A. Harvey, D. B. Reiss and P. Ramond, Nucl. Phys. {\bf B 199}
(1982) 223;
G. K. Leontaris and J. D. Vergados, Phys. Lett. {\bf B 188} (1987) 455;
M. Gronau {\it et al.}, Phys. Rev. {\bf D 37} (1988) 2597;
S. Dimopoulos, L. J. Hall and S. Raby, Phys. Rev. {\bf D 45} (1992) 4192;
K. S. Babu and Q. Shafi, Phys. Lett. {\bf B 294} (1992) 235;
S. Dimopoulos, L. J. Hall and S. Raby LBL-32484/92;
H. Dreiner, G. K. Leontaris and N.D. Tracas, Mod. Phys. Lett.
{\bf A}, to appear;
P. H. Chankowski and Z. Pluciennik, Zurich, 1993, preprint ZU-TH 20/93;
K. S. Babu {\it et al.}, Univ. Delaware, preprint UDHEP-1993-03.
\bibitem{ZEE}A. Zee, Phys. Lett. {\bf B 93} (1980) 389; Phys. Lett.
{\bf B 161} (1985) 141; L. Wolfenstein, Nucl. Phys. {\bf B 175} (1980) 93
\bibitem{PET82}S. T. Petcov, Phys. Lett. {\bf B 110} (1982) 245
\bibitem{LTV}G. K. Leontaris, K. Tamvakis and J. D. Vergados,
Phys. Lett. {\bf B 162} (1985) 153.
\bibitem{slh} M. L. Swartz, Phys. Rev. {\bf D 40} (1989) 1521;
 M. J. S. Levine, Phys. Rev. {\bf D 36} (1987) 1329;
 F. Hoogeveen,  Z. Phys. {\bf C 44} (1989) 259.
\bibitem{BH}R. Barbieri and L. J. Hall, Nucl. Phys. {\bf B 364} (1991) 27
\bibitem{LVV}G. K. Leontaris, C. E. Vayonakis and J.D. Vergados,
Phys. Lett.  {\bf B 285} (1992) 91.
\bibitem{Pul}J. Pulido, Phys. Rep. {\bf 211} (1992) 167.
\bibitem{SOLAR}K. S. Hirata {\it et al.},  Phys. Rev. {\bf Lett. 65}
(1990) 1297;
K. S. Hirata {\it et al.}, Phys. Rev. {\bf D 44} (1991) 2241;
 A. I. Abazov {\it et al.}, (SAGE collaboration), Phys. Rev.
 {\bf Lett. 67} (1991) 3332;
 P. Anselmann {\it et al.}, Phys.  Lett. {\bf B 285} (1992) 376.
\bibitem{SPet} S. T. Petcov, SISSA, 70/93/EP. Talk at the Int. Workshop on
Neutrino telescopes, March 1993, Venice, to appear in the proceedings;
P. I. Krastev and S. T. Petcov, Phys. Lett. {\bf B 299} (1993) 99;
S. A. Bludman {\it et al.}, Phys. Rev. {\bf D 47} (1993) 2220;
N. Hata and P. Langacker, Univ. of Pennsylvania preprint UPR-0570T, May 1993;
G. L. Fogli, E. Lisi and D. Montanino, Preprint CERN-TH 6944/93,
BARI-TH/146-93.
\bibitem{Witten}E. Witten, Phys. Lett. {\bf B 91} (1980) 81.
\bibitem{LV91}G. K. Leontaris and J. D. Vergados, Phys. Lett. {\bf
B 258} (1991) 111.
\bibitem{2loop}S. M. Barr, E. Freire and A. Zee, Phys. Rev. {\bf
Lett. 65} (1990) 2626.
\bibitem{MEG}S. T. Petcov, Sov. J. Nucl. Phys. {\bf 25} (1977) 340;
B. W. Lee, S. Pakvasa, R. E. Shrock and H. Sugawara,
Phys. Rev. {\bf Lett. 38} (1977) 937;
J. D. Bjorken and S. Weinberg, Phys. Rev. {\bf Lett. 38} (1977) 622;
T. P. Cheng and L. F. Li, Phys. Rev. {\bf Lett. 38} (1977) 381;
T. P. Cheng and L. F. Lee. Phys. Rev. {\bf D 16} (1977) 1425;
 B. W. Lee and R. E. Shrock, Phys. Rev. {\bf D 16} (1977) 1444;
 J. D. Bjorken, K. Lane and S. Weinberg, Phys. Rev. {\bf D 16} (1977) 1474;
 A. P. Pich, A. Santamaria and J. Bernabeu, Phys. Lett. {\bf B 148} (1984) 229.
\bibitem{LV88} G. K. Leontaris and J. D. Vergados, in  ref. \cite{NEUT}
\bibitem{MEX}S. T. Petcov, Phys. Lett. {\bf B 110} (1982) 245;
Phys. Lett. {\bf B 115} (1982) 401.
K. Tamvakis and J. D. Vergados, Phys. Lett. {\bf B 155} (1985) 373;
G. K. Leontaris, K. Tamvakis and J. D. Vergados,  ref. \cite{LTV}.
\bibitem{Prima} H. Primakoff, Rev. Mod. Phys. {\bf 31} (1959) 802
\bibitem{G-P} B. Goulard and H. Primakoff, Phys. Rev. {\bf C 10} (1974) 2034
\bibitem{Bern}J. Bernabeu, E. Nardi and D. Tommasini, [UM-TH 93-08],
Nucl. Phys. {\bf B}, to appear; E. Nardi, Phys. Rev. {\bf D 48}
(1990) 1240.
\bibitem{LepQ}G. K. Leontaris and J. D. Vergados, Z. Phys.
              {\bf C 4} (1989) 905;
J. E. Hewett and T. G. Rizzo, Phys. {\bf Rep. 183} (1989) 193;
 R. Arnowitt and P. Nath, Phys. Rev. {\bf Lett 66} (1991) 2708.
\bibitem{LV83}G. K. Leontaris and J. D. Vergados, Nucl. Phys. {\bf B 224}
 (1983) 137.
\bibitem{PeSm}S. Petcov and A. Yu. Smirnof, SISSA 113/93/EP, October 1993.
 \bibitem{BW}J. Bjorken and S. Weinberg, Phys. Rev. {\bf Lett. 38} (1977) 531.
\bibitem{Chang}T. P. Cheng and M. Sher, Phys. Rev. {\bf D 35} (1987) 3484;
D. Chang, W. S. Hou and W. Y. Keung, Phys. Rev. {\bf D48} (1993)217.
\bibitem{BaZe}S. M. Barr and A. Zee, Phys. Rev. {\bf Lett. 65} (1990) 21.
 \bibitem{WS74}S. Weinberg, Phys. Rev. {\bf D 13} (1976) 974;
L. Susskind, Phys. Rev. {\bf D 20} (1979) 2619.
 \bibitem{DS79}S. Dimopoulos and L. Susskind, Nucl. Phys. {\bf B 155}
(1979) 237.
 \bibitem{AKW86}T. Appelquist D. Karabali and L. C. R. Wijewardhana,
Phys. Rev. {\bf Lett. 57} (1986) 957.
\bibitem{UNI}J. Ellis, S. Kelley and D. V. Nanopoulos, Phys. Lett.
{\bf B 249} (1990) 441;
P. Langacker and M. Luo,  Phys. Rev. {\bf D 44} (1991) 817;
U. Amaldi, W. de Boer, and H. F\"urstenau,  Phys. Lett.
{\bf B 260} (1991) 447 ; F. Anselmino, L. Cifarelli and A. Zichichi,
CERN-PPE/92-145 and CERN/LAA/MSL/92-011(July 1992);
G. G. Ross and R. G. Roberts, Nucl. Phys. {\bf B 377} (1992) 571;
 A. Vayonakis, Phys. Lett. {\bf B 307} (1993) 318.
\bibitem{GSW} M. B. Green, J. H. Schwarz and E. Witten,
Superstring Theory, (Cambridge Univ. Press, 1987), Vols. {\bf I, II}.
\bibitem{fvs}J. Ellis and D. V. Nanopoulos, Phys. Lett. {\bf B 110}
(1982) 44;
R. Barbieri and R. Gatto, Phys. Lett. {\bf B 110} (1982) 211;
J.W.F. Valle and G.G. Ross, Phys. Lett. {\bf B 151} (1982) 375.
 \bibitem{LTV86}G. K. Leontaris, K. Tamvakis and
J. D. Vergados, Phys. Lett. {\bf B 171} (1986) 412.
\bibitem{camp} B. Campbell {\it et al.}, Inter. Journ. Mod. Phys.
{\bf A 2} (1987) 831;
S. Kelley {\it et al.}, Nucl. Phys. {\bf B 358} (1991) 27;
\bibitem{KR85}L. J. Hall, V. A. Kostelecky and S. Raby, Nucl. Phys.
 {\bf B 267} (1986) 415.
\bibitem{RGFCNC}F. Gabbiani and  A. Masiero, Phys. Lett. {\bf B 209}
(1988) 289;
G. K. Leontaris, N. D. Tracas and J. D. Vergados, Phys. Lett. {\bf B 206}
(1988) 247;
F. Borzumati and A. Masiero, Phys. Rev. {\bf Lett. 57} (1986) 961.
\bibitem{KLV89}T. S. Kosmas, G. K. Leontaris and J. D. Vergados,
Phys. Lett. {\bf B 219} (1989) 419
 \bibitem{MB} B. R. Greene {\it et al.}, Nucl. Phys. {\bf B 278} (1986) 667;
 {\bf B 292} (1987) 606;
P. Nath and R. Arnowitt, Phys. Rev. {\bf D 39} (1989) 2006;
 L. E. Ibaniez {\it et al.}, Phys. Lett {\bf B 191} (1987) 282;
 A. Font {\it et al.}, Nucl. Phys. {\bf B 331} (1990) 421;
 D. Bailin, A. Love and S. Thomas, Phys. Lett. {\bf B 194} (1987) 385;
 J. A. Casas, E. K. Katehou and C. Munoz, Nucl. Phys. {\bf B 317} (1989) 171;
G. Lazarides, P. K. Mohapatra,
C. Panagiotakopoulos and Q. Shafi, Nucl.Phys. {\bf B 323} (1989) 614
\bibitem{ant} I. Antoniadis, J. Ellis, J. S. Hagelin and D. V. Nanopoulos,
Phys. Lett. {\bf B 194} (1987) 231; Phys. Lett. {\bf B 231} (1989) 65
\bibitem{al} I. Antoniadis and G. K. Leontaris, Phys.
Lett. {\bf B 216} (1989) 333;
I. Antoniadis, G. K. Leontaris and J. Rizos, Phys. Lett. {\bf B 245}
 (1990) 161.
\bibitem{flny}A. E. Farragi, D. V. Nanopoulos and K. Yuan,
Nucl. Phys. {\bf B 335} (1990) 347;
A. E. Faraggi, Nucl. Phys. {\bf B 387} (1992) 239;
F. Gabbiani and A. Masiero, Phys. Lett. {\bf B 209} (1988) 289.
\bibitem{MSTR}V. S. Kaplunovski, Nucl. Phys. {\bf B 307} (1988) 145;
I. Antoniadis, J. Ellis, R. Lacaze and D. V. Nanopoulos, Phys. Lett. {\bf
B 268} (1991) 188;
S. Kalara, J. L. Lopez and D. V. Nanopoulos,
Phys. Lett. {\bf B 269} (1991) 84;
L. E. Iban\~ez, D. L\"ust and G. G. Ross, Phys. Lett. {\bf B 272}
(1991) 251;
 I. Antoniadis {\it et al.}, Phys. Lett. {\bf B 279} (1992) 58;
 G. K. Leontaris and N. D. Tracas, Phys. Lett. {\bf B 291} (1991) 44;
  D. Bailin and A. Love, Phys. Lett. {\bf B 292} (1992) 315.
\bibitem{feln}A. Farragi, J. L. Lopez, D. V. Nanopoulos and K.
Yuan, Phys. Lett. {\bf B 221} (1983) 337;
S. Kelley, J. Lopez, D. V. Nanopoulos and H. Pois, Nucl. Phys.
{\bf B 358} (1991) 27
\bibitem{hkt}J. Hagelin, S. Kelley and Tanaka, MIU-1992, preprint
\bibitem{lt} G. K. Leontaris, Z. Phys. {\bf C 53} (1992) 287,
G. K. Leontaris and N. D. Tracas, Z. Phys. {\bf C 56} (1992) 479
\bibitem{GSS} G. K. Leontaris, Phys. Lett. {\bf B 207} (1988) 447;
 G. K. Leontaris and D. V. Nanopoulos, Phys. Lett. {\bf B 212} (1988) 327;
 G. K. Leontaris and C. E. Vayonakis, Phys. Lett. {\bf B 206} (1988) 271;
 G. K. Leontaris and K. Tamvakis, Phys. Lett. {\bf B 224} (1989) 319;
 S. A. Abel, Phys. Lett.  {\bf B 234} (1990) 113;
 A. E. Faraggi, Phys. Lett. {\bf B 245} (1990) 435;
 G. K. Leontaris and J. D. Vergados, Phys. Lett. {\bf B 258} (1991) 111;
I. Antoniadis, J. Rizos and K. Tamvakis, Phys. Lett. {\bf B 279} (1992) 281;
E. Papageorgiu and S. Ranfone, Phys. Lett. {\bf B 282}
(1992) 89;Phys. Lett. {\bf B 295}
(1992) 79;S. Ranfone, UAB-FT-324; E. Papageorgiu, preprint, Orsay, 1993;
 J. Ellis, J. L. Lopez and D. V. Nanopoulos, Phys. Lett. {\bf B 292}
 (1992) 189.
\bibitem{ENO}J. Ellis, J. L. Lopez, D. V. Nanopoulos and K. A. Olive,
 Phys. Lett. {\bf B 308} (1993) 70.
\bibitem{LV93}G. K. Leontaris and J. D. Vergados, Phys. Lett. {\bf B 305}
  (1990) 242.
\bibitem{CK}C. Kounnas, The Supersymmetry - breaking mechanism
in the string induced no-scale supergravities. CERN-TH preprint, July
1993, and references therein;
F. Zwirner, The quest of low energy supersymmetry and the role of
high energy $e^+e^-$ colliders, CERN-TH.6357/93.
\bibitem{GKL} G. K. Leontaris, [IOA - 95/1993],  Phys. Lett.
{\bf B }, to appear
\bibitem{Ford} K. W. Ford and J. G. Wills, Nucl. Phys. {\bf 35}
(1962) 295;
 J. C. Sens, Phys. Rev. {\bf 113} (1959) 679.
\bibitem{WeiFei} S. Weinberg and G. Feinberg, Phys. Rev.
 {\bf Lett. 3} (1959) 111
\bibitem{Shank}O. Shanker, Phys. Rev {\bf D 20} (1979) 1608
\bibitem{Chiang}H. C. Chiang, E. Oset, T. S. Kosmas, Amand Faessler and
 J. D. Vergados, Nucl. Phys. {\bf A 559} (1993) 526
\bibitem{KV88} T. S. Kosmas and J.D. Vergados, Phys. Lett. {\bf B 215}
(1988) 460.
\bibitem{KVCF}T. S. Kosmas, J. D. Vergados, O. Civitarese and A. Fassler,
 Nucl. Phys. {\bf A}, in press
\bibitem{Vries} H. de Vries, C. W. de Jager and C. de Vries, Atomic Data
 and Nuclear Data Tables, {\bf 36} (1987) 495.
\bibitem{Oset1} H. C. Chiang, E. Oset and P. Fernandez de Cordoba, Nucl. Phys.
 {\bf A 510} (1990) 591.
\bibitem{Oset2} H. C. Chiang, E. Oset, R. C. Carrasco, J. Nieves
and J. Navarro, Nucl. Phys. {\bf A 510} (1990) 573.
\bibitem{Suzu}T. Suzuki, D. F. Measday and J. P. Roalsvig, Phys. Rev.
 {\bf C 35} (1987) 2212.
\bibitem{Oset3}C. Garcia-Recio, J. Nieves and E. Oset, Nucl. Phys.
 {\bf A 547} (1992) 473.
\bibitem{KV92} T. S. Kosmas and J. D. Vergados, Nucl. Phys.
 {\bf A 523} (1992) 72
\bibitem{Ring} P. Ring and P. Schuck, The nuclear many-body problem,
 (Springer-Verlag, 1980)
\bibitem{Row1} D. J. Rowe, Phys. Rev. {\bf 175} (1968) 1283
\bibitem{Row2} D. J. Rowe, Nuclear collective motion,
 (Methuen and CO. LTD., 1970).
\bibitem{Sande}E. A. Sanderson, Phys. Lett. {\bf 19} (1965) 141;
 J. Da Providencia, Phys. Lett. {\bf 21} (1966) 668
\bibitem{PElli}P. J. Ellis, Nucl. Phys. {\bf A 467} (1987) 173
\bibitem{McNe}J. A. McNeil, C. E. Price and J. R. Shepard,
Phys. Rev. {\bf C 42} (1990) 2442
\bibitem{Civi}O. Civitarese, Amand Faessler and T. Tomoda, Phys. Lett.
 {\bf B 194} (1987) 11
\bibitem{KV89} T. S. Kosmas and J. D. Vergados, Phys. Lett. {\bf
B 217} (1989) 19.
\bibitem{BoMo}A. Bohr and B. R. Mottelson, Nuclear Structure, (Benjamin
Inc., 1969) Vol. I, p. 313
\end{thebibliography}
\end{document}